\documentclass[modern]{aastex63}
\newcommand{\kms}{\ensuremath{\,\rm{km\,s^{-1}}}}
\newcommand{\lt}{\ensuremath{<}} 
\newcommand{\gt}{\ensuremath{>}} 
\shorttitle{ Circumstellar Gas Surrounding 51~Oph}
\shortauthors{Jenkins \& Gry}
\usepackage[utf8]{inputenc}
\usepackage{epsfig}
\begin{document}
\title{The Composition, Excitation, and Physical State of Atomic Gas in the Debris Disk 
Surrounding \object{51 Oph}
\footnote{Based on observations with the NASA/ESA Hubble Space Telescope obtained from 
the Data Archive at the Space Telescope Science Institute, which is operated by the Associations 
of Universities for Research in Astronomy, Incorporated, under NASA contract NAS5-26555.~ 
\copyright 2020. The American Astronomical Society. All rights reserved.}}
\author[0000-0003-1892-4423]{Edward B. Jenkins}
\affiliation{Princeton University Observatory,\\ Princeton, NJ 08544}
\author[0000-0003-0788-7452]{Cécile Gry}
\affiliation{Aix Marseille Univ., CNRS, CNES, LAM,\\ Marseille, France}
\correspondingauthor{E. B. Jenkins}
\email{ebj@astro.princeton.edu}
\email{cecile.gry@lam.fr}
\begin{abstract}
We measured 304 absorption features in the ultraviolet and visible spectra of the star 51~Oph, 
which is known to have a debris disk with a high inclination.  We analyzed the relative 
populations of atoms in excited fine-structure and metastable levels that are maintained by 
optical pumping and collisional excitation by electrons, and we found that most of the gas is 
situated at about 6\,AU from the star, has an electron volume density $10^5 \lt n(e)\lt 3\times 
10^6{\rm cm}^{-3}$, and a temperature $T=8000$\,K.  Our interpretations reveal that the gas is 
partly ionized, has a column density of neutral hydrogen equal to $10^{21}\,{\rm cm}^{-2}$, 
and has a composition similar to that of a mildly depleted interstellar medium or that of 
Jupiter-family comets. Compared to results for disks around some other stars, such as 
$\beta$~Pic and 49~Cet, we find surprisingly little neutral carbon.  No molecular features were 
detected, which indicates that our line of sight misses the molecule-rich central plane of the 
disk.  The tilt of the disk is also validated by our being able to detect resonant scattering of the 
starlight by oxygen atoms.
\end{abstract}

\section{Introduction}\label{sec:introduction}
The emergence and growth in our understanding of gas and dust in orbit around stars that are 
approaching or are on the main sequence has been an important new endeavor for both 
observers and theoreticians (as reviewed by Hughes et al. 2018).  This topic has also provided a 
critical link to our knowledge on the formation of extrasolar planets (Vidal-Madjar et al. 1998 ; 
Wyatt 2018) and alterations of element abundances in some stellar atmospheres, such as those 
of white dwarf stars (Jura \& Young 2014).  Discoveries from orbiting observatories, such as the 
{\it Infrared Space Observatory (ISO), Spitzer Space Telescope, Wide Field Infrared Survey 
Explorer (WISE),\/} and {\it Herschel Space Observatory\/} have revealed that approximately 
20\% of both A-type and the F, G, and K-type stars near the main sequence exhibit measurable 
excesses of infrared emission\footnote{An unbiased survey of A-type stars by Thureau et al. 
(2014) indicates a somewhat higher percentage, ranging between 21 and 54\%. A study by 
Moór et al. (2017) of A-type stars with dust-rich disks revealed that 11 out of 16 of them 
exhibited measurable CO emission.} that are well above an extrapolation of the
Rayleigh-Jeans tails for the radiation from the stellar photospheres (Wyatt 2018).  

Coronagraphic,  polarimetric, and interferometric observing facilities working at visible 
wavelengths (e.g., STIS on HST, SPHERE/ZIMPOL on the VLT, and the Keck nulling 
interferometer), as well as radio interferometers (ALMA, SMA, NOEMA, CARMA), the far 
infrared PACS instrument on the {\it Herschel Space Observatory\/}, and millimeter 
observations taken by SCUBA-2 on the JCMT have allowed us to obtain detailed information on 
the strengths and morphologies of the emissions by atoms, molecules, and dust surrounding 
the stars. There is an immense volume of literature covering these observations.   There are 
also many publications that have highlighted the absorption features produced by \ion{Ca}{2}, 
\ion{Na}{1} and sometimes \ion{Fe}{1} that appear at visible wavelengths in the spectra of the 
central stars.  These species have even been imaged in emission for the iconic example of an 
edge-on disk around the star $\beta$~Pic (Olofsson et al. 2001 ; Brandeker et al. 2004).

If a central star is hot enough and close enough to provide a spectrum in the ultraviolet, an 
opportunity to detect with great sensitivity various atomic and molecular constituents in the 
circumstellar medium is presented.  Intensive investigations of the matter surrounding 
$\beta$~Pic got underway with observations using the {\it International Ultraviolet Explorer 
(IUE)\/} (Kondo \& Bruhweiler 1985 ; Lagrange et al. 1987 ; Lagrange-Henri et al. 1988 ; 
Lagrange-Henri et al. 1989).  These early investigations highlighted the existence of atoms in a 
stable component at the velocity of the star, but in addition, there were occasional 
appearances of absorption components at significantly large positive velocities, which were 
interpreted as falling evaporating bodies (FEB) (Vidal-Madjar et al. 1998).  The high frequency of 
detecting the FEB phenomenon has been interpreted to arise from perturbations by planets on 
the orbits of comet-sized objects (Karmann et al. 2001 ; Thébault \& Beust 2001), which 
eventually plunge toward the star.

Following the era of {\it IUE}, observations with the Goddard High Resolution Spectrograph 
(GHRS) (Brandt et al. 1994) and  Space Telescope Imaging Spectrograph (STIS) (Woodgate et al. 
1998) aboard {\it Hubble Space Telescope\/} ({\it HST\/}), together with the {\it Far Ultraviolet 
Spectroscopic Explorer\/} ({\it FUSE\/}) (Moos et al. 2000 ; Sahnow et al. 2000), provided a vast 
improvement in the number and quality of the results.  Absorption features associated with the 
debris disk of $\beta$~Pic have been the most intensively studied ones, but to lesser degrees, 
there have been interpretations of the spectra of other hot stars, such as 49~Cet (Malamut et 
al. 2014 ; Roberge et al. 2014 ; Miles et al. 2016 - STIS spectra), 51~Oph, HD\,256, HD\,42111 
(Lecavelier des Etangs et al. 1997b - GHRS spectra only),  2~And (Cheng \& Neff 2003 - FUSE and 
GHRS), $\sigma$~Her (Chen \& Jura 2003 - FUSE only),  AB~Aur (Roberge et al. 2001 - FUSE and 
STIS), and HD\,109573 (upper limits only) (Chen \& Kamp 2004 - FUSE and STIS), HD\,32297 
(marginal S/N) (Fusco et al. 2013), HD\,141569 (Malamut et al. 2014), HD\,163296 (Tilling et al. 
2012 - STIS medium res. echelle) and HD\,172555
(Grady et al. 2018 - STIS and COS).

The origin of gas accompanying debris disks is a fundamental problem: is it primordial, is it 
ejected from the star (and slowed down by some process), or – more likely -- does it arise from 
the orbiting planetesimals,  comets, or dust grains that are either evaporating atoms or 
liberating gas as they collide with each other?  An examination of the relative abundances of 
different elements, which often differ appreciably from the solar abundances, can provide 
some guidance on this question (Xie et al. 2013).  

A distinctive property of the UV absorption spectra of gas associated with debris disks is the 
appearance of features arising from many different excited fine-structure and metastable levels 
of various atoms and ions.  The most conspicuous absorptions from levels of high excitation are 
those from \ion{Fe}{2}, as we will illustrate later in Figure~\ref{fig:excited_lines2}, which have 
radiative lifetimes ranging from 10 seconds to 50 hours (Quinet et al. 1996) and must be 
populated by radiation pumping  and/or collisional excitations with electrons (the effects of 
collisions with neutral atoms and ions are small by comparison).  For densities below the critical 
densities of the levels, there can be deviations from a straight-line relationship for the level 
populations in an excitation diagram ($\ln(N/g)$ vs. $E/k$), which in turn will reveal clues on 
the environmental parameters that are responsible for the excitations.  Such deviations are 
clearly evident in this type of diagram for the \ion{Fe}{1} level populations derived from a
high-S/N spectrum of $\beta$~Pic at visible wavelengths (Vidal-Madjar et al. 2017).  

Our current study is directed toward using absorption features in the ultraviolet to gain 
information on the gas constituents that surround the star 51~Oph.  This Herbig Ae/Be star has 
been assigned a spectral classification B9.5Ve by Dunkin et al. (1997), and has the parameters 
$M=3.3M_\odot$ (Jamialahmadi et al. 2015), $T_{\rm eff}=10,250\,$K, $\log g=3.57$, 
[M/H]~=~+0.10, and an estimated age $0.7^{+0.4}_{-0.5}\,{\rm Myr}$ (Montesinos et al. 2009).  
The distance to the star is 123\,pc (Arenou et al. 2018 ; Luri et al. 2018). A spectral energy 
distribution with an infrared excess (Malfait et al. 1998), the presence of emissions at 
$10\,\mu{\rm m}$ from silicates (Fajardo-Acosta et al. 1993 ; Meeus et al. 2001), $7.7-
8.2\,\mu{\rm m}$ from PAHs (Keller et al. 2008), several atomic excited atomic fine-structure 
levels (Meeus et al. 2012 ; Dent et al. 2013), and various gas-phase molecules (van den Ancker 
et al. 2001) all present clear evidence of gaseous and solid matter circulating around this star.  
We are not the first to study absorption features in the UV spectrum of this star, and we will 
refer to a number of previous investigations later in the paper. The UV spectrum of 51~Oph 
exhibits a rich assortment of lines from many different elements, and these elements have 
significant populations in those metastable levels that cannot undergo rapid radiative decays.  
We had the good fortune of locating relevant atomic data for \ion{N}{1}, \ion{Fe}{2}, and 
\ion{Ni}{2} that allowed us to investigate in some detail the factors that governed the excited 
level populations.

\section{Data}\label{sec:data}
Table~\ref{tbl:observations} summarizes some key parameters of the observations.  Nearly all 
of the conclusions in this paper are based on observations taken at ultraviolet wavelengths with 
the highest resolution echelle modes of STIS aboard {\it HST\/}.  These data are supplemented 
by observations by the GHRS over very limited wavelength ranges.  The spectra recorded by the 
GHRS are used only to make comparisons with STIS data to check for possible time variability of 
some spectral features.  A limited number of absorption features relevant to our study appear 
at visible wavelengths.  To measure such features, we made use of archived echelle spectra 
recorded by the Ultraviolet and Visual Echelle Spectrograph (UVES) (Dekker et al. 2000) on the 
{\it Kueyen Very Large Telescope\/} ({\it VLT\/}) operated by the European Southern 
Observatory.

Observations of 51~Oph in the far ultraviolet recorded by {\it FUSE\/}  have been reported by 
Roberge et al. (2002).  We confined our use of the {\it FUSE\/} spectra to probe only the 
elements \ion{N}{1}, \ion{Cl}{2}, \ion{P}{2} and a metastable level of \ion{O}{1}, which we could 
not measure in the STIS spectrum.  Finally, we used a spectrum recorded by the {\it Hopkins 
Ultraviolet Telescope\/} ({\it HUT\/}) (Kruk et al. 1995 ; Dixon et al. 2013) to validate our choice 
of parameters for a theoretical stellar spectrum that we used to calculate the optical pumping 
of levels, as will be discussed later in Section~\ref{sec: stellar_flux}.
\begin{deluxetable}{
r	
c	
c	
c	
c	
}
\tablewidth{0pt}
\tablecaption{Observations\label{tbl:observations}}
\tablehead{
\colhead{Instrument, Mode} & \colhead{Aperture} & \colhead{Resolution} &
\colhead{$\lambda$ Range} & \colhead{Observation}\\
\colhead{} & \colhead{arc-sec} & \colhead{$\lambda/\Delta\lambda$} &
\colhead{(\AA)} & \colhead{Date}
}
\startdata
STIS E140H, E230H&$0.2\times 0.2$&114,000&1280.5$-$1902.0&2003 May 26\\
&$0.1\times 0.03$&200,000&1879.5$-$2396.5\\
&&&2576.5$-$2845.5\\
\tableline
GHRS ECH-B&0.25&100,000&2597.2$-$2610.0&1996 Feb 21\\
&&&2793.3$-$2808.0\\
G160M&&20,000&1531.5$-$1567.5\\
\tableline
UVES CD\#2&0.612&72,000&3300$-$4500&2007 May$-$Aug\\
CD\#3&&&4650$-$6650\\
\tableline
FUSE&$4\times20$&15,000&992$-$1187&2001 Apr 29\tablenotemark{a}\\
\tableline
HUT&20&500&800$-$1850&1995 Mar 15\\
\enddata
\tablenotetext{a}{We refrained from using exposures taken in 2000 with a larger entrance 
aperture.  Also, various technical issues discussed by Roberge et al. (2002) made the reduction 
and interpretation of these earlier spectra less attractive.}
\end{deluxetable}

\begin{figure}
\plotone{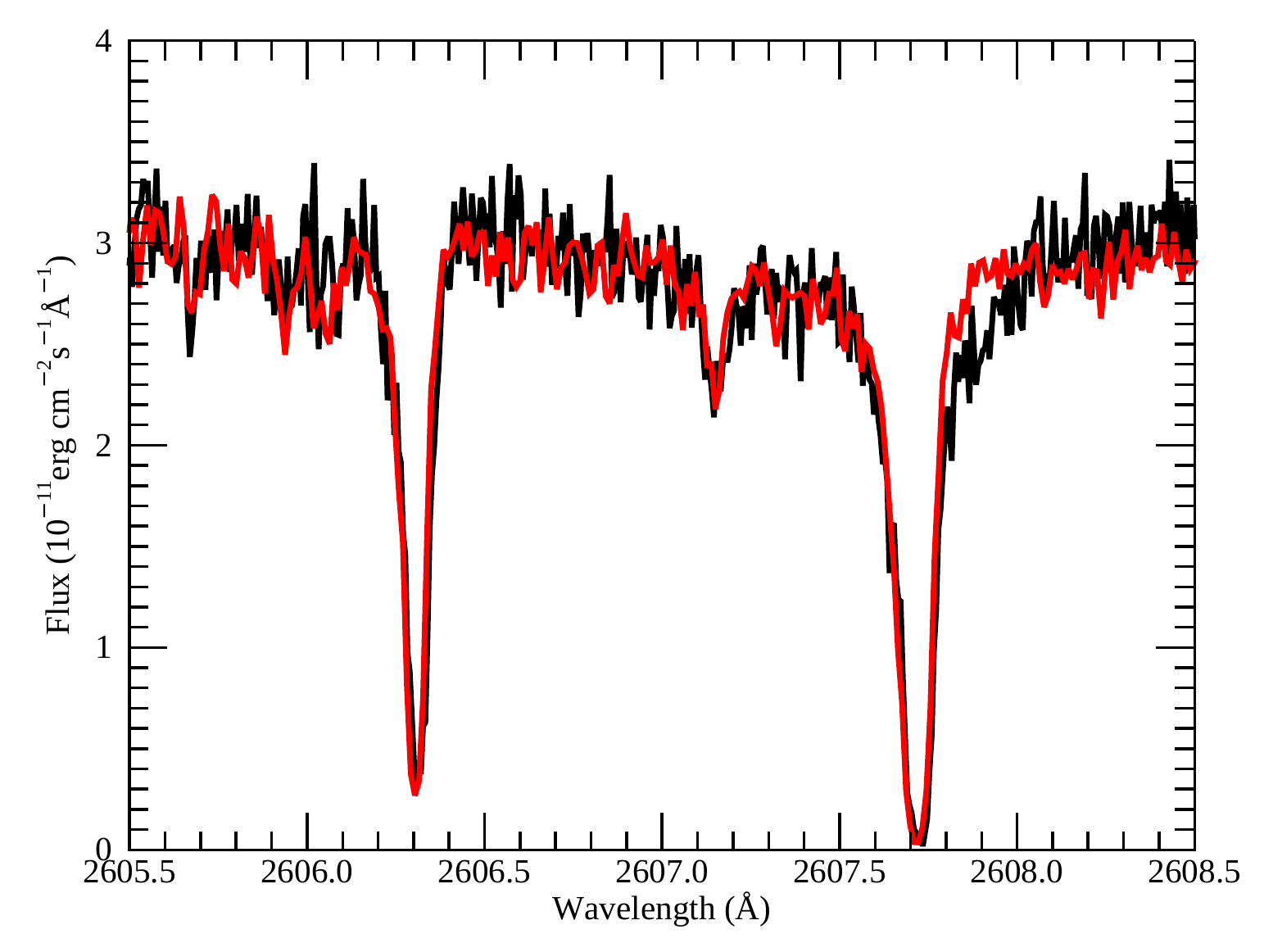}
\caption{Spectra recorded at different times by the GHRS ECH-B mode (black trace) and STIS 
E230H mode (red trace).  The absorption on the left is from the ground state of \ion{Mn}{2}, 
while the two features to the right are from metastable levels of \ion{Fe}{2}.  Except for a 
slightly enhanced absorption on the long-wavelength side of the very strong line line at 
2607.7\,\AA\ from an excited fine-structure level of \ion{Fe}{2} in the GHRS spectrum, the 
spectra appear to be identical.  These absorption features and their identifications can be seen 
in the upper panel of Figure~\protect\ref{fig:excited_lines2}.\label{fig:comparison}}
\end{figure}

It is well established that absorption features arising from circumstellar gas can vary with time, 
with the appearance and disappearance of velocity-shifted features.  A large collection of 
51~Oph spectra recorded by the {\it IUE\/} satellite at different times show occasional 
appearances of features at very high negative and positive velocities, ranging from many tens of 
\kms\ to as much as 100\kms\ (Grady \& Silvis 1993).  All of our observations with STIS were 
taken over a time interval of only two hours, so the possible influence of such variability is 
small.  Nevertheless, we also made use of UVES spectra at visible wavelengths taken at a very 
different time for our analysis of \ion{Na}{1}, \ion{Ca}{1}, \ion{Ca}{2}, \ion{Ti}{2}, \ion{Cr}{2}, 
\ion{Mn}{2}, and \ion{Fe}{1}.

To gain some indication that the features at low velocities that we studied here are not varying 
in time, we compared GHRS observations taken many years earlier than our STIS ones (see 
Table~\ref{tbl:observations}) and found no evidence of any significant changes in the 
appearances of the features.    Figure~\ref{fig:comparison} shows the close match of features in 
the GHRS and STIS spectra taken at nearly the same wavelength resolutions.  Of course, this 
evidence is only circumstantial; we cannot rigorously rule out the existence of some variability 
that could cause features in the UVES spectra to not match those in the STIS ones.  For instance, 
Roberge et al. (2002) reported some very small changes in some profile shapes that occurred 
between their two {\it FUSE\/} exposures separated by 6 days in 2000.

\section{Analysis of the Spectra}\label{sec:analysis}
\subsection{Absorption Features from Atomic Ground-State, Excited Fine-structure, and 
Metastable Levels}\label{sec:excited_features}
Table~\ref{tbl:atomic_lines} in the appendix lists the different features that we analyzed in the 
UV and visible spectra of 51~Oph.  Most were strong enough to appear well above the noise in 
the continuum level and could yield column density values, while others were so weak or 
invisible that we could obtain only upper limits to column densities using a procedure outlined 
below in Section~\ref{sec:upper_limits}. 

Features from the unexcited states of atoms probably have some contamination from 
absorption by the foreground interstellar medium, but all of the lines from the excited levels 
arise from just the circumstellar gas around 51~Oph.  For the ground-state levels, we attempted 
to distinguish contributions from the foreground interstellar medium (ISM) from circumstellar 
ones, using the identifications of distinct radial velocity components as a guide, as we will 
discuss in Section~\ref{sec:velocity_components}.   To accomplish this goal,  we used the 
profile-fitting software called {\tt Owens.f} that was developed in the 1990s by M.~Lemoine 
and the French {\it FUSE\/} science team.  To obtain a more coherent picture of the 
components, the analysis software allows us to fit simultaneously different lines from several 
energy levels and include several species in a single structure of velocity components.  A 
solution emerges where all central velocities and turbulent velocity widths for all species of the 
same component are identical, but with thermal broadening widths that recognize the 
expected variations that change with atomic masses.  In some cases, we sensed that two 
velocity components for purely circumstellar gas were partly resolved, but in others, only one 
was present.  

{\tt Owens} constructs trial theoretical profiles that are then convolved with the instrumental 
line spread function (LSF). The spectra have not been normalized to the stellar continuum; 
instead, the stellar profile is included in the fit as $n+1$ free parameters for an $n$-degree 
polynomial ($n$ is always $\leq$ 3 and often equal to 1).  The LSFs appropriate for the echelle 
gratings E140H and E230H are tabulated in the STIS Instrument Handbook (Riley et al. 2018).  
For the observations of 51~Oph, E140H data (1460\,\AA\ to 1896\,\AA) were observed with the 
$0.2\times 0.2$\,arc-sec aperture, and E230H data (1874 to 2846\,\AA) through the
high-resolution slit with dimensions $0.1\times 0.03$\,arc-sec.  For both slits, the LSFs are 
composed of a broad and a narrow component whose full widths at half maxima (FWHM) are 
derived by performing a double-Gaussian fit to the tabulated LSFs.  As a result, we adopted for 
the low-amplitude, broad component an FWHM of 4.9 pixels and 5.1 pixels for E140H and 
E230H data, respectively, and for the taller narrow component, an FWHM of 1.26 pixels for 
E140H and 1.45 pixels for E230H.  For the UVES data, we have adopted a Gaussian LSF as 
specified by Figure~2.7 of the UVES User Manual (Sbordone \& Ledoux 2019): 
FWHM~=~2.4\,pixels ($\sim 2.4\lambda/193,000$) for the blue side and FWHM~=~3.5\,pixels 
($\sim 3.5\lambda/250,000$) for the red side. 

The wavelength accuracy of STIS across exposures should be $0.2-0.5$ pixels, according to the 
STIS Instrument Handbook.   Because we use spectra from several facilities with different 
characteristics, we introduce a relative velocity shift as a free parameter for each wavelength 
window included in the fit to account for the possible wavelength displacement of one window 
relative to the others.  The derived velocity shifts have a dispersion of about 0.5\,\kms, 
effectively slightly less than 0.5 STIS pixels (0.66\,\kms\ for the E230H data). However, we note 
the existence of a general shift of 0.6$-$0.8\kms\ between lines in the E230H spectrum and 
lines in the E140H spectrum, and another 0.2\kms\ shift relative to the UVES spectra. Because 
most of the neutral lines are observed with the E140H  spectrum (and UVES) whereas most of 
the \ion{Fe}{2} and \ion{Ni}{2} lines are in the E230H spectrum, this could suggest a slight 
velocity shift between neutrals and ions. However, we think it is not the case because the same 
velocity shift is observed for the few \ion{Fe}{2} lines observed in the E140H spectrum.  In 
short, we regard that the absolute velocity in the STIS spectra can have an error of at least 
0.8\kms, and we conclude that the main disk component velocity does not differ significantly 
for all studied elements. 

\begin{figure}
\epsscale{1.2}
\plotone{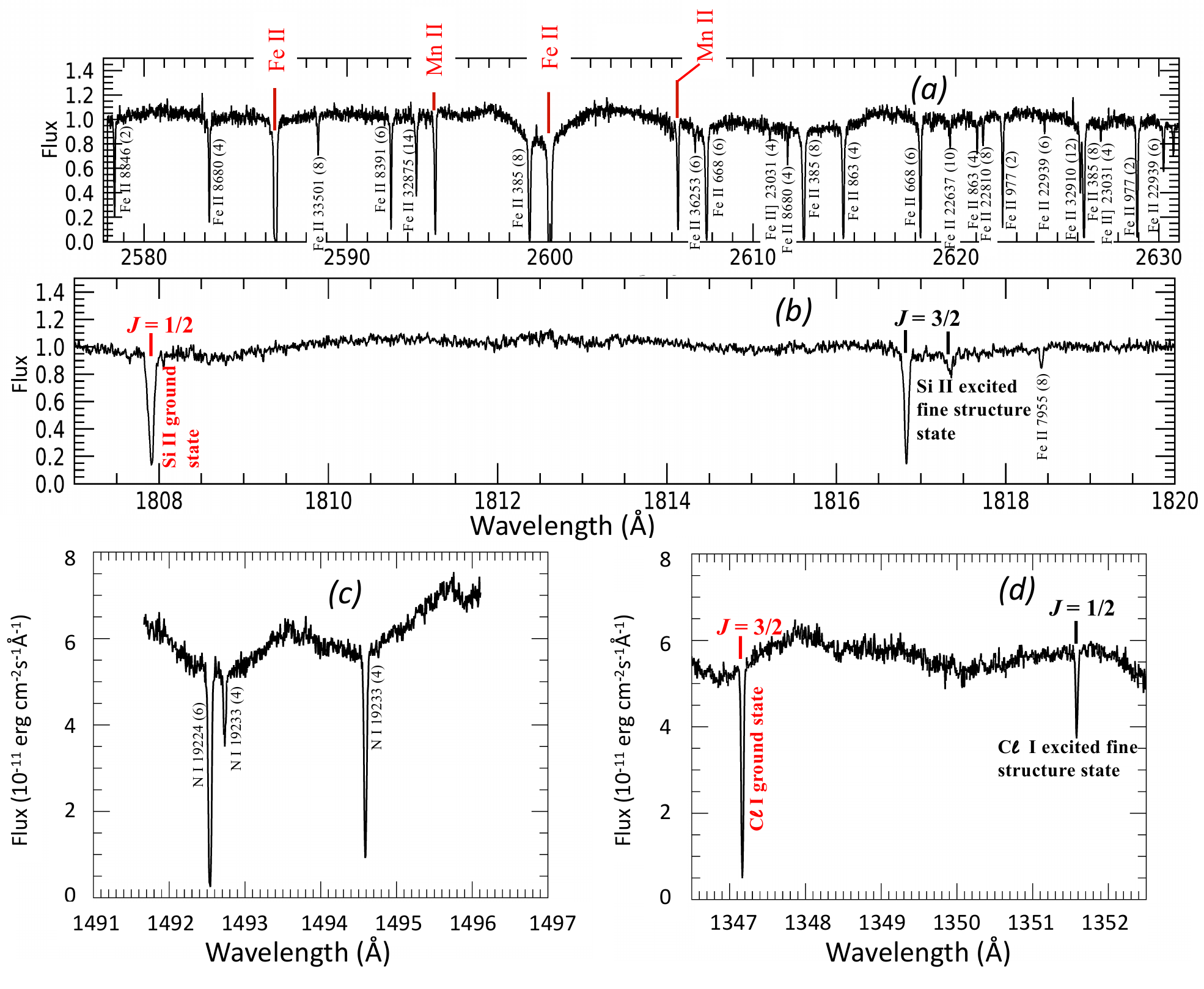}
\caption{Noteworthy circumstellar features in the spectrum of 51~Oph. Absorptions from 
metastable levels are identified with the relevant atom or ion, along with the excitation energy 
(in ${\rm cm}^{-1}$) and the degeneracy of the lower level.  Features identified in red text 
indicate absorptions from the ground states, which may suffer some contamination from 
foreground interstellar gas.  Panel {\it (a)\/} shows a portion of the spectrum dominated by 
lines from \ion{Fe}{2} metastable levels.  Panels {\it (b)-(d)\/} show features from the two
fine-structure levels of \ion{Si}{2}, the metastable levels of \ion{N}{1}, and the two
fine-structure levels of \ion{Cl}{1}, respectively.\label{fig:excited_lines2}}
\end{figure}

Figure~\ref{fig:excited_lines2} shows some representative pieces of the STIS echelle spectrum 
with line identifications.  Table~\ref{tbl:col_dens} lists our results for the column densities.  We 
estimated our measurement uncertainties for the lines using the $\Delta \chi^2$ method 
described by Hébrard et al. (2002) .  We then evaluated the uncertainties in $\log N$ by 
combining the $f$-value and measurement uncertainties in quadrature.  Many of the column 
density outcomes were based on a collection of features with different uncertainties in their 
$f$-values.  In such cases, we gave the most weight to the $f$-value uncertainties of the lines 
that were neither very weak nor strongly saturated, as these medium-strength lines are the 
ones that are most influential in determining the column densities.  In addition, we made a 
worst-case assumption that the uncertainties of many $f$-values derived from a single source 
had systematic errors that all had the same direction and amplitude (i.e., we did not consider 
that errors for different lines would tend to cancel each other).

\subsection{Determinations of Upper Limits for Column Densities}\label{sec:upper_limits}

Several important species show either extremely marginal detections or no discernible 
features.  For such cases, we did not attempt to evaluate profile fits.  Instead, we carried out 
formal measurements of equivalent widths $W_\lambda$ over the expected spans of the 
features and assumed that the column densities scaled in direct proportion to $W_\lambda$, 
i.e. the absorptions were completely unsaturated.  Our initial outcomes for upper limits were 
expressed in terms of a formal value for $W_\lambda$ plus a $1\sigma$ positive excursion.  
However, our final expressions went beyond these determinations and stated column density 
upper limits in terms of a modification of this initial upper bound that made use of a Bayesian 
prior that acknowledges that negative real line strengths are not allowed.  This avoids the 
absurd result that arises when random positive intensity fluctuations or a continuum placement 
that was too low creates a raw answer for $W_\lambda$  that is $\leq -1\sigma$ uncertainty in 
its value.  The details of this calculation and expositions on how it behaves are given in 
Appendix D of Bowen et al. (2008).  This method yields a seamless transition to a conventional 
expression of a measured value plus its $1\sigma$ upper error bar when the outcomes progress 
to higher levels of significance. 
\newpage
\startlongtable
\begin{deluxetable}{
c	
c	
c	
c	
}
\tablewidth{0pt}
\tablecolumns{4}
\tablecaption{Column Densities\label{tbl:col_dens}}
\tablehead{
\colhead{Species} & \colhead{Excitation} & \colhead{g} & \colhead{$\log N$}\\
\colhead{} & \colhead{Energy (${\rm cm}^{-1}$)} & \colhead{} & \colhead{$({\rm cm}^{-2})$}
}
\startdata
%
%
\ion{C}{1}&    0& 1&12.33\tablenotemark{a}\\ 
&   43& 5&$<$12.15\\ %
\ion{C}{2}&    0& 2&$<$18.21\\ 
\ion{N}{1}&    0& 4&16.60$\pm 0.20$\tablenotemark{b}\\ 
&19224& 6&14.03$\pm 0.10$\\ 
&19233& 4&13.82$\pm 0.06$\\ 
&28838& 2&12.79$\pm 0.08$\\ 
&28839& 4&13.16$\pm 0.08$\\ 
\ion{O}{1}&    0& 5&17.37$\pm 0.06$\\ 
&15868& 5&$>$14.30\\ 
\ion{Na}{1}&    0& 2&10.47$\pm 0.04$\\ 
\ion{Mg}{1}&    0& 1&$<$12.09\\ 
\ion{Si}{1}&    0& 1&$<$11.52\\ 
\ion{Si}{2}&    0& 2&15.27$\pm 0.15$\\ 
&  287& 4&15.37$\pm 0.15$\\ 
\ion{P}{1}&    0& 4&$<$11.63\\ 
\ion{P}{2}&    0& 1&13.20$\pm 0.03$\tablenotemark{b,c}\\ 
&  164& 3&12.73$\pm 0.13$\tablenotemark{c}\\ 
&  469& 5&13.11$\pm 0.04$\tablenotemark{c}\\ 
\ion{S}{1}&    0& 5&$<$11.74\\ 
\ion{Cl}{1}&    0& 4&13.60$\pm 0.14$\\ 
&  882& 2&12.86$\pm 0.09$\\ 
\ion{Cl}{2}&    0& 5&$<$13.64\\ 
\ion{Ca}{1}&    0& 1&$<$ 8.45\\ 
\ion{Ca}{2}&    0& 2&11.02$\pm 0.06$\tablenotemark{d}\\ 
\ion{Ti}{2}&    0& 4&10.91$\pm 0.04$\\ 
\ion{Cr}{2}&    0& 6&12.97$\pm 0.06$\\ 
&11961& 2&11.44$\pm 0.30$\\ 
&12032& 4&12.04$\pm 0.11$\\ 
&12147& 6&12.11$\pm 0.11$\\ 
&12303& 8&12.29$\pm 0.06$\\ 
&12496&10&12.39$\pm 0.06$\\ 
&19528& 2&11.34$\pm 0.06$\\ 
&19631& 4&11.61$\pm 0.12$\\ 
&19797& 6&11.78$\pm 0.06$\\ 
&20024& 8&11.93$\pm 0.06$\\ 
&25046& 6&11.62$\pm 0.16$\\ 
\ion{Mn}{2}&    0& 7&13.10$\pm 0.03$\\ 
&14325& 9&11.86$\pm 0.06$\\ 
&14593& 7&11.75$\pm 0.10$\\ 
&14901& 3&11.63$\pm 0.10$\\ 
&27547&13&11.50$\pm 0.28$\\ 
\ion{Fe}{1}&    0& 9&$<$11.08\\ 
\ion{Fe}{2}&    0&10&14.50$\pm 0.09$\\ 
&  384& 8&14.30$\pm 0.09$\\ 
&  667& 6&14.07$\pm 0.09$\\ 
&  862& 4&13.94$\pm 0.09$\\ 
&  977& 2&13.55$\pm 0.20$\\ 
& 1872&10&14.05$\pm 0.16$\\ 
& 2430& 8&13.76$\pm 0.07$\\ 
& 2837& 6&13.64$\pm 0.05$\\ 
& 3117& 4&13.48$\pm 0.06$\\ 
& 7955& 8&13.95$\pm 0.12$\\ 
& 8391& 6&13.68$\pm 0.12$\\ 
& 8680& 4&13.37$\pm 0.05$\\ 
& 8846& 2&13.10$\pm 0.07$\\ 
&15844&10&13.15$\pm 0.20$\\ 
&16369& 8&13.11$\pm 0.20$\\ 
&18360& 4&12.76$\pm 0.36$\\ 
&20340&12&13.04$\pm 0.13$\\ 
&20516& 6&12.69$\pm 0.21$\\ 
&20805&10&12.83$\pm 0.13$\\ 
&21251&14&13.14$\pm 0.17$\\ 
&21711& 8&13.01$\pm 0.19$\\ 
&22637&10&12.91$\pm 0.19$\\ 
&22810& 8&12.85$\pm 0.21$\\ 
&22939& 6&12.83$\pm 0.19$\\ 
&23031& 4&12.55$\pm 0.20$\\ 
&23317& 6&12.24$\pm 0.20$\\ 
&25428&12&12.74$\pm 0.09$\\ 
&25787& 4&12.38$\pm 0.11$\\ 
&25805&10&12.92$\pm 0.15$\\ 
&26170&12&12.81$\pm 0.05$\\ 
&26352&10&12.75$\pm 0.12$\\ 
&26932& 2&11.99$\pm 0.30$\\ 
&27314& 8&12.57$\pm 0.05$\\ 
&27620& 6&12.46$\pm 0.07$\\ 
&30388&10&12.47$\pm 0.08$\\ 
&30764& 8&12.36$\pm 0.08$\\ 
&32875&14&12.52$\pm 0.08$\\ 
&32909&12&12.45$\pm 0.08$\\ 
&33466&10&12.44$\pm 0.30$\\ 
&33501& 8&12.16$\pm 0.05$\\ 
&36126& 4&12.15$\pm 0.31$\\ 
&36252& 6&11.99$\pm 0.10$\\ 
\ion{Co}{2}& 3350&11&11.34$\pm 0.30$\\ 
& 4028& 9&11.38$\pm 0.30$\\ 
& 4560& 7&11.29$\pm 0.30$\\ 
\ion{Ni}{2}&    0& 6&13.22$\pm 0.07$\\ 
& 1506& 4&12.80$\pm 0.12$\\ 
& 8393&10&13.12$\pm 0.06$\\ 
& 9330& 8&12.95$\pm 0.05$\\ 
&10115& 6&12.71$\pm 0.06$\\ 
&10663& 4&12.54$\pm 0.06$\\ 
&13550& 8&12.62$\pm 0.06$\\ 
&14995& 6&12.56$\pm 0.06$\\ 
&23108& 6&11.93$\pm 0.30$\\ 
&23796& 4&11.65$\pm 0.30$\\ 
&24788& 4&11.57$\pm 0.36$\\ 
&25036& 6&11.86$\pm 0.30$\\ 
&29070& 4&11.27$\pm 0.36$\\ 
&32499&10&11.40$\pm 0.36$\\ 
&32523& 8&$<$11.78\\ 
\tablebreak
\ion{Cu}{2}&    0& 1&12.28$\pm 0.10$\\ 
&21928& 7&11.72$\pm 0.09$\\ 
&22847& 5&11.50$\pm 0.20$\\ 
\ion{Zn}{1}&    0& 1&10.65$\pm 0.05$\\ 
\ion{Zn}{2}&    0&13&13.10$\pm 0.06$\\ 
\enddata
\tablenotetext{a}{Unlike other lines that we list in this table, we have determined from the  
discordant radial velocity of this feature that it arises purely from the foreground interstellar 
medium (ISM).  The meaningful result for \ion{C}{1} in the circumstellar gas can be understood 
from our upper limit for the $3{\rm P}_2$ state at an excitation of $43\,{\rm cm}^{-1}$, which 
leads to $\log N({\rm C~I})\lt12.40$ for the sum of the three fine-structure levels of the ground 
state; see the discussion in Section~\protect\ref{sec:free_electrons}.}
\tablenotetext{b}{This ground-state column density was determined from FUSE data, where we 
do not have the velocity resolution to separate the circumstellar contribution from that arising 
from the foreground ISM.  Hence, this value represents the sum of the two sources.}
\tablenotetext{c}{This stated error refers to the measurement uncertainty alone, since the 
uncertainty in the $f$-value is not available.}
\tablenotetext{d}{This value is consistent with $\log N=10.99 (+0.09,-0.04)$ for the 
circumstellar \ion{Ca}{2} determined by Crawford et al. (1997).}
\end{deluxetable}

\section{Overview of Velocity Components}\label{sec:velocity_components}

\begin{figure}
\plotone{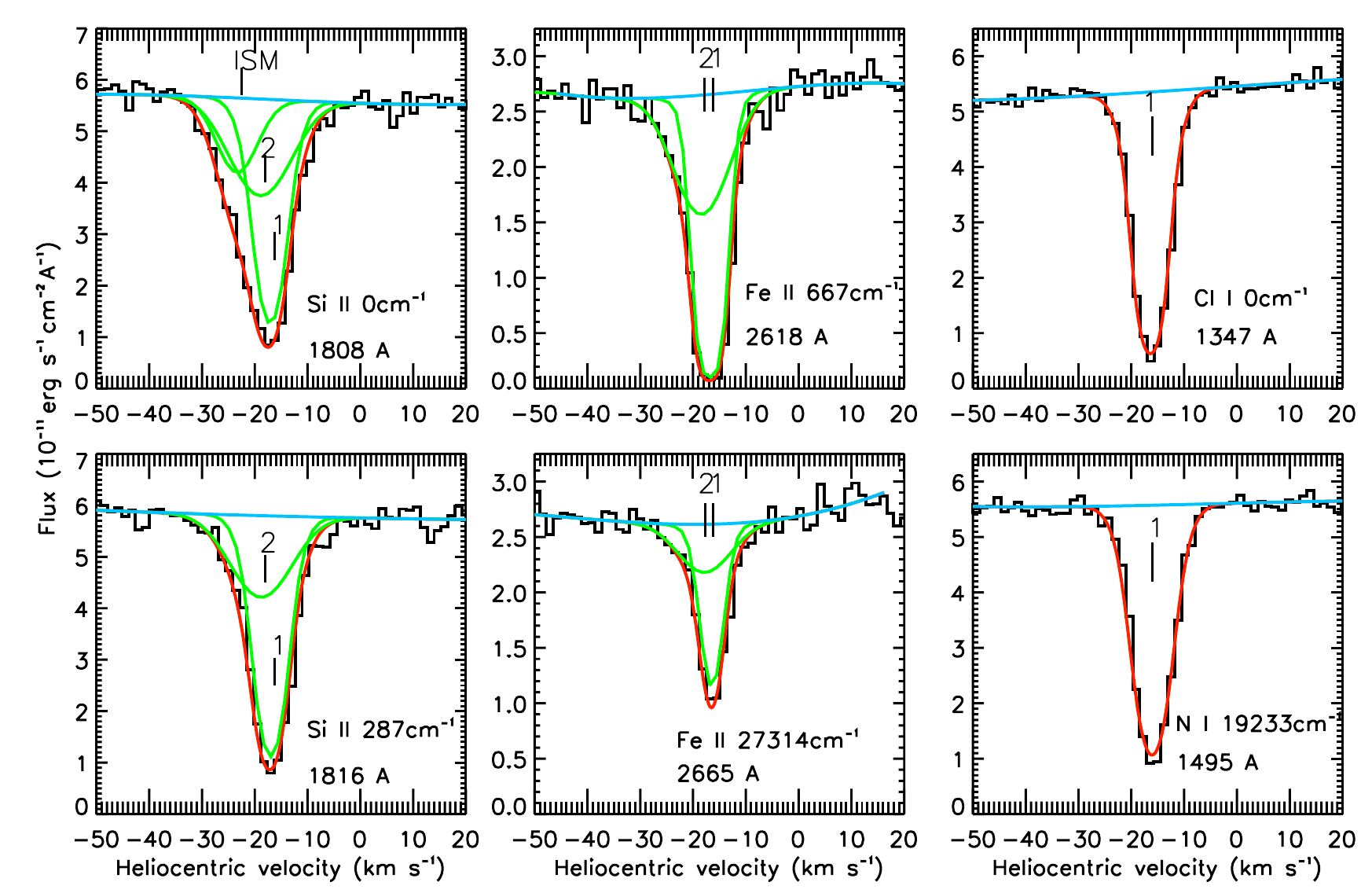}
\caption{Examples of the absorption line profiles arising from gas in the circumstellar disk of 
51~Oph.  Excited lines for ions can be fitted with two components (excited lines of \ion{Si}{2} 
and \ion{Fe}{2}) consisting of a strong component that has width consistent with a thermal 
broadening at $T_1=8000\,$K and turbulent broadening $b_{\rm 1,turb}=1.9\kms$, which is 
accompanied by a weaker one that is broad ($b_2\gt 6.5\,\kms$ for both thermal and turbulent 
broadening) and displaced to slightly more negative velocities.  For neutral atoms (panels on 
the right), we detect only the strong component.  The central velocities of the components are 
marked with the identification numbers discussed in the text.  For absorption out of the ground 
state of \ion{Si}{2} (upper left panel), there is an additional contribution from the foreground 
ISM.  In all of the panels, the black, histogram-style curves represent the observations, the red 
lines are the best fits, and the green lines depict the separate fits to the two individual 
components.  Stellar continua are shown in blue.\label{fig:fits}}
\end{figure}

In the spectrum of 51~Oph at all wavelengths, all circumstellar features from all of the
singly-ionized species show one predominant absorption at  $-16.2\pm 0.5$\kms, which we call 
Component~1.  In the strongest lines, another absorption is clearly visible, broader, shallower, 
and displaced by $-1.6\pm 0.2$\,\kms\ relative to the main component. We refer to it as 
Component~2. Therefore, a two-component fit is performed for all ions and gives a good fit to 
the lines as shown in a few examples of fitted profiles for a selection of species in  
Figure~\ref{fig:fits}.  However, because Component~2 is very broad ($b\gt 6.5$\,\kms\ in all 
cases) and kinematically very close to Component~1, in faint lines, its absorption is very shallow 
and difficult to distinguish from that of Component~1, even when its absorption amplitude is 
somewhat higher than the noise. Therefore, for species (or energy levels) that do not have 
strong lines, the distribution of matter between the components is not reliable. On the other 
hand, for the abundant energy levels that do have strong lines as well as unsaturated lines, 
where the distribution of matter in the two components is more reliable, no significant 
differences have been noted in the relative populations. So, although the fits are done with two 
components, we decided to list the results as the total column density in the two circumstellar 
components taken as a whole. 

The profiles of the neutral atoms \ion{N}{1} and \ion{Cl}{1} that do have relatively strong lines 
in some of their metastable or fine-structure levels do not show a second, broad component. 
The profiles of the strongest \ion{O}{1} lines (1302, 1304 \&\ 1306\,\AA) are very difficult to 
analyze because of their strong saturation, as we will show in Section~\ref{sec:allowed_OI}. 
However, there is evidence for some broad contribution, but at a  small level, apparently of 
order 1\% of Component~1. This suggests that neutrals are proportionally far less abundant in 
the broad component.

Except for \ion{Ca}{2} and \ion{Na}{1}, we found that for both neutrals and ions, our fits to 
Component~1 yielded turbulent velocities $b_{\rm turb}$ equal to $1.9\pm 0.17\,\kms$, 
assuming that the contribution to the total width caused by thermal broadening was consistent 
with $T=8000$\,K, as we found from our analysis of the metastable populations (which will be 
presented in Section~\ref{sec:outcomes}).  For \ion{Ca}{2} and \ion{Na}{1} we derived slightly 
higher velocity widths: $b_{\rm turb}=3.4\,\kms$.

Crawford et al. (1997) have reported the observation of the line of sight to 51~Oph in the Ca~K 
line recorded at very high resolution ($R\sim 860,000$) with the Ultra High Resolution Facility 
at the Anglo-Australian Telescope. They had recognized their reddest component as being 
probably circumstellar because of its velocity coincidence with an excited \ion{Fe}{2} line from a 
metastable level in the visible. It had a measured velocity of $-$15.8$\pm$0.2\,\kms, in good  
agreement with the \ion{Ca}{2} lines of our Component~1 in the UVES data.  Their outcome for 
the width of this \ion{Ca}{2} feature, $b=3.3(+0.7,-0.3)\,\kms$, is compatible with ours.

For the unexcited states, our ability to identify and disentangle the circumstellar contributions 
from interstellar ones was greatly facilitated by the interpretation of Crawford et al (1997). As 
shown in their Figure~1 and Table~2, four absorption components are detected in addition to 
the circumstellar one at $-$15.8\,\kms: two very narrow, very cold, components at $-$20.3 and 
$-$21.2\,\kms, and two broader components at $-$25 and $-$29\,\kms, corresponding to 
velocities of the diffuse local gas (the predicted velocity of the local cloud that surrounds the 
Sun in all directions is $-$27.6\,\kms\ toward 51~Oph according to Gry \& Jenkins (2014)). We 
indeed retrieved this velocity structure in the two \ion{Ca}{2} lines in our UVES spectra, but 
with our lower resolution, each pair of interstellar components appeared as a single absorption, 
at $v_3\,\sim\,-21.0$\,\kms\ for the cold gas  ($b\sim1.3$\,\kms) and $v_4\,\sim\,-
26.4$\,\kms\ for the warm local gas ($b\sim4.7$\,\kms). (The subscript ``2" is kept for the 
second disk component and is not included in the \ion{Ca}{2} model, because it is undetectable 
in the faint \ion{Ca}{2} lines.)

We subsequently apply this interstellar velocity model to fit all ground-state features. 
For some of them, the second broad disk component must be considered, which may have 
some overlap with the absorption of interstellar Component~3, resulting in a somewhat 
increased uncertainty.  Nevertheless, to a good approximation, we were able to determine the 
disk column density of all atoms in their ground states, except for \ion{N}{1} and \ion{P}{2} 
because the unexcited states of these species were only available in the FUSE spectrum, which 
does not have the resolution to separate the circumstellar from the interstellar velocity 
components.  We point out that for \ion{C}{1} no disk component is detected, the only feature 
clearly visible is due to the cold interstellar Component~3. Component~3 also dominates the 
absorption in \ion{Na}{1}, however, a disk component is also detected.  For the faint \ion{O}{1} 
line at 1355\,\AA, only the circumstellar feature is detected, which allowed us to obtain a 
reliable measurement for the disk $N$(\ion{O}{1}), which would be impossible with the sole 
strong 1302\,\AA\ line.

\section{Lack of CO Absorption}\label{sec:CO}

The wavelength coverage of the spectrum recorded by STIS for 51~Oph covers the locations of 
many absorption bands of CO, the strongest of which are in the $A\,^1\Pi-X\,^1\Sigma^+$ 
system (Morton \& Noreau 1994).  For the vibrational bands that range from $0-0$ to $10-0$, 
we see no evidence of any absorption features at the correct locations, apart from a few 
random coincidences with features from \ion{Fe}{2} in metastable levels.  This lack of CO UV 
absorption is in stark contrast to observations of far infrared emission from the rotation and 
vibration bands of CO that surrounds this star (van den Ancker et al. 2001 ; Thi et al. 2005 ; 
Berthoud et al. 2007 ; Tatulli et al. 2008).   It is difficult for us to quantify an upper limit for the 
column density of CO because our ability to detect this molecule depends on the broadening 
caused by rotational excitation of the ground vibrational state, which can be large.  
Nevertheless, even if much of the CO is so highly excited that the UV features are broadened 
and many of the molecules are in excited vibrational states, our detection threshold is 
considerably below the column density estimates of order $10^{20}-10^{21}{\rm cm}^{-2}$ in 
the models of Thi et al. (2005) and Berthoud et al. (2007).  We also failed to detect features 
from vibrationally excited H$_2$, such as those found by Meyer et al. (2001) toward 
HD\,37903.  Our lack of molecular absorption is consistent with the inability of Roberge et al. 
(2002)  to find any H$_2$ absorption features out of the ground vibrational state in a far-UV 
spectrum of 51~Oph recorded by FUSE. Therefore, we conclude that the line of sight to the star 
does not pass through a  zone near the central plane of the disk where there is enough 
shielding from H$_2$ and CO-dissociating radiation to allow appreciable concentrations of 
these molecules to accumulate.  Essentially, we are not viewing the disk exactly edge on, even 
though the inclination angle could be not too far from 90\arcdeg.  This conclusion is similar to 
that of Roberge et al. (2014), who failed to detect CO absorption in the spectrum of 49~Ceti, 
which has a circumstellar disk that exhibits strong CO emission (Hughes et al. 2017).

\section{Interpretation of the Excited Level Populations}\label{sec: excited_level_populations}

The excited fine-structure levels of the ground electronic states and the metastable states at 
higher energies of the different atomic species are primarily populated by either inelastic 
collisions with electrons and neutral atoms, or by optical pumping by the light from the central 
star (or possibly a combination of all three).  For now, we will dismiss the importance of 
collisions with H atoms, and we will justify our neglect of H-atom interactions later in 
Section~\ref{sec:HvsE}.

If the electron density at a temperature $T$ is above the critical density for exciting the levels 
of a particular atom, we would expect the level populations to achieve a thermodynamic 
equilibrium, where a plot that shows a relationship for $\ln (N_i/g_i)$ vs. $E_i/k$ to be a 
straight line with a slope equal to $-1/T$ for all levels $i$.  At lower electron densities or within 
a dilute radiation field, deviations from this trend can be influenced by differing collision 
strengths and radiative decay rates of various levels, from which one can attempt to solve for 
representative values of $n(e)$, $T$, and the strength of optical pumping.  We acknowledge 
that the electron densities and temperatures probably vary throughout different zones in the 
circumstellar environment that contain appreciable concentrations of gas, so the outcomes of 
any analyses represent averages weighted according to the local densities of the atoms being 
investigated.  We interpret the strength of optical pumping in terms of the distance of the gas 
from the center of the star $R_g$, but the answers that we quote actually represent $\langle 
R_g^{-2}\rangle^{-1/2}$ (or to be more precise, $4\langle W\rangle$, where a radiation dilution 
factor $W$ is defined in Eq.~\ref{eqn: W} as shown later in 
Section~\ref{sec:radiation_dilution}), again weighted according to atomic densities. 

In the following subsections, we will discuss the concepts that are important for the 
interpretations of the excited level populations of \ion{N}{1}, \ion{Fe}{2}, and \ion{Ni}{2}, where 
we have measurements of good quality for the column densities of many different metastable 
states and for which we were able to find a broad range of relevant atomic data that allowed us 
to solve for the relative populations.  While our principal effort will be to focus our attention on 
these three species,  we will also present a rudimentary interpretation of an excited level of 
\ion{O}{1} caused by inelastic collisions with electrons and H atoms to demonstrate the relative 
importance of these two collision partners.

\subsection{Optical Pumping}\label{sec: optical_pumping}
\subsubsection{Stellar Flux Model}\label{sec: stellar_flux}

To estimate the effects of optical pumping, we must start with a representation of the flux from 
the central star at all of the relevant wavelengths for transitions that can populate the upper 
electronic levels of different atoms.  We rely on models for the stellar fluxes, as the wavelength 
coverage of our observations is not broad enough to cover all possible transitions.  However, 
we can use UV observations to guide our choice for the best match of an effective temperature.  
The best observation for this purpose is that obtained by {\it HUT}, which provided continuous 
coverage of the critical wavelength range (see Table~\ref{tbl:observations}) where small 
differences in temperature have the strongest effect.  The best match to the observations arose 
from the mean of the 11,000 and 10,000\,K effective temperature models from 
UVBLUE\footnote{Available from \url{http://www.bo.astro.it/~eps/uvblue/uvblue.html}} 
(Rodriguez-Merino et al. 2005), and we adopted a surface gravity $\log g = 4.0$.  Both of these 
parameters are consistent with the values listed for 51~Oph by Dunkin et al. (1997) and 
Montesinos et al. (2009).   For $\lambda \lt 1300$\,\AA\ we used the UVBLUE spectrum and 
joined it with the model stellar spectrum of Allende Prieto et al. (2018)\footnote{Available from
\url{http://cdsarc.u-strasbg.fr/viz-bin/qcat?J/A+A/618/A25}} that had the same parameters, so 
that we could obtain fluxes out to 6500\,\AA.  While Dunkin et al. (1997) determined that the 
projected rotation velocity $v\sin i=267\pm 5\kms$ for 51~Oph based on stellar spectral 
features at visible wavelengths, the lines appearing in the STIS spectrum seem more consistent 
with approximately 200\kms.  We expect that this difference arises from gravity darkening, 
which de-emphasizes equatorial fluxes at short wavelengths.  Shortward of the Balmer break, 
we smoothed the model spectrum with a kernel scaled to a rotation velocity of 200\kms, while 
for longer wavelengths, we increased the assumed velocity to 267\kms.

In principle, an additional contribution to the radiation field can arise from the emission from a 
shock within a magnetic accretion column.  For 51~Oph, this contribution over wavelengths 
that are relevant for the optical pumping is small (Mendigutía et al. 2011), and we can ignore it. 

\subsubsection{Radiation Dilution Factor}\label{sec:radiation_dilution}

The rapid rotation of 51~Oph distorts the photosphere into an oblate spheroid.  Interferometric 
measurements by Jamialahmadi et al. (2015) indicated that the major axis $\theta_{\rm 
eq}=0.6\pm 0.05$\,mas and the minor axis $\theta_{\rm pol}=0.42\pm 0.01$\,mas, which is 
equivalent to $R_{\rm eq}=8.08\pm 0.7 R_\sun$ and $R_{\rm pol}=5.66\pm0.23 R_\sun$ if the 
star is at a distance $d=123$\,pc (Arenou et al. 2018 ; Luri et al. 2018).  As viewed in the 
equatorial plane at a large distance from the star, the solid angle occupied by the star’s 
photosphere should be equivalent to that of a sphere with a radius 
$R_*=[(8.08)(5.66)]^{\onehalf} R_\sun=6.76 R_\sun = 0.0314\,{\rm AU}$.  At a distance $R_g$ 
from the center of the star, the radiation has a dilution factor $W$ given by
\begin{equation}\label{eqn: W}
W=\onehalf\left\{ 1-\left[1-\left({R_*\over R_g}\right)^2\right]^{\onehalf} \right\}
\end{equation}
(Viotti 1976).

\subsubsection{The Attenuation of Radiation for Strong Transitions}\label{sec:attenuation}

Figures ~\ref{fig:comparison} and \ref{fig:excited_lines2} show absorption features that reach 
zero intensity, indicating that there is sufficient gas around 51~Oph to create deficiencies of flux 
in the cores of the strongest lines.  Therefore, we must recognize that the starlight is shielded 
for many transitions that can be influential in the optical pumping of the levels.  The amount of 
this shielding varies from virtually nothing at small distances from the star to an almost 
complete loss of photons for atoms at great distances.  When photons are absorbed, they can 
be reradiated at the same or lower energies when the excited levels decay to ones at lower 
energies.  We invoke the simplifying assumption that, except for stimulated emission, such 
photons are lost and do not cause further pumping elsewhere.  This concept is probably valid 
for gas confined to a thin disk around the star, because the photons from the spontaneous 
decays are emitted isotropically.

An average level of attenuation for all of the atoms may be computed for a given transition by 
measuring the equivalent width of the absorption and comparing it to that which one would 
expect if the velocity dispersion were so large that the central depth of the line would not be 
far below the continuum level.  Instead of using actual equivalent width measurement 
outcomes, we simplified our computations by invoking a standard Gaussian curve-of-growth 
behavior for all equivalent widths and assumed a single velocity dispersion parameter $b$ that 
approximated the composite behavior of the components described in 
Section~\ref{sec:velocity_components}.  A spot check of equivalent widths for \ion{N}{1} and 
\ion{Fe}{2} indicated that good values for $b$ are 4.5 and $4\kms$, respectively.  Hence, we 
can derive values of the central optical depths of the lines,
\begin{equation}\label{eqn: tau0}
\tau_0=1.5\times 10^{15}Nf\lambda/b~,
\end{equation}
where $N$ is expressed in ${\rm cm}^{-2}$, $\lambda$ in \AA, and $b$ in \kms.  It then follows 
that we can compute the saturation factor s.f. (the ratio of the equivalent widths of a saturated 
to unsaturated line),
\begin{equation}\label{eqn: s.f.}
{\rm s.f.}={2F(\tau_0)\over \sqrt{\pi}\tau_0}~,
\end{equation}
where
\begin{equation}\label{eqn: F(tau_0)}
F(\tau_0)={\sqrt{\pi}\over 2}\sum_{n=1}^\infty {(-1)^{n-1}\tau_0^n\over n{\rm !}\sqrt{n}}~.
\end{equation}
When we computed pumping rates, we multiplied the stellar fluxes by the appropriate values 
of s.f. for all of the transitions of \ion{N}{1} and \ion{Fe}{2}.  All of the lines of \ion{Ni}{2} were 
weak and thus did not need corrections for saturation.

\subsubsection{Radiative Transition Rates}\label{sec: transition_rates}

The level populations are influenced by the effects of absorption, stimulated emission, and 
spontaneous decay.  We follow a formulation given by Draine (2011, pp 53-55) and express the 
rates of these processes in terms of a dimensionless local photon occupation number for the 
starlight,
\begin{equation}\label{eqn: n_gamma}
n_\gamma = {c^3\over 8\pi h \nu^3}u_\nu\, ,
\end{equation}
where the energy density
\begin{equation}\label{eqn: u_nu}
u_\nu = {4\pi\over c}WB_\nu(T)({\rm s.f.})\, .
\end{equation}
The $W$ term in this equation is the dilution factor defined in Eq.~\ref{eqn: W}, and 
$B_\nu(T)$ is the radiation intensity at a frequency $\nu$ for a Planck distribution with a 
temperature $T$, and s.f. was defined in Eq.~\ref{eqn: s.f.}.  It then follows that for interactions 
between any given pair of levels, the rate of downward transitions per unit volume is given by 
the product of the number density $n_u$ of atoms in the upper level and the sum of 
spontaneous and stimulated emission rates,
\begin{equation}\label{eqn: u_to_l}
{dn_\ell\over dt} = n_u A_{u\ell}(1+n_\gamma)\, ,
\end{equation}
where $A_{u\ell}$ is the Einstein $A$ coefficient for the spontaneous decay from level $u$ to a 
lower level $\ell$.\footnote{The $A$ coefficients are derived from The Atomic Line List v 
2.05b21 \url{http://www.pa.uky.edu/~peter/newpage/}.  Not all possible transitions have 
values of $A$ listed; missing values are generally semi-forbidden transitions or allowed ones at 
very long wavelengths.  We have no choice but to assume their contributions are small and thus 
can be neglected.} The reverse process that repopulates the upper level is given by
\begin{equation}\label{eqn: l_to_u}
{dn_u\over dt} = n_\ell {g_u\over g_\ell}A_{u\ell} n_\gamma\, ,
\end{equation}
where $g_u$ and $g_\ell$ are the degeneracies of the two levels.
For $T$ that applies to $B_\nu(T)$, we adopted 10,000\,K and modified the radiation density to 
reflect the small deviations of the star’s model flux from this distribution. 

\subsection{Collisional Excitation and De-excitation by Electrons}\label{sec: 
collisional_excitation}

We operate under the premise that collisions with electrons dominate over those with neutral 
atoms.  We will justify this approach by citing in Section~\ref{sec:HvsE} some specific examples 
that demonstrate that the collisions by neutral partners have reaction rates that are 
considerably weaker than those associated with electrons.

For \ion{N}{1} we used the collision strengths $\Omega(T)$ defined by the fits by Draine (2011, 
Appendix F) to the calculations by Tayal (2006).  We obtained values for $\Omega(T)$ that 
applied to \ion{Fe}{2} and \ion{Ni}{2} from Tayal \& Zatsarinny (2018) and Cassidy et al. (2010), 
respectively.  Collision rate constants $C_{u,\ell}(e,T)$ for de-excitation of an upper level $u$ to 
a lower one $\ell$ are given by
\begin{equation}\label{eqn: C_u,l(e,T)}
C_{u,\ell}(e,T)={h^2\Omega(T)\over g_u (2\pi m_e)^{3/2}(kT)^{1/2}}={8.63\times 10^{-
6}\Omega(T) \over g_uT^{1/2}}\,{\rm cm}^3{\rm s}^{-1}\, .
\end{equation}
The reverse process (i.e., excitation from a low level to a higher one) is related to 
$C_{u,\ell}(e,T)$ through the principle of detailed balancing,
\begin{equation}\label{eqn: C_l,u(e,T)}
C_{\ell,u}(e,T)={g_u\over g_\ell}\exp(-\Delta E/kT) C_{u,\ell}(e,T) \, ,
\end{equation}
where $\Delta E$ is the energy separation of the two levels. 

\subsection{Special Considerations for Nitrogen}\label{sec:nitrogen}

We will show later that, along with electrons, there are significant amounts of both ionized 
nitrogen and neutral hydrogen in the gas (Sections~\ref{sec:HvsE} and \ref{sec:neutral_cl}, 
respectively).  This raises the prospect that for neutral nitrogen atoms there may be additional 
ways to populate the metastable levels, which we will investigate in the following two 
subsections.

\subsubsection{Charge Exchange}\label{sec:chg_exch}

The ionization fractions of N and H are coupled to each other through charge exchange 
reactions.  In particular, the reaction ${\rm N}^++{\rm H}^0\rightarrow {\rm N}^0+{\rm H}^+$ 
could create ${\rm N}^0$ atoms not only in the ground $^2p^3\,^4{\rm S}^o$ state but also in 
the excited $2p^3\,^2{\rm D}^o$ levels (with excitation energies $E=19224$ and $19233\,{\rm 
cm}^{-1}$).  However, we compared the strength of the collisional excitation rate constants 
against  the charge exchange rate constants to the excited level of N$^0$ calculated by Lin et al. 
(2005), and we found that the former dominates over the latter by a factor $\gt 10^3$ for 
$4000\lt T\lt 10^4$\,K.  Therefore, we feel that it is safe to ignore the contribution of charge 
exchange in populating the $2p^3\,^2{\rm D}^o$ level of N$^0$.

\subsubsection{Recombination}\label{sec:recombination}

Ignoring other processes, the volume density of neutral nitrogen atoms in an excited state 
$n({\rm N}^{0*})$ (where * denotes either $^2$D or $^2$P) is given by the equilibrium 
condition
\begin{equation}\label{eqn: N0*}
n({\rm N}^{0*})[A_{u,\ell}+C_{u,\ell}(e,T)n(e)]=n({\rm N}^+)n(e)\alpha_{\rm partial}
\end{equation}
where $\alpha_{\rm partial}$ is the recombination rate from the ground state of N$^+$ to the 
relevant state of N$^{0*}$.  Knowing that
\begin{equation}\label{eqn:N(N^0)}
n({\rm N}^0)\approx n({\rm N}^+){n({\rm H}^0)\over n(e)}~,
\end{equation}
we can divide both sides of Eq.~\ref{eqn: N0*} by $ n({\rm N}^0)$ and obtain the expression
\begin{equation}\label{eqn: ratio_N}
{n({\rm N}^{0*})\over n({\rm N}^0)}={n(e)^2\alpha_{\rm partial}\over n({\rm H}^0) 
[A_{u,\ell}+C_{u,\ell}(e,T)n(e)]}~.
\end{equation}
From our findings that will be presented in Section~\ref{sec:neutral_cl}, we can substitute 
$2.8n(e)$ for a lower limit for $n({\rm H}^0)$ and simplify this expression to the form
\begin{equation}\label{eqn: ratioN}
{n({\rm N}^{0*})\over n({\rm N}^0)}\leq{n(e)\alpha_{\rm partial}\over 2.8 
[A_{u,\ell}+C_{u,\ell}(e,T)n(e)]}~.
\end{equation}
Later (Section~\ref{sec:outcomes}), we will show that the temperature $T=8000\,$K is strongly 
favored.  At this temperature, the partial recombination coefficients $\alpha_{\rm partial}$ to 
the two $^2$D metastable states equal $1.89\times 10^{-14}{\rm cm}^3{\rm s}^{-1}$, which 
was obtained from the compilation at \url{https://open.adas.ac.uk/adf08}, and for the two 
$^2$P states $\alpha_{\rm partial}=8.45\times 10^{-15}{\rm cm}^3{\rm s}^{-1}$.  For what we 
will later consider to be the largest value for the electron density, $n(e)=3\times 10^6{\rm 
cm}^{-3}$, we obtain the numerical result for Eq.~\ref{eqn: ratioN} $n({\rm N}^{0*})/n({\rm 
N}^0)=1.7\times 10^{-6}$ for the $^2$D states (independent of $n(e)$ because the $A_{u,\ell}$ 
term is so small) and $1.1\times 10^{-7}$ for the $^2$P states.  The respective observational 
counterparts to these two values are $4.4\times 10^{-3}$ and $5.2\times 10^{-4}$.  While in 
this development we have disregarded collisional exchanges between $^2$P and $^2$D states, 
we can still safely assert that the recombinations are of little consequence to our analysis.

While the 3.5 orders of magnitude may seem like a very large margin of safety (i.e., observed 
excited fractions vs. our computed values arising from recombination), in our particular case it 
may be reduced to only 2 orders of magnitude (Ahmad Nemer, private communication) by an 
additional recombination mechanism called Rydberg Enhanced Recombination (RER), which in 
some cases can dominate over other recombination routes at temperatures below about 
$10^4$\,K (Nemer et al. 2019).

\subsection{Solutions for the Level Populations}\label{sec:solutions_level_populations}

We return to the general interpretations of the metastable excitations and derive solutions for 
the equilibrium concentrations of the $n$ electronic levels (the ground state and its excited 
fine-structure levels, plus all higher metastable levels including ones we could not see) by 
solving $n$ simultaneous linear equations.   For \ion{N}{1} $n=159$, \ion{Fe}{2} $n=340$, and 
\ion{Ni}{2} $n=686$.   The occupation fractions of these levels that we wish to derive are 
expressed in the terms contained within a column vector {\bf f} which is multiplied by an 
$n\times n$ matrix {\bf R} consisting of the radiative and collisional rates to form an
$n$-element column vector ${\bf b}=[0,0,0\ldots , 1]$, i.e.,
\begin{equation}\label{eqn: linear_eqns}
{\bf R \cdot  f=b}~.
\end{equation}
The first $n-1$ elements of {\bf b} set the requirement that for each row of {\bf R} the sum of 
the products of the rates and the unknown quantities in {\bf f} balance each other so that a 
given constituent is in equilibrium with the others.  Setting the last element in {\bf b} and all 
elements in the last row of {\bf R} equal to 1 insures that the sum of all constituent fractions is 
equal to 1.

Terms in the diagonal elements of {\bf R} represent the sums of losses of level $i$ that populate 
other levels $j$, which consist of spontaneous radiative decays, stimulated emission, upward 
and downward collision rates, and absorption,
\begin{equation}\label{eqn: R_diagonal}
{\bf R}_{i,i}=-\sum_j[ A_{ij,(i\gt j)}(1+n_\gamma)+ C_{i,j}(e,T)n(e)+{g_i\over g_j}n_\gamma 
A_{ij,(i\lt j)}]\, .
\end{equation}
Terms above the diagonal of {\bf R} represent gains for levels $j$ from higher levels $i$, which 
include spontaneous radiative decays, stimulated emissions, and collisional de-excitations 
\begin{equation}\label{eqn: R_above-diagonal}
{\bf R}_{i,j}=A_{ij}(1+n_\gamma)+C_{i,j}(e,T)n(e)~;~(i\gt j)\, .
\end{equation}
Terms below the diagonal of {\bf R} represent gains for levels $j$ from lower levels $i$, which 
include absorptions and excitations:
\begin{equation}\label{eqn: R_below-diagonal}
{\bf R}_{i,j}={g_j\over g_i}A_{ji} n_\gamma+C_{i,j}(e,T)n(e)~;~(i\lt j)\, .
\end{equation}

\subsection{Analysis Approach}\label{sec:analysis_approach}
\subsubsection{Foreground Contamination}\label{sec:foreground_contamination}
 
Our ability to isolate components with different radial velocities in the STIS and UVES spectra 
and identify foreground contributions from the ISM helped us to focus on measuring just the 
contributions from the circumstellar gas.  Nevertheless, we felt that it was important to 
acknowledge that for the unexcited levels of the atoms, our determinations might have been 
contaminated by contributions from some unrecognized portion of the ISM that had a velocity 
that overlapped that of the circumstellar gas, as we indicated in 
Section~\ref{sec:velocity_components}.  For this reason, we elected to not include the ground 
levels of \ion{Fe}{2} and \ion{Ni}{2} in our analysis, because these two elements have many 
other levels that we can analyze.  However, for \ion{N}{1} we were able to measure only 4 
excited levels, and thus we regarded the zero-excitation state as being critical for analyzing 
effects of collisional excitations and optical pumping.  Unfortunately,  for determining the 
column density of the unexcited level of \ion{N}{1}, we could access only a single unsaturated 
feature that was detected in a FUSE spectrum, where the velocity resolution was insufficient to 
isolate the interstellar from the circumstellar contributions.  For this reason, we included our 
basic measurement of the ground state but lengthened the downward error bar to $\Delta\ln 
N=-1.4$ below this value, which allows a reasonable margin for uncertainties in the estimated 
amount of contamination.  (This contamination is likely to be small, as the ratio of interstellar 
plus circumstellar N to what we regarded as purely circumstellar O is about the same as the 
solar abundance ratio.)  The main benefit of including the ground state of \ion{N}{1} is to 
disallow solutions that require a column density much higher than the observed one.

\subsubsection{H Atom vs. Electron Excitations}\label{sec:HvsE}

Next, we will explore the issue of collisional excitations by H atoms and how they compare to 
those by electrons in populating the upper levels.  We start with a discussion about \ion{O}{1} 
to serve as an illustrative example, and later we will briefly touch upon \ion{N}{1}  and 
\ion{Fe}{2}.

We measured \ion{O}{1} only in its lowest fine-structure level with a degeneracy $g=5$, so our 
determination of the total column density of this atom should be multiplied by 9/5 to account 
for the realistic premise that at the large electron and hydrogen densities and temperatures of 
the circumstellar material (see Sections~\ref{sec:outcomes} and \ref{sec:free_electrons}) the 
occupations of the atoms in all three fine-structure levels are proportional to their degeneracies 
to within our accuracy of determining $N$(\ion{O}{1} in the unexcited level.  Hence, we 
estimate that the total column density of \ion{O}{1} that we must consider for the $^3{\rm 
P}_{0,1,2}$ state is about $4.2\times 10^{17}\,{\rm cm}^{-2}$.

Krems et al. (2006) addressed the problem of interpreting the emission lines at 6300 and 
6364\,\AA\ in different astrophysical contexts and published calculations of rate constants for H 
atom collisions that could populate the $^1{\rm D}_2$ upper level $u$ that has an excitation of 
15868\,${\rm cm}^{-1}$.  As indicated in Table~\ref{tbl:atomic_lines}, there is an allowed 
transition out of this level at 1152\,\AA.  The FUSE spectrum shows that this line is strongly 
saturated and it is poorly resolved by the instrument, so we are only able to derive a lower limit 
for the column density equal to $2\times 10^{14}\,{\rm cm}^{-2}$.  For our adopted value of 
$N$(\ion{O}{1}) in the $^3{\rm P}_{0,1,2}$ state, we find that the ratio $n({\rm O}_u)/n({\rm 
O}_\ell)$ for the population of the upper level $u$ relative to that of the lower level $\ell$ is 
given by $n({\rm O}_u)/n({\rm O}_\ell)\gt 5\times 10^{-4}$.  We can perform a simple 
calculation for the equilibrium arising from collisions with H atoms alone that includes both 
upward and downward transitions, governed by the rate constants $C_{\ell,u}({\rm H},T)$ and 
$C_{u,\ell}({\rm H},T)$, plus spontaneous radiative decays (at a rate $A_{u,l}=8.6\times 10^{-
3}\,{\rm s}^{-1}$ for the sum of the two most important transitions).  In this simple treatment, 
we ignore cascades from even higher levels and the effects of optical pumping.  Therefore, we 
state that in equilibrium
\begin{equation}\label{eqn: H_equib}
n({\rm O}_\ell)n({\rm H}) C_{\ell,u}({\rm H},T)=n({\rm O}_u)[n({\rm H})C_{u,\ell}({\rm 
H},T)+A_{u,\ell}]
\end{equation}
which leads to a solution for the density of neutral H atoms
\begin{equation}\label{eqn: H_dens}
n({\rm H})={[n({\rm O}_u)/n({\rm O}_\ell)]A_{u,\ell}\over C_{\ell,u}({\rm H},T)-[n({\rm 
O}_u)/n({\rm O}_\ell)]C_{u,\ell}({\rm H},T)}~.
\end{equation}
 From results that will be presented later on the outcomes from \ion{N}{1}, \ion{Fe}{2}, and 
\ion{Ni}{2}, we adopt a temperature $T=8000$\,K.  At this temperature, $C_{\ell,u}({\rm 
H},T)=3.52\times 10^{-14}\,{\rm cm}^3{\rm s}^{-1}$ and $C_{u,\ell}({\rm H},T)=1.10\times 10^{-
12}\,{\rm cm}^3{\rm s}^{-1}$ (Krems et al. 2006).  If we adopt a value for $n({\rm O}_u)/n({\rm 
O}_\ell)$ equal to its lower limit of $5\times 10^{-4}$, the result that we find using 
Eq.~\ref{eqn: H_dens} is $n({\rm H})=1.2\times 10^8\,{\rm cm}^{-3}$.

We now compare the outcome for neutral hydrogen collisions with a similar calculation for 
excitations from just electrons, where the collision strength $\Omega=0.235$ at $T=8000$\,K 
(Zatsarinny \& Tayal 2003).  For the collision rate constants, we apply Eqs.~\ref{eqn: C_u,l(e,T)} 
and \ref{eqn: C_l,u(e,T)} to obtain $C_{\ell,u}(e,T)=1.45\times 10^{-10}\,{\rm cm}^3{\rm s}^{-
1}$ and $C_{u,\ell}(e,T)=4.53\times 10^{-9}\,{\rm cm}^3{\rm s}^{-1}$ and where $g_u=5$ but 
$g_\ell=9$ applies to the combined degeneracy of all 3 fine-structure levels of the ground 
$^3{\rm P_{0,1,2}}$ state.  If we apply Eq.~\ref{eqn: H_dens} with the substitution of $e$ for H, 
we obtain $n(e)=2.9\times 10^4\,{\rm cm}^{-3}$.

In this approximate treatment, we find that for \ion{O}{1} we have derived lower limits for 
either $n(e)$ or $n({\rm H})$ (because $n({\rm O}_u)/n({\rm O}_\ell)$ is a lower limit).   It is 
apparent that for an electron concentration $x_e=n(e)/n({\rm H})=1.4\times 10^{-4}$ the 
electrons have an excitation capability that is equivalent to those of the H atoms.  It is 
important to note that this statement is not strongly dependent on whether or not $n({\rm 
O}_u)/n({\rm O}_\ell)$ is above the lower limit that we derived.

Heavy-element atoms that have first-ionization potentials below that of H should be 
predominantly ionized (an exception is Cl, which will be discussed in 
Section~\ref{sec:neutral_cl}).  After allowing for some depletions of these elements to form 
solids, we would expect that, in the absence of any other source of electrons, $x_e$ should be 
about $1.5\times 10^{-4}$, which is close to the value that would equalize the excitation rates 
of \ion{O}{1} by electrons and neutral H atoms.  

It is probably more realistic to assert that the value of $x_e$ is much higher than $1.5\times 
10^{-4}$.  There appears to be significant ionization of H caused by some combination of 
radiation beyond the Lyman limit emitted by the star; soft X-rays emitted by gas in an accretion 
shock, for which there is evidence of one for 51~Oph (Mendigutía et al. 2011), and by cosmic 
rays (Padovani et al. 2018).  Support for this idea comes from the FUSE spectrum of 51~Oph 
that shows very strong, saturated absorption features from the three fine-structure levels in 
the ground state of \ion{N}{2}.  (We are unable to derive column densities because the features 
are strongly saturated and the signal-to-noise ratio is very low at the wavelengths covering the 
1085\,\AA\ multiplet of \ion{N}{2}.)  The existence of a significant amount of \ion{N}{2} 
indicates that there are many additional electrons contributed by the ionization of hydrogen, 
because the ionizations of N and H are coupled to each other by charge exchange reactions (Lin 
et al. 2005).  We will return to this topic in more detail in Section~\ref{sec:HI_HII}.  Thus, in 
short, we can justify interpreting the excitation of the metastable level of \ion{O}{1}  in terms of 
just the electron density.

For \ion{Fe}{2} we will be invoking a more comprehensive approach for understanding the ways 
to populate the many different metastable levels that we observed.  To sense the relative 
importance of H atom collisions compared to those from electrons, we evaluated for the lowest 
10 excited levels the quantity $C_{\ell,u}({\rm H},8000\,{\rm K})/C_{\ell,u}(e,8000\,{\rm K})$ 
from the ground state, using the rate constants for H collisions as calculated by Yakovleva et al. 
(2019).  We found a median value for this ratio equal to $1.4\times 10^{-4}$, a numerical 
outcome that duplicates the result that we found in our simple treatment for populating an 
excited level of \ion{O}{1}.

The derivations of the collisional rate constants for \ion{N}{1} by Amarsi \& Barklem (2019) 
indicate that within the context of their analysis, the rates that involve the lowest four excited 
levels are virtually zero because the level crossings occur at internuclear separations that are 
too small and beyond the regime of their model.  Other rate constants across levels above 
these levels are many orders of magnitude below the $C_{\ell,u}(e,T)$ values that we used, 
except for a few levels adjacent to each other in energy.

\subsubsection{Excitations of \ion{N}{1}, \ion{Fe}{2}, and \ion{Ni}{2}}\label{sec:NFeNi_exc}

We now move on to the consideration of more detailed calculations for \ion{N}{1}, \ion{Fe}{2}, 
and \ion{Ni}{2}.  For any one of these elements, our goal is to determine acceptable matches of 
the observed level populations to the expected equilibria that are influenced by both collisional 
exchanges and optical pumping.  The collisions are governed by the local physical quantities 
$n(e)$ and $T$, while the strength of the pumping to upper levels is controlled by the distance 
from the star $R_g$.  In this particular context, we consider the collective abundances of all 
levels as a nuisance parameter that we will need to marginalize.  In total, we must explore the 
goodness of fit between the data and our calculations over a range of 4 free parameters.  We 
undertake this task by employing a Markov Chain Monte Carlo (MCMC) analysis with a Gibbs 
sampling.  We assign a relative probability of any trial based on the probability of obtaining a 
worse fit to the data for a $\chi^2$ distribution with the degrees of freedom set to the number 
of observed levels  (we do not subtract the number of free parameters because for any sample, 
we are not actively solving for a minimum value for $\chi^2$).  We assumed uniform priors over 
reasonable ranges for the logarithmic variables, but subject to two restrictions: (1) $\log T \lt 
4.3$ based on the fact that the absorption features would be too wide if this constraint were 
violated, and (2) $\log R_g\gt -1.4$, i.e., $R_g\gt R_*$.

\begin{figure}[b]
\epsscale{1.35}
\plotone{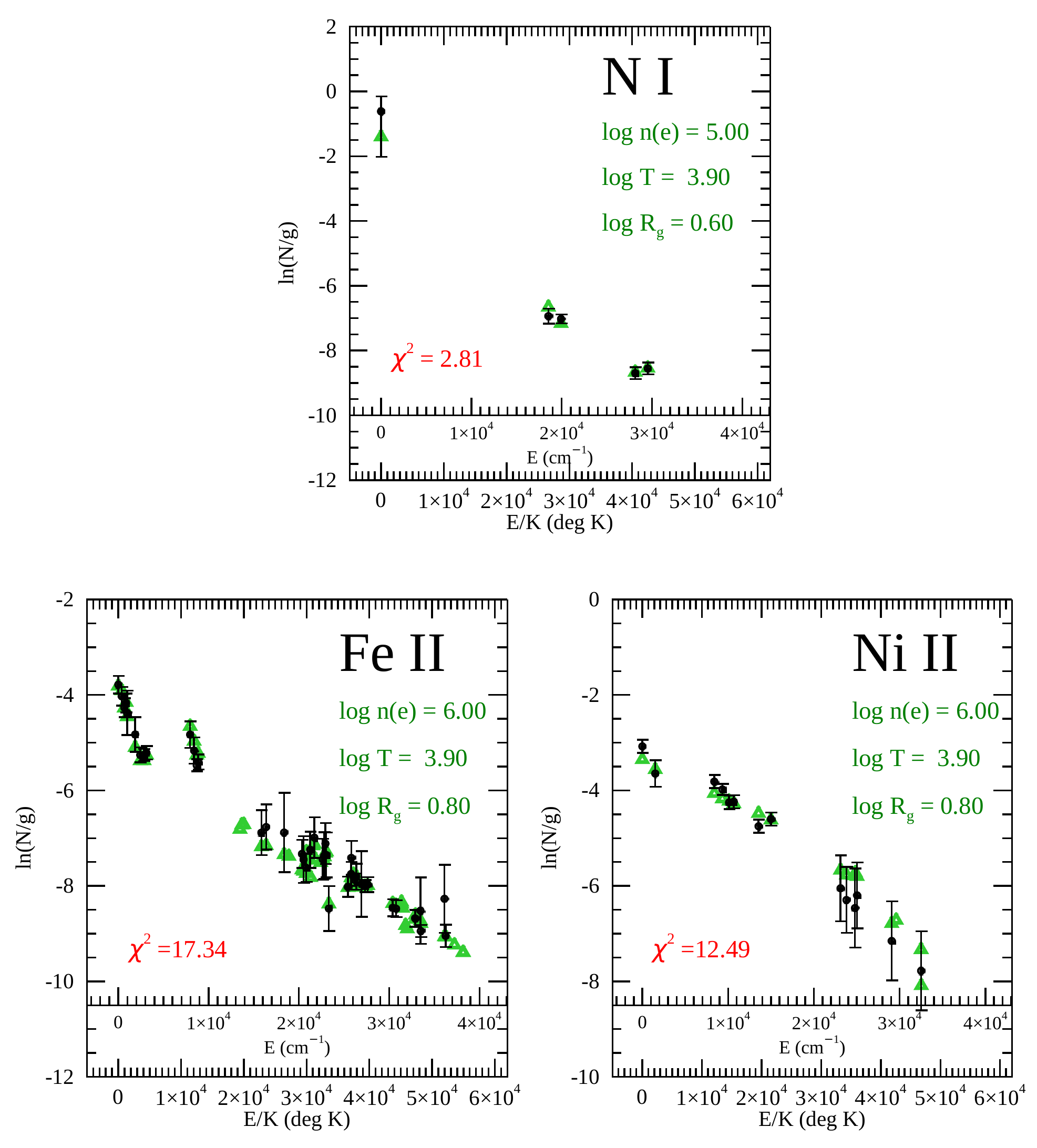}
\caption{The populations of various fine-structure and metastable levels of \ion{N}{1}, 
\ion{Fe}{2}, and \ion{Ni}{2} divided by their degeneracies, as a function of the excitation 
energies.  Values of $\ln(N/g)$ are relative and not absolute.  The parameters shown in the 
panels represent physical conditions that give reasonably good matches between observations 
(black points with error bars) and theory (green triangles).  The two pairs of excited levels with 
nearly the same energy for \ion{N}{1} are artificially separated from each other by a small 
amount in a horizontal direction for clarity.  The unreduced values of $\chi^2$ are indicated in 
each panel.\label{fig:threeBpanels}}
\end{figure}
\begin{figure}[h]
\epsscale{1.75} 
\plottwo{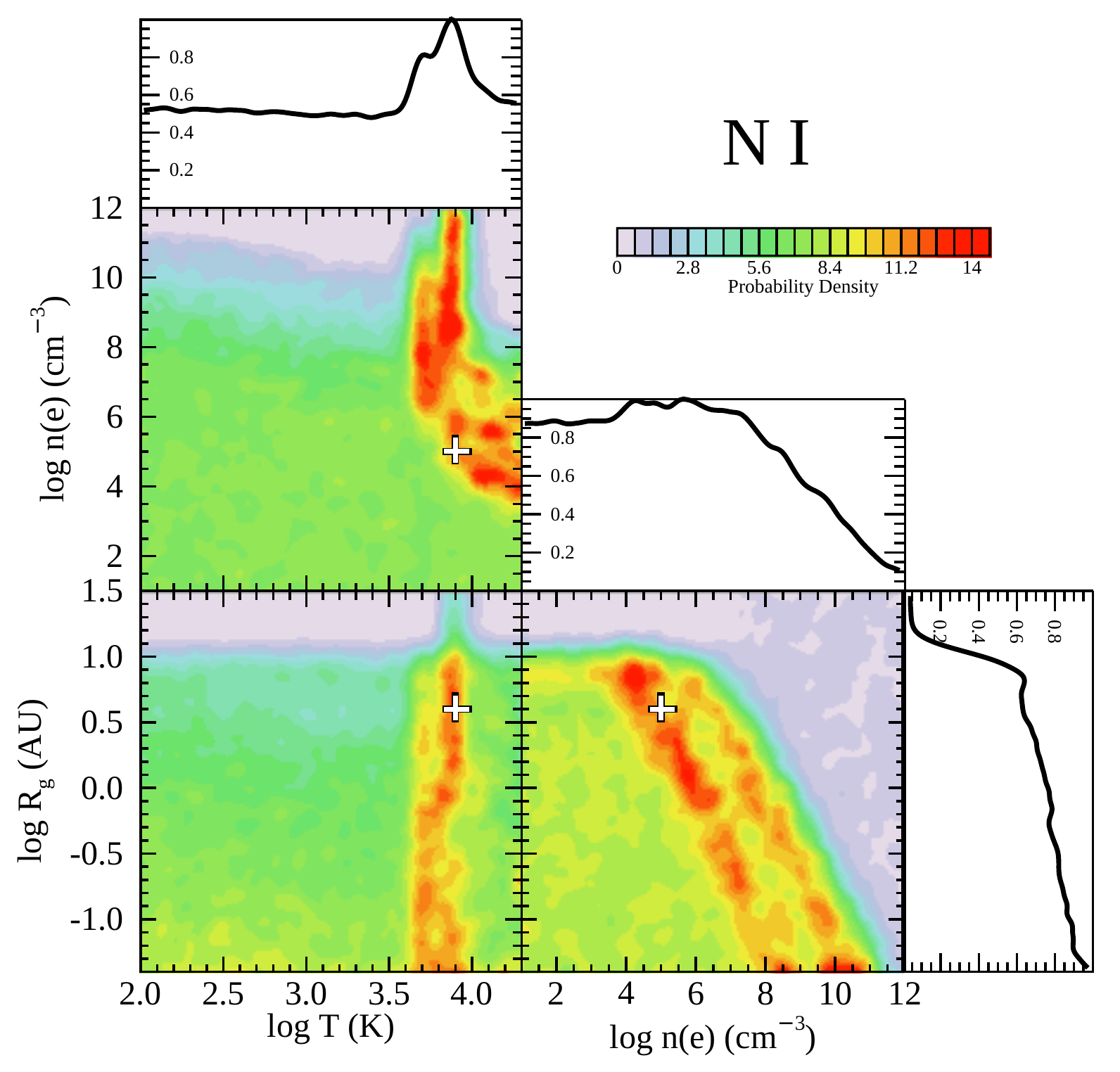}{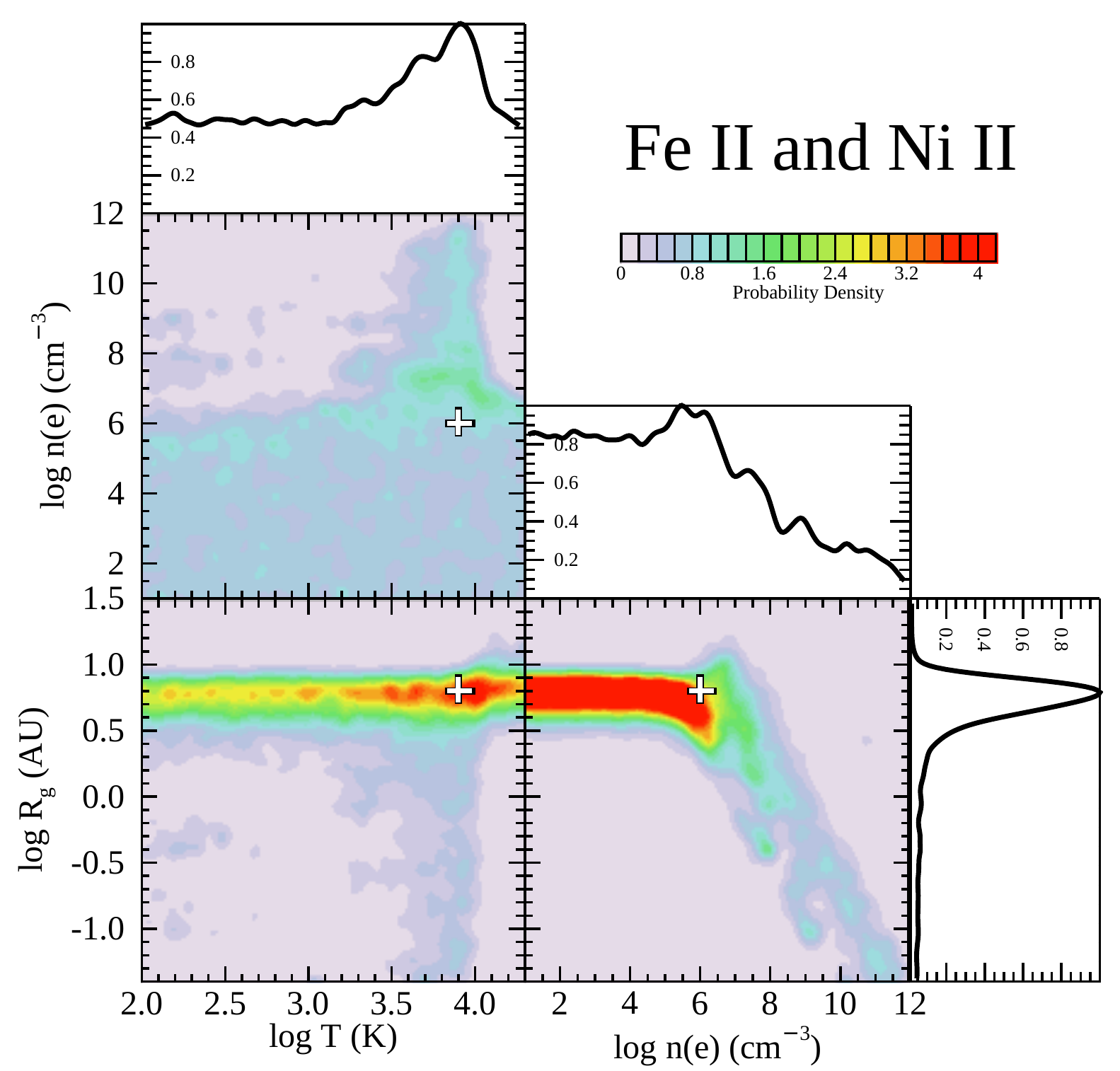}
\caption{Outputs from the MCMC analyses that indicate the relative probabilities of the 
fundamental parameters that govern the populations of the excited levels of \ion{N}{1}, 
\ion{Fe}{2}, and \ion{Ni}{2}.  The white crosses indicate the locations of the parameter choices 
shown in Figure~\protect\ref{fig:threeBpanels}.\label{fig:corner_plots}}
\end{figure}

We have organized our investigation so that levels of \ion{Fe}{2} and \ion{Ni}{2} are analyzed 
jointly, but the levels of \ion{N}{1} are treated separately.  The rationale for this approach is 
that Fe and Ni are refractory elements that probably share a common location, whereas N is a 
volatile element that may be distributed differently within the circumstellar environment.

Figure~\ref{fig:threeBpanels} shows examples where the conditions exhibit favorable matches 
between the observed level populations and the predictions arising from the analysis discussed 
in Section~\ref{sec:solutions_level_populations}.  As will be clear from the results presented in 
the next section, these conditions are not the only ones that give satisfactory fits to the data.

\subsection{Outcomes}\label{sec:outcomes}

Figure~\ref{fig:corner_plots} illustrates the coverages of high probabilities in the $\log n(e)$, 
$\log T$ and $\log R_g$ parameter space, which are depicted by the densities of successful 
trials in the MCMC runs.  As expected, the outcomes for the most probable values of these 
parameters for \ion{N}{1} and \ion{Fe}{2}+\ion{Ni}{2} sometimes differ from each other, yet 
there is still a broad range of choices where they overlap.  This overlap suggests that at least 
some, or perhaps most, of the neutral and ionized constituents probably coexist within a 
common volume, rather than being separated from each other and situated in separate zones. 
We find that \ion{N}{1} favors a temperature $T=8000$\,K ($\log T=3.9$), and part of the 
outcomes for \ion{Fe}{2}+\ion{Ni}{2} are consistent with this result.  Aside from temperature, 
the four metastable levels of \ion{N}{1} constrain the possible combinations of the other 
conditions rather poorly.  The results from \ion{Fe}{2}+\ion{Ni}{2} indicate rather well the 
distance to the star, $R_g = 6$\,AU ($\log R_g = 10^{0.8}$), and we can rule out values of $n(e)$ 
that are significantly larger than $3\times 10^6{\rm cm}^{-3}$.  For these two ions, arbitrarily 
low values for the electron density $n(e)$ and temperature $T$ are possible, which indicates 
that excitations due to optical pumping dominate over those from collisions by electrons.    
From our interpretation of our lower limit for the relative population of \ion{O}{1} in the 
$^1{\rm D}_2$ level that we derived in Section~\ref{sec:HvsE}, we concluded that $n(e)\geq 
2.9\times 10^4\,{\rm cm}^{-3}$, which is consistent with the regions exhibiting high 
probabilities for the excitations of \ion{N}{1}, \ion{Fe}{2}, and \ion{Ni}{2}.

\section{Other Neutral Atoms and Their Ionization Equilibria}\label{sec:other_neutral}

\subsection{Recombinations with Free Electrons}\label{sec:free_electrons}

In previous sections, we discussed our findings on the circumstellar abundances of the neutral 
forms of two elements, O and N.  These elements are plentiful, and the logarithm of the ratio of 
their abundances $\log N({\rm O~I})-\log N({\rm N~I})=17.63-16.60=1.03$, which is not far from  
the solar abundance ratio 0.86 (Asplund et al. 2009).\footnote{The difference from the solar 
abundance ratio decreases a bit when we consider that $\log ({\rm N/O})=\log ({\rm N}^0/{\rm 
O}^0)+0.07$, as we will show later in Section~\protect\ref{sec:HI_HII}.}  For the neutral forms 
of species that should be predominantly ionized, a very different picture emerges, as we discuss 
below.

We now explore the implications for electron densities by comparing the column densities $N$ 
of elements $X$ in their neutral states $N(X^0)$ to those of their singly-ionized states 
$N(X^+)$.  The abundances of these two forms are influenced by an equilibrium between the 
starlight photoionization of the neutral atoms at a rate $\Gamma$ balanced against the sum of 
the radiative recombination (rr) and dielectronic recombination (dr) of the ions with free 
electrons with a combined rate constant $\alpha_{\rm rr+dr}$ at 8000\,K.  

We consider the formula for the ionization equilibrium for the element $X$,
\begin{equation}\label{eqn: n(e)}
n(e)={\Gamma N(X^0)\over \alpha_{\rm rr+dr} N(X^+)}~,
\end{equation}
where we have substituted observed column densities $N$ for the local densities $n$ and set
\begin{equation}\label{eqn:Gamma}
\Gamma={4W\over hc}10^{-8}\int_{\rm 920\AA}^{hc/E({\rm IP})}\sigma_\lambda \lambda 
F_\lambda d\lambda~.
\end{equation}
The dilution factor that we defined in Eq.~\ref{eqn: W} $W=(R_*/R_g)^2/4$ when $R_g\gg 
R_*$.  For $R_g=6$\,AU, $W=6.8\times 10^{-6}$.  This value is considerably larger than the 
interstellar UV and visible radiation field at our location in the Galaxy, which is described in 
terms of $W=10^{-14}$ for a blackbody spectrum with $T=7500$\,K added to $W=10^{-13}$ for 
$T=4000$\,K (Mathis et al. 1983) and even lower effective values of $W$ for very hot stars 
(Mezger et al. 1982).  Hence, we can ignore the effects of the general radiation emitted by stars 
elsewhere.  The upper limit for the integral  $hc/E({\rm IP})$ is the wavelength of the ionization 
edge of the element in question ($1.24\times 10^4/E({\rm IP})$ if $E({\rm IP})$ is expressed in 
eV), and  $\sigma_\lambda$ is its ionization cross section.  The stellar flux at the surface of the 
star $F_\lambda$ is expressed in the form ${\rm erg~cm}^{-2}{\rm s}^{-1}{\rm \AA}^{-1}$, 
which we defined in Section~\ref{sec: stellar_flux}, and the factor $10^{-8}$ in the equation 
converts from \AA\ to cm.  The lower integration limit of 920\,\AA\ corresponds to the cutoff in 
the stellar flux caused by high members of the Lyman series and Lyman limit continuum 
absorptions.

\begin{deluxetable}{
l	
c	
c	
c	
c	
c	
c	
c	
}
\tablewidth{0pt}
\tablecolumns{8}
\tablecaption{Electron Densities from Ionization Equilibria at 8000\,K\label{tbl:n(e)}}
\tablehead{
\colhead{Elem.} & \colhead{$\Gamma$ (6\,AU)} & \colhead{$\langle \sigma v\rangle$} & 
\colhead{$\alpha_{\rm rr+dr}$} & \colhead{$\alpha_g$\tablenotemark{a}} & 
\colhead{$N(X^0)$} & \colhead{$N(X^+)$} & \colhead{$n(e)$\tablenotemark{b}}\\
\colhead{$X$} & \colhead{$({\rm s}^{-1}$)} & \colhead{$({\rm cm}^3{\rm s}^{-1}$)} & 
\colhead{$({\rm cm}^3{\rm s}^{-1}$)} & \colhead{$({\rm cm}^3{\rm s}^{-1}$)} & 
\colhead{$({\rm cm}^{-2})$} & \colhead{$({\rm cm}^{-2})$} & \colhead{$({\rm cm}^{-3})$}\\
\colhead{(1)} &\colhead{(2)} &\colhead{(3)} &\colhead{(4)} &\colhead{(5)} &\colhead{(6)} 
&\colhead{(7)} &\colhead{(8)}
}
\startdata
C &3.4e-03&6.7e-16&9.1e-13&1.9e-16&\lt 2.5e+12&2.3e+17\tablenotemark{c}&\lt 4.0e+04\tablenotemark{d}\\
Na&1.4e-02&4.5e-11&1.8e-13&3.3e-18&3.0e+10&6.5e+14\tablenotemark{c}&3.6e+06\\
Mg&6.8e-02&2.1e-12&4.7e-12&9.0e-17&\lt 1.2e+12&1.4e+16\tablenotemark{c}&\lt 1.3e+06\\
Si&4.7e+00&9.3e-13&1.7e-12&7.9e-17&\lt 3.3e+11&4.2e+15&\lt 2.2e+08\\
P &2.3e-01&3.0e-14&1.1e-12&\nodata&\lt 4.3e+11&3.4e+13&\lt 2.5e+09\\
S &1.6e-01&1.0e-14&4.7e-13&1.1e-16&\lt 5.5e+11&6.3e+15\tablenotemark{c}&\lt 2.9e+07\\
Cl&2.5e-05&6.0e-16&1.2e-12&\nodata&4.7e+13&\lt 4.4e+13&\gt 7.0e+07\tablenotemark{e}\\
Ca&1.1e+00&1.3e-11&6.2e-12&7.8e-18&\lt 2.8e+08&1.1e+11&\lt 4.4e+08\\
Fe&5.6e-01&4.0e-13&1.5e-12&5.7e-17&\lt 3.3e+11&1.3e+15&\lt 9.3e+07\\
Zn&2.0e-02&\nodata&3.9e-11&\nodata&4.5e+10&1.5e+13&1.5e+06\\

\enddata
\tablerefs{$\sigma_\lambda$ used for computing $\Gamma$: C (Verner et al. 1996); Si, Fe: 
NORAD-Atomic-Data website,
\url{https://norad.astronomy.osu.edu/}; Mg: (Wang et al. 2010); S (Bautista et al. 1998);
Ca: (Verner et al. 1996); Na, P, Cl, Zn: (Verner \& Yakovlev 1995), Collisional ionization from 
electron impacts $\langle \sigma v\rangle$: (Voronov 1997), Recombination $\alpha_{\rm 
rr+dr}$: C, Na, Mg: (Badnell 2006); Si: (Nahar 2000); P, Cl: (Landini \& Monsignori Fossi 1991); S, 
Ca: (Shull \& Van Steenberg 1982); Fe: (Nahar et al. 1997), Zn: (Mazzitelli \& Mattioli 2002), 
Grain recombination $\alpha_g$ for all elements where a value can be calculated: (Weingartner 
\& Draine 2001b).}
\tablenotetext{a}{Derived using $\psi=8.8\times 10^4$, $T=8000$\,K, and the use of the fitting 
equation, Eq.~8 of Weingartner \& Draine (2001b) for the element in question.  See the text for 
details.}
\tablenotetext{b}{Solutions to Eq.~\protect\ref{eqn: n(e)}.}
\tablenotetext{c}{This value is not measured; instead, we had to estimate it by relating it to O 
and N and using a solar abundance ratio (Asplund et al. 2009).  If this element is depleted 
relative to O and N, the value or upper limit for $n(e)$ will be higher.}
\tablenotetext{d}{The abundance of \ion{C}{2} is probably lower than the value given in 
Column~7.  For this reason, the upper limit shown here is too low.  See 
Section~\protect\ref{sec:deficiency_carbon}.}
\tablenotetext{e}{This outcome is misleading for reasons outlined in 
Section~\protect\ref{sec:neutral_cl}.}
\end{deluxetable}

Numbers that are relevant for evaluating $n(e)$ based on the ratios of neutrals to ions of 
various elements are listed in Table~\ref{tbl:n(e)}. Values of $\Gamma$ listed in Column~(2) of 
the table are based on a radial distance from the star $R_g=6$\,AU (i.e., $10^{0.8}$, which was 
favored by the \ion{Fe}{2}+\ion{Ni}{2} metastable excitations; we recall that the stellar radius 
$R_*=0.0314$\,AU from Section~\ref{sec:radiation_dilution}).  Equation~\ref{eqn: n(e)} ignores 
the effects of collisional ionizations by electron impacts.  At $T=8000$\,K, indicated by the 
\ion{N}{1} excitations, we find the effects of such collisions to be very small compared to 
$\Gamma$.  For instance, we list the rate constants $\langle \sigma v\rangle$ for such 
collisions in Column~(3) of the table;   $\langle \sigma v\rangle n(e)\ll \Gamma$ for the largest 
value of $n(e)$ indicated by the metastable excitations.  At $T=8000$\,K, the dielectronic 
recombinations are substantial for the elements Mg, Ca, and Zn, and the inferred values for 
$n(e)$ change rapidly as a function of temperature for Mg and Ca.  The quantity $\alpha_g$ in 
Column~(5) will be discussed in Section~\ref{sec:dust}.

We start by comparing the abundances of neutral and singly-ionized carbon.  For the former, 
we stated in Table~\ref{tbl:col_dens} a measurement\footnote{Our determination is 
considerably lower than that of Lecavelier Des Etangs et al. (1997a) of $\log N=13.08$ based on 
a GHRS G160M spectrum taken at a different time – see Table~\protect\ref{tbl:observations}).} 
$\log N({\rm C~I})=12.33$, but we conclude that the origin of this \ion{C}{1} absorption is 
interstellar based on two arguments: first, it has a velocity of $-3.5\kms$ with respect to the 
other strongest circumstellar features (see Section~\ref{sec:velocity_components}), and 
second, we see no evidence of absorption out of either of the two excited fine-structure levels.  
In a very dense and hot gas, the populations of all three fine-structure levels should be in 
proportion to their degeneracies.  Thus, in our attempt to identify \ion{C}{1} associated with 
the circumstellar gas, it is best to try to measure features out of the $^3{\rm P}_2$ level at 
$43\,{\rm cm}^{-1}$, which has a degeneracy equal to 5 (compared to 3 and 1 for the other two 
levels).  After considering relative signal-to-noise ratios and the groupings of lines within the 
different multiplets, we chose to use the two lines listed in Table~\ref{tbl:atomic_lines} for this 
level.  The absorptions cannot be seen in the spectrum, so we derived an upper limit equal to 
$1.4\times 10^{12}\,{\rm cm}^{-2}$ using the method described in 
Section~\ref{sec:upper_limits}.  We then multiply this upper limit by 9/5 to account for 
\ion{C}{1} in the other two levels to arrive at a total upper limit $N$(\ion{C}{1})$\lt 2.5\times 
10^{12}\,{\rm cm}^{-2}$.  

For \ion{C}{2}, we must make a preliminary estimate for its column density because our 
determination $\log N({\rm C~II})\lt 18.21$ from a semiforbidden transition is quite high (the 
allowed transition at 1334.5\,\AA\ is very badly saturated and not useful).  If we assume that 
the ratio of C to O or N in the circumstellar gas is identical to the solar ratio (Asplund et al. 
2009), we arrive at an approximate value $\log N({\rm C~II})\approx 17.37$.  At a radial 
distance from the star $R_g=6\,$AU, our determination of $\Gamma$ in Eq.~\ref{eqn:Gamma} 
gives a value $3.4\times 10^{-3}\,{\rm s}^{-1}$ when we use the ionization cross sections 
provided by Verner et al. (1996).  For carbon ions, $\alpha_{\rm rr+dr}=9.1\times 10^{-
13}\,{\rm cm}^3{\rm s}^{-1}$ at $T=8000$\,K (Badnell 2006).  An application of Eq.~\ref{eqn: 
n(e)} to our upper limit for $N$(\ion{C}{1}) yields $n(e)\lt 4.0\times 10^4\,{\rm cm}^{-3}$.  This 
limit is below the example for $n(e)$ that we chose to represent for the excitation of \ion{N}{1} 
in Fig.~\ref{fig:threeBpanels}, but setting $n(e)$ equal to the limit does not significantly 
degrade the match with the data and is consistent with our lower limit $n(e)\geq 2.9\times 
10^4\,{\rm cm}^{-3}$ that we derived from the population of the \ion{O}{1} metastable level in 
Section~\ref{sec:HvsE}.

Our calculation for $\Gamma$ only considered direct photoionization by the flux from the star.  
From the presence of \ion{N}{2} we know that ionized hydrogen must be abundant.  We will 
argue later (Section~\ref{summarysec:electron_dens}) that H must be ionized by photons that 
are far more energetic than this atom’s ionization potential.  When this happens, the electrons 
that are released can produce secondary ionizations of other elements, including C.  Because 
we did not include secondary ionizations in our calculations, it is possible that the outcome of 
our calculation of the upper limit for $n(e)$ is too low because we underestimated the true 
total value of $\Gamma$.  

In principle, another route for ionizing the carbon atoms is through the charge exchange 
reaction ${\rm C}^0+{\rm H}^+\rightarrow {\rm C}^++{\rm H}^0$. However, this process is 
quantitatively unimportant since even for a very large proton density, say perhaps $10^7\,{\rm 
cm}^{-3}$, the effect of this reaction is 7 orders of magnitude weaker than the photoionization 
rate because its rate constant is only $1.7\times 10^{-17}\,{\rm cm}^3{\rm s}^{-1}$ at 
$T\approx 10^4$\,K (Butler \& Dalgarno 1980). (In contrast to carbon, we find that charge 
exchange reactions that involve chlorine are important, as we will show in 
Section~\ref{sec:neutral_cl}).

We summarize in Column (8) of Table~\ref{tbl:n(e)} the values of $n(e)$ that we derived for C 
and various other elements using Eq.~\ref{eqn: n(e)}.  For \ion{Fe}{1} the combined 
degeneracies of the other four fine-structure levels of this atom $\sum g=16$ compared to 
$g=9$ of the lowest level could mean that our upper limit for all of the \ion{Fe}{1} should be 
increased by a factor of 25/9 to $3.3\times 10^{11}\,{\rm cm}^{-2}$ if the excitation 
temperature is large. Our limit for $N$(\ion{Fe}{1}) is considerably lower than the amount that 
appears in the spectrum of $\beta$~Pic  (Vidal-Madjar et al. 2017 ; Kiefer et al. 2019).  The 
column density for \ion{Fe}{2} includes all of the levels that we observed plus reasonable 
estimates for the populations of intermediate levels that we did not observe (to be discussed 
later in Section~\ref{sec:element_abundances}).

While most of the upper limits for $n(e)$ are consistent with our derivation based on C, there 
are two major inconsistencies with the C result.  The most prominent disagreement is the 
outcome of a lower limit for $n(e)$ derived from chlorine.  In Section~\ref{sec:neutral_cl} we 
will explain why we can disregard this result.  The outcomes for Na and Zn are substantially 
higher than that for C, which probably indicates that the actual $N$(\ion{C}{2}) is lower than 
our assumed value.  We will discuss the possibility that neutral carbon atoms might be expelled 
by an outward radiation pressure arising from the stellar flux in 
Section~\ref{sec:deficiency_carbon}.  That discussion will rely on conclusions derived here on 
the ionization and recombination of carbon atoms and ions and how these processes should 
affect any outward migration of the atoms.
\newpage

\subsection{Interactions with Dust Grains}\label{sec:dust}

Up to now, we have not considered recombinations of the ions with electrons on the surfaces 
of dust grains (Snow 1975 ; Weingartner \& Draine 2001b).  The efficiency of this process 
depends on the local density of grains and their charge.  Our discussion will proceed within the 
framework of the dust-to-gas ratio being identical to that of the general ISM.  However, we 
note that in reality it may be lower in the circumstellar atomic gas region that we sampled – at 
least for grains in the size range that is important for reddening.  The color excess E($B-V$) for 
51~Oph has been reported to be 0.04 by Malfait et al. (1998) and $0.09\pm 0.06$ by Manoj et 
al. (2006).   Using our measured column densities of \ion{N}{1} and \ion{O}{1} and assuming 
that their depletions are consistent with a moderate density ISM (i.e., $F_*\approx 0.5$ in the 
characterization of Jenkins (2009)), we would expect to find that $N({\rm H})=1.0\times 
10^{21}\,{\rm cm}^{-2}$, as we will explain in more detail in Section~\ref{sec:HI_HII}.  This 
value is greater than what we would have expected from the general result of Diplas \& Savage 
(1994) for the ISM, where $\langle N({\rm H~I)/E}(B-V)\rangle=5\times 10^{21}{\rm cm}^{-
2}{\rm mag}^{-1}$, which would yield $N({\rm H~I})\approx 2~{\rm or}~4.5\pm3\times 
10^{20}{\rm cm}^{-2}$ for the two determinations of E($B-V$).  This disparity, which is made 
even worse by the presence of ionized hydrogen, may indicate that some fraction of the grains 
have been driven out of the circumstellar atomic gas by an outward radiation pressure.  
Nevertheless, in the discussion that follows, we will frame our arguments within a perspective 
that the relative proportion of grains in the gas and their size distribution are the same as in the 
general ISM.

The charge of the grains is regulated by a balance between the loss of electrons due to 
photoelectric emission, which is governed by the local radiation density, and the capture of free 
electrons, which depends on $n(e)$ (Weingartner \& Draine 2001a).  In a formalism developed 
by Weingartner \& Draine (2001b) for the neutralization of ions by grains in the general ISM, a 
rate coefficient $\alpha_g(\psi[G,T,n(e)],T)$ is normalized to the local hydrogen density $n({\rm 
H})$ and depends on the parameter
\begin{equation}\label{eqn: psi}
\psi=G\sqrt{T}/n(e)
\end{equation}
 and $T$, where $G$ is the radiation density relative to that in the ISM as defined by Mathis et 
al. (1983).  The flux of 51~Oph at its surface corresponds to $G\approx 10^{14}$ for photons 
with $\lambda\lt 5000$\,\AA.  In the panel that shows  $\log R_g$ vs. $\log n(e)$ for \ion{N}{1} 
in Figure~\ref{fig:corner_plots}, the right-hand edge of the diagonal, red-colored region 
showing the most probable combinations corresponds to $n(e)=10^8R_g^{-2}$, where $n(e)$ is 
expressed in cm$^{-3}$ and $R_g$ in AU.  Here, we have chosen the highest set of values for 
$n(e)$ to explore the most favorable condition for promoting the neutralization of ions by their 
interactions with grains.  Because $G\approx 10^{14}(R_*/R_g)^2$, we obtain the simple 
outcome for the lowest value $\psi=8.8\times 10^4$ (i.e., one that is the most conservative 
from the standpoint of giving the highest values for $\alpha_g$).  

We do not have a firm number for a representative value of $n({\rm H})$, but if we assume that 
the column density $N({\rm H~I})\approx 10^{21}\,{\rm cm}^{-2}$ (from our measurements of 
\ion{N}{1} and \ion{O}{1}) and the thickness of the zone is no smaller than about $0.1R_g$, we 
obtain the estimate $n({\rm H~I})=6.7\times 10^8\,{\rm cm}^{-3}R_g$.  When we combine this 
limit with our condition $n(e)\lt 10^8R_g^{-2}$, we obtain an outcome for the electron fraction 
$x_e=n(e)/n({\rm H})\lt 1/(6.7R_g^3+1)$, which equals $6.9\times 10^{-4}$ if $R_g=6$\,AU.  As 
long as $x_e$ is not less than $\alpha_g/\alpha_{\rm rr+dr}$, we can declare that 
recombinations onto grains are less important than recombinations with free electrons. (Even if 
$x_e$ is smaller than this example, one should recall that $x_e\gg 1.5\times 10^{-4}$ from the 
discussion in Section~\ref{sec:HvsE}.)  The results shown in Table~\ref{tbl:n(e)} indicate that for 
all of the elements, $\alpha_g\ll x_e\alpha_{\rm rr+dr}$, so we can ignore the influence of 
grains in the ionization equilibrium.

\subsection{An Interpretation of the Large Abundance of Neutral 
Chlorine}\label{sec:neutral_cl}

The ionization balance for chlorine is vastly different from the examples that applied to the 
other elements listed in Table~\ref{tbl:n(e)}. We find a very strong presence of \ion{Cl}{1} ($\log 
N=13.67$ for the sum of the two fine-structure levels that we observed), but we were unable to 
detect an absorption feature out of the singly-ionized form of Cl.  The outcome shown in the 
table indicates that the lower limit for $n(e)$ is strongly inconsistent with the values or limits 
derived for C, Na, Mg, S, and Zn.  The only ways to reduce the inferred value of $n(e)$ are 
either to lower the temperature to well below $10^{3.8}$\,K (thus increasing $\alpha_{\rm 
rr+dr}(T)$) or to consider that $R_g\gg 6$\,AU.  The probabilities indicated in 
Figure~\ref{fig:corner_plots} reveal that the former possibility is disfavored by the \ion{N}{1} 
excitations and the latter is ruled out by the \ion{Fe}{2}+\ion{Ni}{2} excitations.

We now reach a point where we must consider alternate means for neutralizing Cl.  One of 
them might be the reaction ${\rm Cl^++H_2\rightarrow HCl^++H}$ followed by the dissociative 
recombination with electrons ${\rm HCl^+}+e\rightarrow {\rm H}+{\rm Cl}$ to form neutral 
hydrogen and chlorine (Jura 1974 ; Smith et al. 1980).  The possible importance of this channel 
is supported by observations of the ISM that show a strong correlation between 
$N$(\ion{Cl}{1}) and $N({\rm H_2})$ and evidence that in a number of cases $N({\rm Cl~I})\gt 
N({\rm Cl~II})$ (Moomey et al. 2011).  However, we regard this process as unimportant for the 
material surrounding 51~Oph, because Roberge et al. (2002) found that $N({\rm H_2})\lt 
7\times 10^{13}\,{\rm cm}^{-2}$ in the $J=0$ and $J=1$ states in the ground vibrational level, 
which is far lower than typical amounts of ${\rm H_2}$ that were found in ISM sight lines that 
showed enhancements of \ion{Cl}{1}.

Still another pathway for neutralizing Cl ions is through charge exchange with neutral hydrogen, 
${\rm Cl}^++{\rm H}^0\rightarrow {\rm Cl}^0+{\rm H}^+$.  The reaction is endothermic, but not 
strongly so because the first-ionization potential of Cl is 12.97\,eV, which is only slightly below 
that of hydrogen.  The rate constant for this reaction is 
\begin{equation}\label{C_+0}
C_{+,0}=6.2\times 10^{-11}\left( {T\over 300}\right)^{0.79}\exp(-6920/T)\,{\rm cm}^3{\rm s}^{-
1}
\end{equation}
and the reverse reaction rate is
\begin{equation}\label{C_0+}
C_{0,+}=9.3\times 10^{-11}\left( {T\over 300}\right)^{0.73}\exp(-232/T)\,{\rm cm}^3{\rm s}^{-1}
\end{equation}
(Pradhan \& Dalgarno 1994).  At $T=8000$\,K, $C_{+,0}=3.5\times 10^{-10}\,{\rm cm}^3{\rm 
s}^{-1}$ and $C_{0,+}=9.9\times 10^{-10}\,{\rm cm}^3{\rm s}^{-1}$.  We now examine the 
equilibrium of the two chlorine ionization states that takes into account these reactions, along 
with recombinations with free electrons and starlight photoionizations (at rates specified in 
Table~\ref{tbl:n(e)}):
\begin{equation}\label{eqn: chg_ex_equlib}
n({\rm Cl}^+)[n(e)\alpha_{\rm rr+dr}+n({\rm H}^0)C_{+,0}]=n({\rm Cl}^0)[\Gamma({\rm 
Cl})+n({\rm H}^+)C_{0,+}]~.
\end{equation}
When solving this equation for $n({\rm H}^0)$ we find that the $n(e)\alpha_{\rm rr+dr}$ term 
is insignificant and obtain the numerical result:
\begin{equation}\label{eqn: n(H0)}
n({\rm H}^0)={2.5\times 10^{-5}+9.9\times 10^{-10} n({\rm H}^+)\over 3.5\times 10^{-10} 
[n({\rm Cl}^+)/n({\rm Cl}^0)]}\,{\rm cm}^{-3}~.
\end{equation}
Because our observation indicates that $n({\rm Cl}^+)/n({\rm Cl}^0)\leq 1$, the above 
expression gives a lower limit:
\begin{equation}\label{eqn: n(H0)_outcome}
n({\rm H}^0)\geq 7.1\times 10^4\,{\rm cm}^{-3}+2.8 n({\rm H}^+)~.
\end{equation}
This outcome demonstrates that even with the large electron densities indicated by the 
metastable level populations, the neutral fraction of hydrogen is significant, which is a 
conclusion that is consistent with our observed abundances of \ion{N}{1} and \ion{O}{1}.  The 
fact that we could not obtain a column density for \ion{N}{2} and compare it with \ion{N}{1} 
means that our knowledge on the neutral versus ionized hydrogen abundances is limited to our 
finding expressed here and the less definitive estimate $n({\rm H}^0)\lt 4\times 10^9\,{\rm 
cm}^{-3}$ at $R_g=6$\,AU derived in Section~\ref{sec:dust}.
\section{The Allowed Transitions of \ion{O}{1} to the $^3S_1$ Level}\label{sec:allowed_OI}

\begin{figure}[b]
\plotone{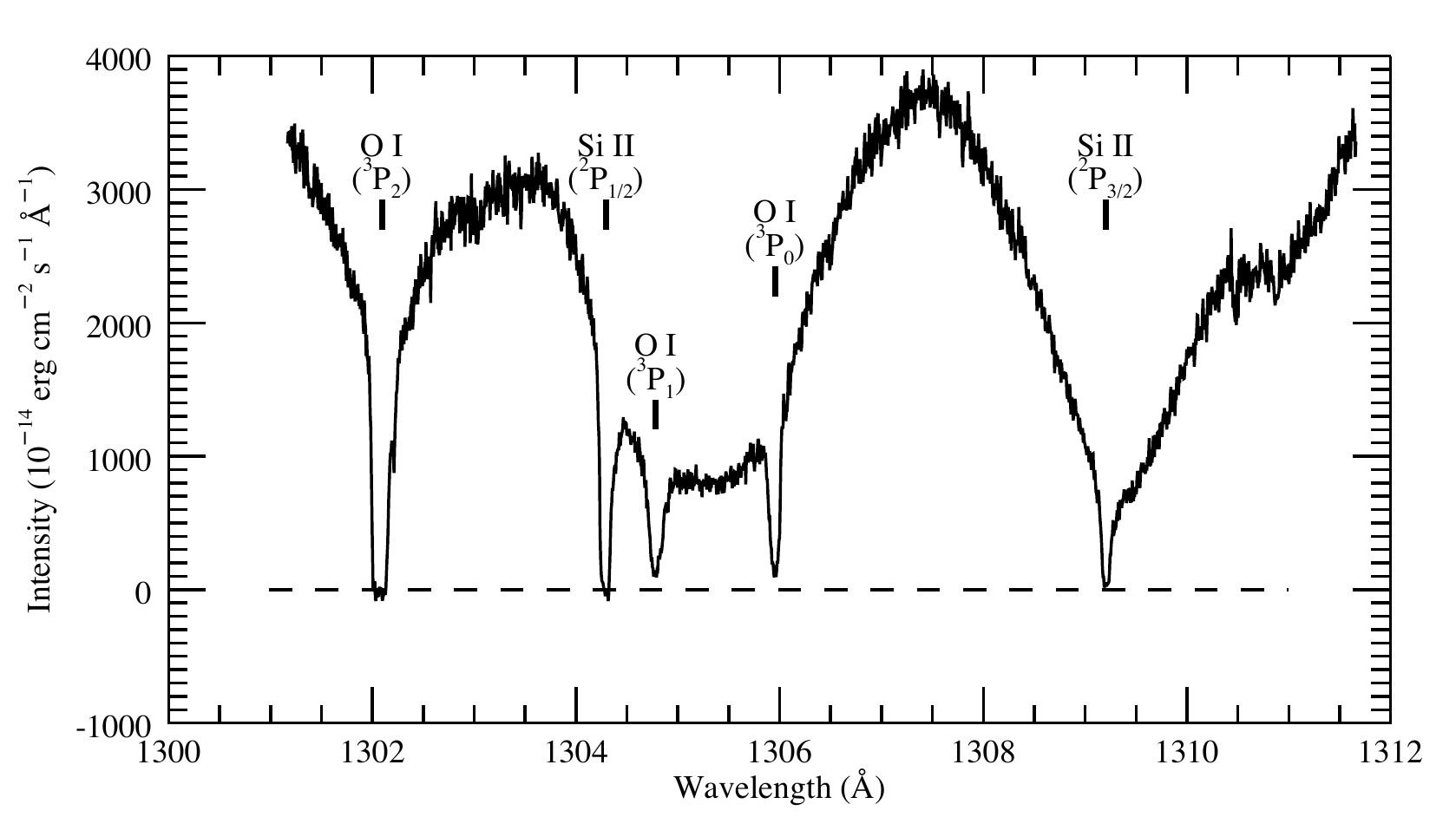}
\caption{A segment of the spectrum of 51~Oph covering a wavelength interval that includes 
features from \ion{O}{1} arising from three different fine-structure states, along with those of 
\ion{Si}{2} from two fine-structure levels.\label{fig:OI_wideplot}}
\end{figure}

Figure~\ref{fig:OI_wideplot} shows the spectrum of 51~Oph over a wavelength interval that 
covers the very strong features from three fine-structure levels of the ground electronic state of 
\ion{O}{1}.  The levels $^3P_1$ and $^3P_0$ have excitation energies equal to 158 and 
$227\,{\rm cm}^{-1}$, respectively,  above the $^3P_2$ state.  The absorption features from 
the excited levels of \ion{O}{1} have cores that do not reach zero intensity, unlike the 
absorptions from the ground states of both \ion{O}{1} and \ion{Si}{2}.  The $^2P_{3/2}$ excited 
fine-structure level of \ion{Si}{2} also does not reach zero intensity, but it is closer to zero than 
the features of excited \ion{O}{1}. 
\begin{figure}
\plotone{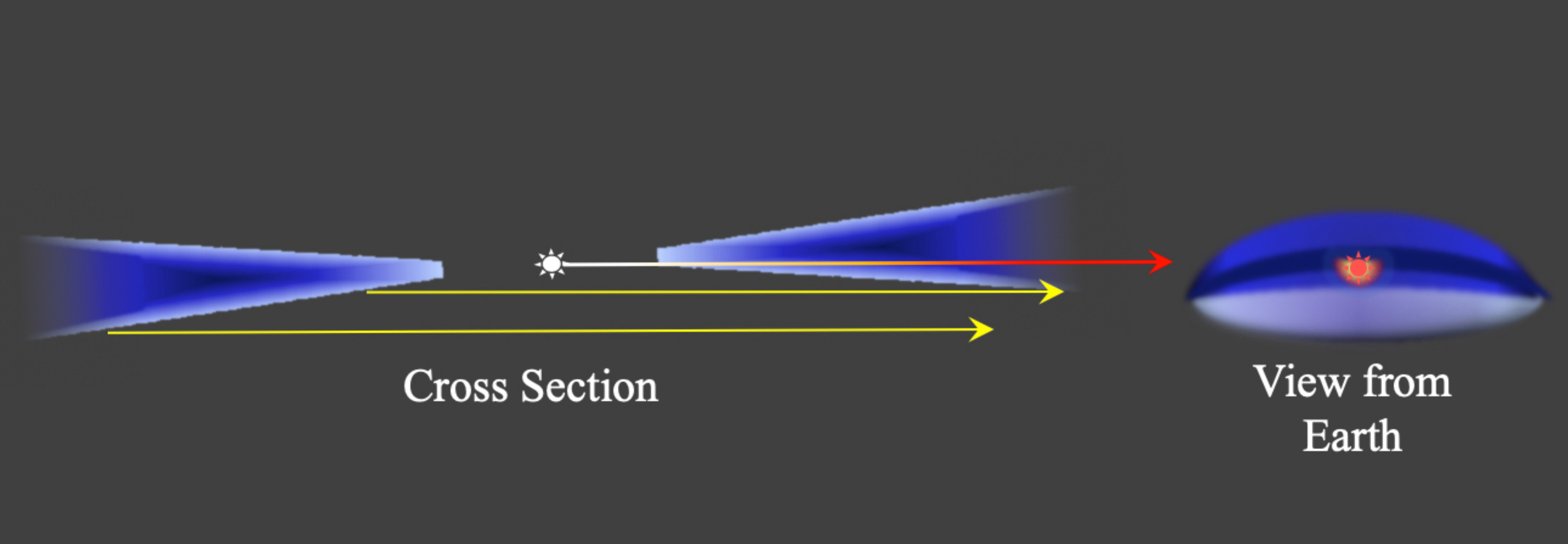}
\caption{A schematic illustration of a condition that can create the emission that is superposed 
on the cores of the absorption lines of excited \ion{O}{1}.  Light passing directly from the 
central star to the observer (red arrow) is intercepted by atoms in the disk and thus results in 
an absorption feature.  The disk has a wedge-shaped cross section that permits light from the 
star to illuminate the edges of the disk.  Photons that are within the passbands of the \ion{O}{1} 
features are resonantly scattered and can then reach the observer (yellow arrows) if the tilt of 
the disk is sufficient to prevent this light from being intercepted by the outer portion of the 
disk.\label{fig:cartoon}}
\end{figure}

While one might envision that the profiles do not reach zero intensity because the absorbing 
medium does not completely cover the disk of star seen in projection, we propose that a more 
likely explanation is that some of the flux from the star is resonantly scattered by oxygen atoms 
in the far side of the circumstellar disk.\footnote{We can dismiss the notion that the emission 
could be caused by telluric \ion{O}{1} emission, because if this were so, we would find the same 
effect in the unexcited \ion{O}{1} line at 1302\,\AA.}  If some of this emission were able to 
bypass the foreground absorption in the near side of the disk because it was somewhat 
displaced in projection, it would create excess flux contributions which fill in positive intensities 
superposed on the two saturated line cores.  Portions of the emitting gas that subtend less than 
$20R_{100}{\rm AU}$ should be admitted through the $0\farcs 2\times 0\farcs 2$ entrance 
aperture of the spectrograph, where $R_{100}=1.23\pm 0.04$ (Arenou et al. 2018 ; Luri et al. 
2018) is the distance of the star from us in units of 100\,pc. This emission effect is not seen in 
the feature from the $^3{\rm P}_2$ unexcited state because it is blocked by foreground 
\ion{O}{1} in the interstellar medium, whereas the interstellar absorption from the two excited 
levels is extremely weak or nonexistent. 

A plausible configuration for our being able to view the emission from resonant scattering is 
depicted in Figure~\ref{fig:cartoon}.  If the disk has a wedge-shaped cross section and the 
vertex falls short of the star, both sides can be illuminated by radiation from the star. We know 
from the nondetection of CO discussed in Section~\ref{sec:CO} that we are not viewing the disk 
edge on.  If the tilt of the disk with respect to the observer is sufficient, light scattered by atoms 
in the rear portion could bypass gas in the foreground part that would otherwise absorb it.  
Under the right conditions, other geometries that could also create conditions for this 
phenomenon might include circumstellar rings or warped disks.
\begin{figure}
\plotone{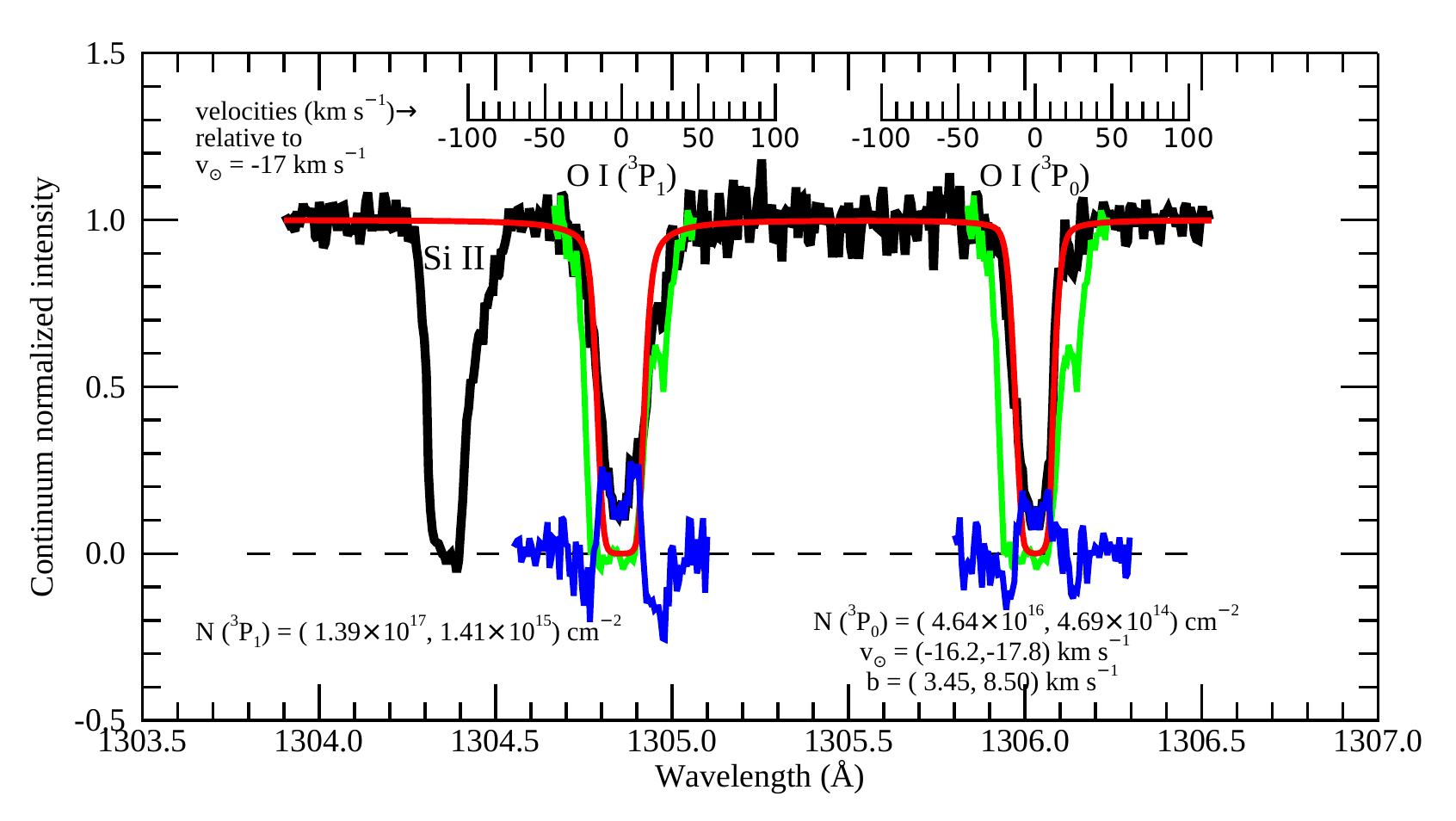}
\caption{Portion of the spectrum of 51~Oph that shows the two features of \ion{O}{1} in the 
excited fine-structure levels.  The thick, black trace shows the intensities that have been 
normalized to our best reconstruction of the stellar continuum.  The wavelength and velocity 
scales shown in this plot have been shifted to a reference frame of $-17\kms$.  For comparison, 
the two green traces show the shape of the absorption out of the unexcited level but displaced 
by the differences in transition wavelengths. The ISM absorption in the unexcited level makes 
its profile broader on the blue wing.  The absorption feature in the three \ion{O}{1} lines at 
+26.9\kms is telluric. The red trace shows our reconstruction of the absorption features 
according to our model discussed in the text, the parameters for which are shown below the 
zero level.  The blue trace shows the difference between the observed intensities and this 
absorption model, which we interpret as the resonantly scattered light from the disk whose 
radial velocities span $\pm 20\kms$ on either side of the line center.\label{fig:OI_plot}}
\end{figure}

Figure~\ref{fig:OI_plot} shows the spectrum in the vicinity of the excited lines after the 
intensities have been normalized to the stellar continuum (black trace).  To estimate the 
strengths and shapes of the emission profiles, we must first estimate what the absorption 
would look like in the absence of emission.  To recreate this absorption, we used our value for 
$N$(\ion{O}{1}) in the $^3P_2$ level that we derived from the weak intersystem line at 
1355.598\,\AA, as listed in Table~\ref{tbl:col_dens}, and assumed that each fine-structure level 
was populated in proportion to its degeneracy, which is approximately valid if the medium is 
dense and has an excitation temperature is of order 8000\,K.  Next, we used the velocity 
parameters for our neutral gas absorptions that we described in 
Section~\ref{sec:velocity_components} to construct the expected profiles.  The red trace in the 
plot shows the absorption profiles that should arise from this model.  The effect of smoothing 
by the STIS line spread function is very small for such wide absorption features, but it has been 
included.  As a check on the validity of this reconstruction, we require that the two profiles do 
not extend beyond the velocity span of the absorption from the unexcited $^3P_2$ level, which 
should be stronger because it has a higher degeneracy than either of the two excited levels.  
The green traces show the absorption by the unexcited level superposed on the two excited 
level absorptions.  In both cases, the red traces fall inside or on the green ones.

The blue traces in the figure show the differences between the observed intensities and our 
reconstruction of what the absorption features should look like in the absence of any emission.  
The broadening of the emission should arise from the Keplerian motion of gas around the star, 
and the increased levels of intensity away from the line center are probably created by limb 
brightening along the tangent points.  Both of the emission profiles reach zero intensity at 
about 20\,\kms away from the line center, and this velocity offset is consistent with the 
expected Keplerian velocity $v_{\rm K}=\sqrt{GM_*/R_g}=22\kms$ at $R_g=6$\,AU from the 
star, where the star has a mass equal to $M_*=3.3M_\odot$ (Jamialahmadi et al. 2015).  The 
intensity dip seen on the right-hand side of the $^3{\rm P}_1$ profile is caused by telluric 
\ion{O}{1} absorption, which we predict to occur at a heliocentric velocity of +9.9\kms\ at the 
time the observation was carried out (or +26.9\kms\ for the scale shown in 
Fig.~\ref{fig:OI_plot}). If the gas were optically thin, we would expect the $^3P_1$ emission 
intensity to be three times as strong as that from the $^3P_0$ state, but here it appears to be 
only about twice as strong.  High optical depths would tend to equalize the two intensities, but 
at some expense to the limb-brightening effect.

We touch briefly on the possibility that some of the emission might arise from Bowen 
fluorescence (Bowen 1947) produced by optical pumping by L$\beta$ irradiation to the levels in 
the $2p^33d^3{\rm D}^o_{1,2,3}$ states, which then decay to produce radiation at 8446\AA.  
Eventually, there is a cascade to the $2s^22p^3s^3{\rm S}_1$ state that produces 1304 and 
1306\AA\ photons as it decays to the ground fine-structure levels.  Broad H$\alpha$ emission 
arising from a shock in an accretion flow has been observed for 51~Oph (Manoj et al. 2006 ; 
Mendigutía et al. 2011).  If the medium is optically thin, the predicted value of $I_{{\rm 
H}\alpha}/I \lambda 8446$ is 7500 (Strittmatter et al. 1977), but it can be lower if the emitting 
region is optically thick.  For the $^3{\rm P}_0$ feature, the average strength of the emission 
over the 40\kms\ span centered on the systemic velocity ($v_\odot=-17\kms$) is 0.10 times the 
continuum intensity over the same interval.  We compute that the prediction for the 
fluorescence would result in about one order of magnitude less flux than what we observe, but 
a higher intensity could result if the region is optically thick or there is a sufficient flux of 
L$\beta$ photons from the star. 

\section{The Pattern of Element Abundances}\label{sec:element_abundances}

\begin{figure}[t]
\plotone{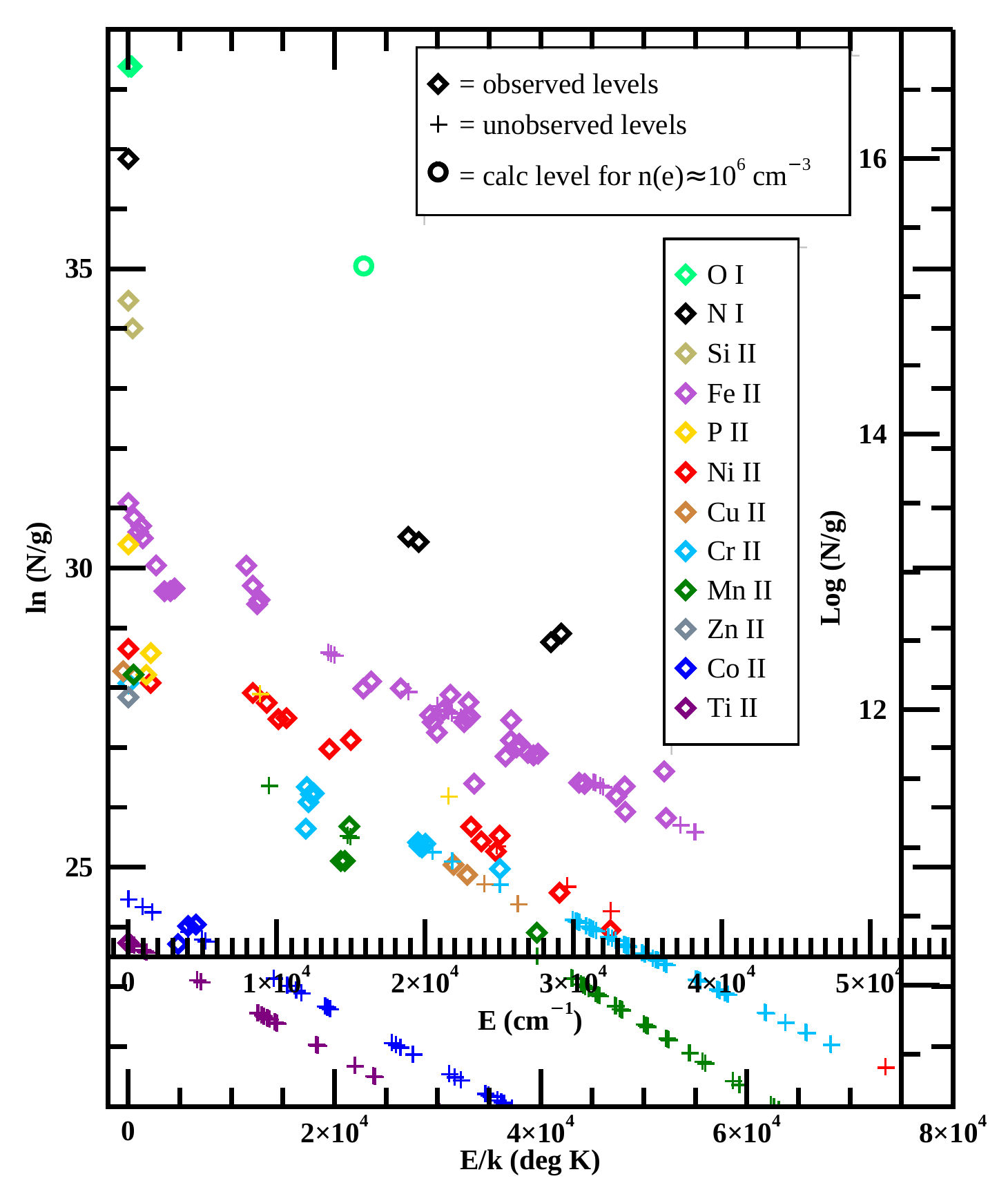}
\caption{A plot that shows all of the column densities of elements in their preferred ionization 
stages for all of the levels that we observed, expressed in the representation $\ln(N/g)$, where 
$N$ is the column density and $g$ is the degeneracy of the level, as a function of excitation 
energy above the ground state.   Levels that had measured column densities are depicted with 
diamond symbols, while our estimates for unobserved states are shown by plus markers.  See 
the text for details on the excited level of O that is shown by a circle.  Some markers have been 
moved in a horizontal direction by a small amount to prevent overlaps with other 
ones.\label{fig:all_elements}}
\end{figure}

Figure~\ref{fig:all_elements} is a representation of $\ln(N/g)$ vs. excitation energy for the key 
elements covered in this study.  For many of the elements, there are levels whose absorptions 
we did not observe and yet would be expected to be populated.   For the elements Cr, Mn, Fe, 
Ni, and Cu, we estimated the locations for the unobserved levels in this diagram by placing 
them on a straight-line best fit to the measured levels.  Unseen levels that were considered to 
have populations worthy of note were those that had excitation energies below $80,000\,{\rm 
cm}^{-1}$ and that did not have electric dipole or intercombination lines to lower levels that 
would cause rapid radiative decays.  The elements N, O, Si, and Zn did not have any such 
unseen levels.  For the elements P, Ti, and Co, the observations were so incomplete that we had 
to adopt a representative slope, and we chose one that was equal to the mean of the 
\ion{Fe}{2} and \ion{Ni}{2} fits.  The situation for O is a special one; aside from the
fine-structure levels, the only noteworthy excited level is the one at $15,868\,{\rm cm}^{-1}$ 
that we know has an appreciable column density, but for which we could derive only a lower 
limit to its value.  The circle symbol in the diagram corresponds to our expectation for 
$n(e)=10^6\,{\rm cm}^{-3}$ and $T=8000$\,K.

\begin{deluxetable}{
l	
c	
c	
c	
}
\tablewidth{0pt}
\tablecolumns{4}
\tablecaption{Observations of the Elements\label{tbl:elements}}
\tablehead{
\colhead{} & \colhead{Log Total} & \colhead{Fraction} & \colhead{Excitation}\\
\colhead{Element} & \colhead{Column Density} & \colhead{Observed} & 
\colhead{Temperature (K)}\\
\colhead{(1)} & \colhead{(2)} & \colhead{(3)} & \colhead{(4)}\\
}
\startdata
N&16.60&1.00&5180\\
O&17.63&0.98\tablenotemark{a}&\nodata\\
Si&15.62&1.00&890\tablenotemark{b}\\
P&13.61&0.85&450\tablenotemark{b}\\
Ti&12.24&0.05&\nodata\\
Cr&13.49&0.61&11,990\\
Mn&13.27&0.78&9120\\
Fe&15.15&0.94&3090\tablenotemark{b}, 11,840\tablenotemark{c}\\
Co&13.28&0.27&\nodata\\
Ni&13.81&0.99&10,210\\
Cu&12.49&0.88&9700\\
Zn&13.61&1.00&\nodata\\
\enddata
\tablenotetext{a}{We could measure only the ground $^3{\rm P}_2$ level, but the other two 
fine-structure levels should be populated in proportion to their degeneracies.  While we 
observed an absorption out of the metastable level at an excitation of $15868\,{\rm cm}^{-1}$ 
we could only obtain a lower limit for its column density.}
\tablenotetext{b}{Fine-structure levels only.}
\tablenotetext{c}{Metastable levels only.}
\end{deluxetable}

We may be mildly understating the abundances of nitrogen and oxygen, as some fraction of 
these elements may be singly ionized, i.e., a condition similar to that of hydrogen.  We know 
from the FUSE observations discussed in Section~\ref{sec:HvsE} that this is the case for 
nitrogen, but we are not able to quantify the relative ionization.  Later, in 
Table~\ref{tbl:relative_ionizations} within Section~\ref{sec:HI_HII}, we will show our estimates 
for the ratios of ions to neutrals for these elements.  The other elements probably have no 
appreciable concentrations in their doubly ionized form, but we have no firm evidence that this 
is so.

Table~\ref{tbl:elements} shows some outcomes of our measurements of the elements.  
Column~(2) lists the sum of observed and unobserved column densities, while Column~(3) 
shows the fraction of this combined column density that was actually observed, which can be 
used as a measure of the reliability of the total column density.  The excitation temperatures 
given in Column~(4) are derived from the negative inverses of the slopes of the level 
populations shown in Fig.~\ref{fig:all_elements}.  From the conclusions that we derived in 
Section~\ref{sec:outcomes}, it should be clear that these excitation temperatures should not be 
regarded as actual physical temperatures.  We list these temperatures only for the purpose of 
empirical comparisons with results for the circumstellar populations around other stars.  In 
particular, we caution against using the populations of the fine-structure levels of \ion{Si}{2} 
and \ion{P}{2} as an alternate means of determining $n(e)$ and $n({\rm H}^0)$ because the 
uncertainties in the column densities are large enough to allow their excitation temperatures to 
be equal to the kinetic temperature.  For instance, the critical density for the excitation of the 
\ion{Si}{2} levels is as low as $n(e)\approx 10^4{\rm cm}^{-3}$ (Silva \& Viegas 2002, Fig.~8).

\begin{figure}
\epsscale{0.8}
\plotone{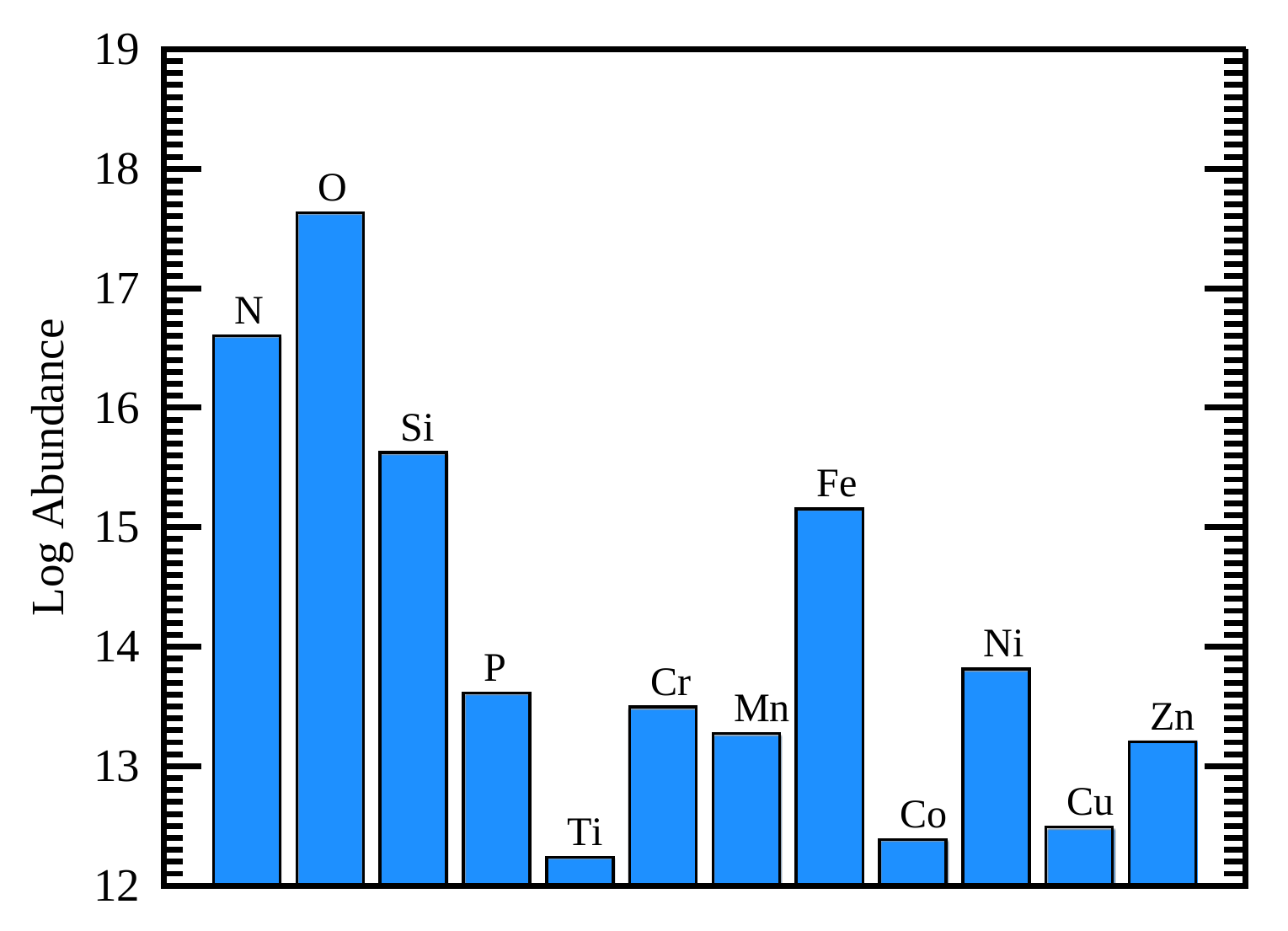}
\caption{Total column densities of the preferred ionization stages of the circumstellar elements 
associated with 51~Oph.\label{fig:barplot_abund}}.
\end{figure}
\begin{figure}
\epsscale{1.2}
\plotone{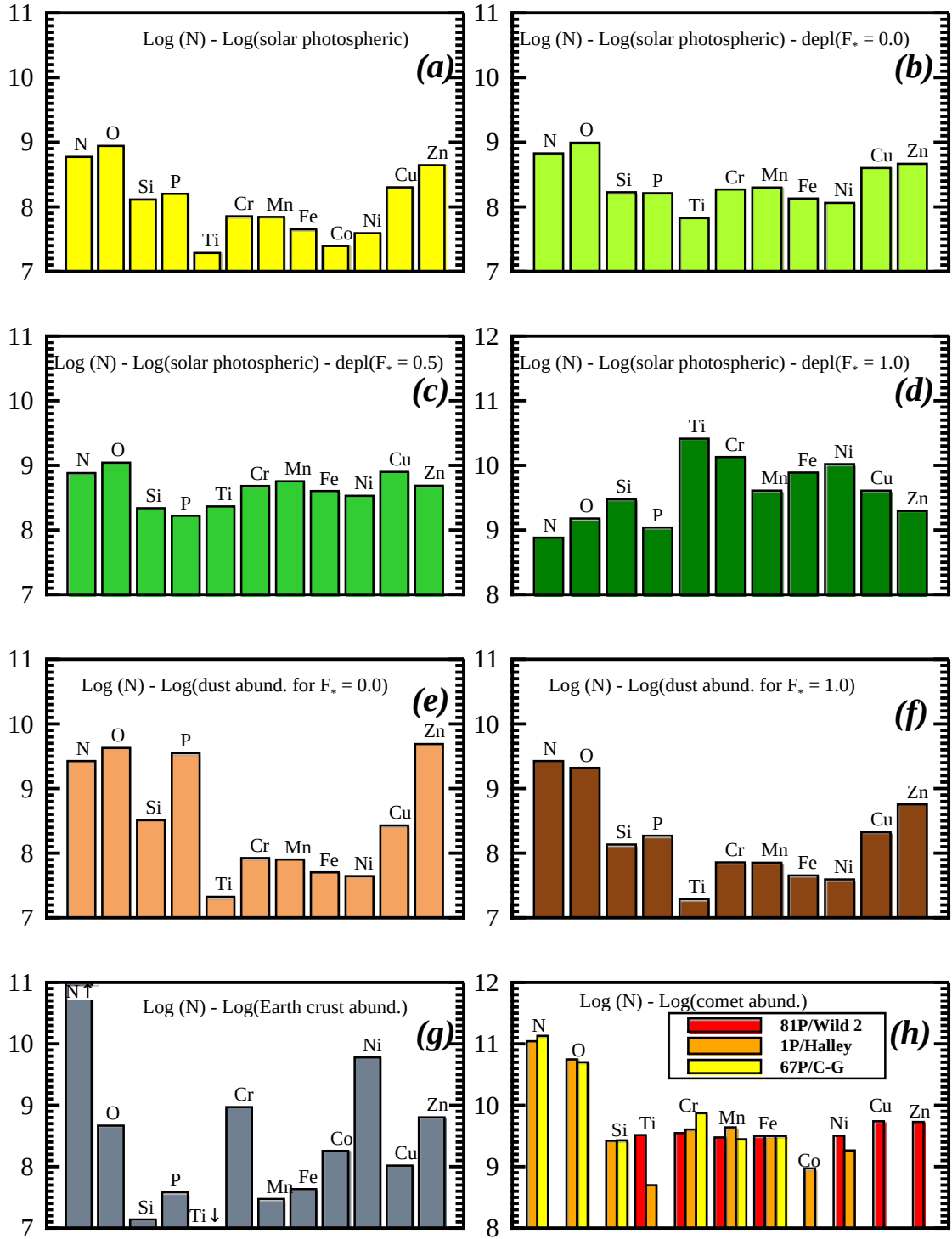}
\caption{Logarithms of our total abundance measurements given in 
Table~\protect\ref{tbl:elements} after subtracting the logarithms of various choices for 
constituents in the circumstellar gas associated with 51~Oph.  Panel $(a)$ depicts a subtraction 
of pristine solar abundance gas.  Panels ($b-d$) represent subtractions for depleted gas-phase 
ISM material with three levels of severity of such depletions.  Panels ($e$ and $f$) represent 
various choices for the conversions of ISM dust solids into gaseous form, as described in the 
text.  Panel ($g$) represents the subtraction of a mixture of elements in the Earth’s crust, and 
Panel ($h$) does the same for samples of dust from three different comets.  In Panel $(h)$, we 
have adjusted the overall abundances for each of the three choices so that they match each 
other for Fe.\label{fig:barplot_comparisons}}
\end{figure}

We present the pattern of the total logarithmic column densities in 
Figure~\ref{fig:barplot_abund}.  Our next step will be to explore some possible interpretations 
of these abundances.  Figure~\ref{fig:barplot_comparisons} presents a number of choices.  In 
panel $(a)$ of this figure, we show a representation of our total logarithmic column densities 
listed in Table~\ref{tbl:elements} after subtracting the logarithm of the solar photospheric 
abundances (on a scale H~=~12) (Asplund et al. 2009) to test the proposition that the gas that 
we measured is dominated either by mass loss from the star or by interstellar material that has 
virtually no depletions of atoms onto dust grains.  The lack of uniformity in the bar heights 
indicates that this is not a favorable interpretation.  Next, we apply the same test for gas that 
has a composition that is similar to that of the general ISM with three possible choices for the 
severity of atomic depletions.  These choices are represented by the expected depletions for 
the parameter $F_*$ (Jenkins 2009), which ranged from values of 0.0 (i.e., very light depletions) 
in Panel $(b)$, 0.5 (moderate depletions) in Panel $(c)$, and 1.0 (heavy depletions) in Panel 
$(d)$.\footnote{The depletions of P and Zn have been modified from the original description by 
Jenkins (2009) to account for the revised $f$-values for transitions of P by Brown et al. (2018) 
and Zn by Kisielius et al. (2015) that replace earlier ones reported by Morton (2003).} The 
moderate depletions ($F_*=0.5$) seem to show the best uniformity in the bar heights.

The lower four panels of Fig.~\ref{fig:barplot_comparisons} test the proposition that the 
original gas that has condensed out of the ISM to form the circumstellar disk has been expelled 
and been replaced by atomic constituents arising from different possible forms of solid material 
that have been converted into gaseous form by collisions or ablation.  Recall that from our 
finding stated in Section~\ref{sec:CO} about the absence of CO features in our spectrum, we 
concluded that we must be sampling gas at some distance from the midplane of the disk.  This 
gas might have been subject to erosion by photoevaporation or a stellar wind and replaced by 
elements arising from destructive processes for solids in the midplane.  Panels $(e)$ and $(f)$ 
represent our abundances after subtracting the pattern of abundances of ISM dust for $F_*=0$ 
and 1, respectively.  The bar heights are quite uneven in these panels.  The same kind of 
presentation is presented for Earth’s crust abundances (Anderson 1989 p. 150), which could 
stand for material ejected from the outer portions of chemically differentiated planets by 
collisions (Panel $(g)$).  We have also tested the pattern that could emerge from the 
destruction of objects that have chemical makeups similar to those of CI chondritic meteorites 
(Lodders 2003).  However, except for volatile elements, the abundances in such primitive 
meteorites closely follow the solar photospheric ones, so we would simply be duplicating what 
we see in Panel $(a)$, except for the elements N and O.  Finally, in Panel $(h)$ we investigate 
the relative differences between our abundances and those found for three different 
determinations of some relative abundances for cometary dust: (1) dust from Comet 
81P/Wild~2 collected in the aerogel on the {\it Stardust\/} mission (Flynn et al. 2006), (2)  
material observed with the impact-ionization time-of-flight mass spectrometer in the PUMA-1 
experiment on the {\it VEGA-1\/}  spacecraft, which encountered Comet 1P/Halley (Jessberger 
et al. 1988), and (3) the sampling of particles from Comet 67P/Churyumov-Gerasimenko by the 
COmetary Secondary Ion Mass Analyzer (COSIMA) on the {\it Rosetta\/} mission (Bardyn et al. 
2017).  Comets 81P/Wild~2 and 67P/Churyumov-Gerasimenko are Jupiter-family comets (with 
a low orbital inclination and perturbed by Jupiter into a short period), and 1P/Halley is a
long-period comet with a high inclination.

From the choices that we presented in Fig.~\ref{fig:barplot_comparisons}, we conclude that 
ISM-type gas that has undergone moderate levels of depletion gives a satisfactory match to the 
abundance pattern that we observed.  Over a limited range of refractory elements that were 
reported, the results for 81P/Wild~2 and 67P/Churyumov-Gerasimenko also seem to show a 
reasonably good uniformity of bar heights, but for Halley and 67P our abundances of N and O 
seem too large.  The ISM dust solids and the Earth’s crust material outcomes show significantly 
uneven bar heights in their respective panels.

\section{Estimates for the Column Densities of H~I and H~II}\label{sec:HI_HII}

Even though we cannot directly observe hydrogen in its neutral or ionized forms, we can make 
use of the good match between the abundance pattern of different elements to that of a mildly 
depleted gas-phase ISM to propose that the volatile elements N and O can be used as proxies 
for H after correcting for abundance differences.  Again referring to the ISM, we conclude that 
the circumstellar N and O should be depleted by about 0.1~dex below their protosolar 
abundances.  From the evidence presented in Section~\ref{sec:outcomes}, we adopt the 
viewpoint that the neutral hydrogen atoms and protons are distributed in a common volume, 
rather than being in separate locations.

The ionization fractions of N and O are coupled to that of H by charge exchange reactions with 
large rate constants.  As we will discuss in Section~\ref{sec:discussion}, the ionization levels for 
all three elements must be maintained by photoionizations caused by energetic radiation in 
excess of the flux from the star at wavelengths below the Lyman limit (and perhaps 
supplemented to some extent by cosmic-ray ionization).

In order to have a good understanding about the strength of the coupling of the relative 
ionizations of nitrogen and oxygen to that of hydrogen, we must know the volume densities 
$n({\rm H}^0)$ and $n({\rm H}^+)~(\approx n(e))$. Our constraints on these quantities cover 
some broad ranges in values.  From the metastable excitations and the ratios of some neutral 
atoms to their ionized counterparts, we feel that it is reasonable to adopt $10^5\lt n(e) \lt 
3\times 10^6\,{\rm cm}^{-3}$.  For neutral hydrogen, we described in 
Section~\ref{sec:neutral_cl} how the abundance of neutral chlorine compared to an upper limit 
to its ionized form led to an expression given Eq.~\ref{eqn: n(H0)_outcome}, but we must 
remember that the specified value for $n({\rm H}^0)$ is a lower limit.  We have no direct way 
to constrain how far above this limit the density could go, other than to argue that if it were to 
reach $\sim 10^8{\rm cm}^{-3}$, our determination of the column density $N({\rm H}^0)$ 
(which we will derive shortly) divided by $n({\rm H}^0)$ would shrink to a thickness $0.1R_g$ 
(i.e., 0.6\,AU), which we can deem to be a reasonable lower limit.

To determine the ionization balance of nitrogen, we can use the charge exchange rate 
constants $C_{+,0}=1.11\times 10^{-13}\,{\rm cm}^3{\rm s}^{-1}$ and $C_{0,+}=5.49\times 
10^{-14}\,{\rm cm}^3{\rm s}^{-1}$ derived by Lin et al. (2005) for $T=8000$\,K.  The 
recombination coefficient for nitrogen is $\alpha=1.1\times 10^{-12}\,{\rm cm}^3{\rm s}^{-1}$ 
at $T=8000$\,K (Shull \& Van Steenberg 1982 ; Nussbaumer \& Storey 1983).  The value of 
$\Gamma({\rm N}^0)$ is about ten times the value of $\Gamma({\rm H}^0)=\alpha^{(2)}({\rm 
H})n(e)^2/n({\rm H}^0)$ after one considers additional secondary ionization processes for both 
elements (Jenkins 2013).  Because the region is optically thick to Lyman limit photons, we adopt 
a recombination rate to levels $n=2$ and higher $\alpha^{(2)}({\rm H})=3.09\times 10^{-
13}{\rm cm}^3{\rm s}^{-1}$ at $T=8000$\,K.  We now can substitute N for Cl in Eq.~\ref{eqn: 
chg_ex_equlib} and solve for $n({\rm N}^+)/n({\rm N}^0)$ to obtain the estimates listed in 
Table~\ref{tbl:relative_ionizations} and compare them to their hydrogen counterparts $n({\rm 
H}^+)/n({\rm H}^0)$, also listed in the table.  We now define an ionization correction factor
\begin{equation}
Y({\rm N})=\log\left[ {1+n({\rm N}^+)/n({\rm N}^0)\over 1+n({\rm H}^+)/n({\rm H}^0)}\right]
\end{equation}

\begin{deluxetable}{
L	
C
C
C
C
}[b]
\tablewidth{0pt}
\tablecolumns{5}
\tablecaption{Relative Ionizations\label{tbl:relative_ionizations}}
\tablehead{
\colhead{} & \colhead{min[$n(e)$]} & \colhead{min[$n(e)$]} & \colhead{max[$n(e)$]} & 
\colhead{max[$n(e)$]}\\
\colhead{Quantity} & \colhead{min[$n({\rm H}^0)$]} & \colhead{max[$n({\rm H}^0)$]} & 
\colhead{min[$n({\rm H}^0)$]} & \colhead{max[$n({\rm H}^0)$]}\\
}
\startdata
n(e)~({\rm cm}^{-3})&10^5&10^5&3\times 10^6&3\times 10^6\\
n({\rm H}^0)~({\rm cm}^{-3})&3.5\times 10^5&10^8&8.4\times 10^6&10^8\\
n({\rm H}^+)/n({\rm H}^0)&0.285&10^{-3}&0.36&0.03\\
n({\rm N}^+)/n({\rm N}^0)&0.63&5.5\times 10^{-4}&0.82&0.0315\\
Y({\rm N})&0.10&0.00&0.13&0.00\\
n({\rm O}^+)/n({\rm O}^0)&0.25&9\times 10^{-4}&0.31&3\times 10^{-2}\\
Y({\rm O})&-0.01&0.00&-0.02&0.00\\
\enddata
\end{deluxetable}

From our observation of \ion{N}{1}, we evaluate our estimate for $\log N({\rm H~I})$ using the 
relation
\begin{equation}\label{eqn: N(HI)}
\log N({\rm H~I})=\log N({\rm N~I})-\log ({\rm N/H})_\odot+0.1+Y({\rm N})=20.93^{+0.20}_{-
0.23}
\end{equation}
where we use our determination of $\log N({\rm N~I})$ listed in Table~\ref{tbl:elements}, and 
the protosolar abundance $\log ({\rm N/H})_\odot=7.87-12$ (Asplund et al. 2009).  The term 
0.1 in the equation is a correction that applies to our estimate of the depletion of N.  The stated 
value given in Eq.~\ref{eqn: N(HI)} is for $Y({\rm N})=0.10$, which recognizes that the FUSE 
spectrum exhibited strong absorptions of the \ion{N}{2} multiplet near 1085\,\AA, as we had 
indicated earlier in Section~\ref{sec:HvsE}.  The uncertainty for the result includes both the 
range in possible $Y({\rm N})$ values and the uncertainty in the measurement of $\log N({\rm 
N~I})$ added in quadrature. 

For oxygen, we may repeat the calculation that we carried out for nitrogen, again adopting 
rates at $T=8000$\,K.  Here, we make use of Draine’s (2011, p.~155) characterization of the 
charge exchange reaction rates calculated by Stancil et al. (1999) between O$^+$ and the three 
fine-structure levels of O$^0$, yielding $C_{+,0}=1.85\times 10^{-9}\,{\rm cm}^3{\rm s}^{-1}$ 
for the three levels and $C_{0,+}=(1.64,~0.84,~3.91)\times 10^{-9}\,{\rm cm}^3{\rm s}^{-1}$, for 
the levels $J=(2,~1,~0)$ of O$^0$.   As with nitrogen, of $\Gamma({\rm O}^0)\approx 
10\Gamma({\rm H}^0)$.  For oxygen, $\alpha=4.36\times 10^{-13}\,{\rm cm}^3{\rm s}^{-1}$ 
(Shull \& Van Steenberg 1982 ; Nussbaumer \& Storey 1983).

An application of Eq.~\ref{eqn: N(HI)} for O with $\log ({\rm O/H})_\odot=8.73-12$ and again 
assuming a depletion correction of 0.1\,dex yields the result $\log N({\rm H~I})=21.00\pm 
0.06$.  The uncertainty for this result is dominated by the measurement uncertainty for 
$N({\rm O~I})$; possible values of $Y({\rm O})$ given in the last row of 
Table~\ref{tbl:relative_ionizations} are close to zero and small compared to the uncertainty in 
$N({\rm O~I})$.  Because the determination of the column density of H~I based on O is 
substantially more precise than that for N, we adopt this value but note that the outcome for N 
is consistent with that for O within our uncertainties of $\log({\rm N~I})$ and $Y({\rm N})$.

If we treat the expression for $n({\rm H}^0)$ in Eq.~\ref{eqn: n(H0)_outcome} as an equality, 
rather than a limit, and then evaluate a characteristic length scale $\ell=N({\rm H~I})/n({\rm 
H}^0)$ we obtain the result $\ell \leq 7.9$ and 190\,AU for $n(e)=3\times 10^6$ and 
$10^5\,{\rm cm}^{-3}$, respectively.  Because both of these values for $\ell$ are larger than our 
representative value $R_g=6$\,AU, the inequality in Eq.~\ref{eqn: n(H0)_outcome} must apply 
to some extent.

Again we can use Eq.~\ref{eqn: n(H0)_outcome} to arrive at a column density for ionized 
hydrogen $\log N({\rm H~II})\leq \log N({\rm H~I})+\log [n({\rm H}^+)/n({\rm H}^0)]=20.45$ 
and 20.56 for $n(e)=10^5$ and $3\times 10^6\,{\rm cm}^{-3}$, respectively.

\section{A Deficiency of Carbon}\label{sec:deficiency_carbon}

\subsection{Abundance Estimate}\label{sec:Cabund_est}

Unlike the preferred ionization stages of N and O, whose neutral forms we could measure using 
weak transitions to obtain column densities, the dominant phase of C (\ion{C}{2}) exhibits only 
two transitions: (1) an allowed transition at 1334.5\,\AA\ that is far too strong to be useful for 
measuring $N$(\ion{C}{2}) and (2) an intersystem line at 2325.4\,\AA\ that is so weak that it 
provides an upper limit for $N$(\ion{C}{2}) that is far too high to be of any use.  To gain an 
understanding about the abundance of carbon, we must direct our attention to the neutral 
form of this element and attempt to relate it to the total abundance of carbon in the gas phase.

As we discussed in Section~\ref{sec:free_electrons}, our inability to detect an absorption from 
one of the excited fine-structure levels of \ion{C}{1} sets a limit on the abundance of neutral 
carbon $\log N({\rm C~I})\lt 2.5\times 10^{12}\,{\rm cm}^{-2}$.  This limit for the amount of 
\ion{C}{1} is substantially lower than the findings by Roberge et al. (2000)  that $N({\rm 
C~I})=(2-4)\times 10^{16}\,{\rm cm}^{-2}$ for $\beta$~Pic and $4.5\times 10^{15}\,{\rm cm}^{-
2}$  for 49~Cet (Roberge et al. 2014).  A comparison of the 49~Cet result against ours is 
especially noteworthy, since they found considerably more \ion{C}{1} than our upper limit, yet 
their sum of results for $N({\rm O~I})$ in the $^3{\rm P}_1$ and $^3{\rm P}_0$ levels equaled 
$1.0\times 10^{14}\,{\rm cm}^{-2}$ compared to our result for the same sum of $1.9\times 
10^{17}\,{\rm cm}^{-2}$. (The A1V spectral type of 49~Cet is only slightly later than that of 
51~Oph.)  From this disparity in the observations of the neutral forms of C and O, we might 
conclude that $R_g$ and/or $n(e)$ for the gas around 49~Cet is larger than what we found for 
51~Oph, thus favoring a shift in the ionization equilibrium toward a higher neutral fraction for 
carbon, or alternatively, the ratio of the C to O abundance in the gas surrounding 49~Cet is 
larger than that associated with 51~Oph.

A contributing factor that drives a deficiency of \ion{C}{1} is the high value of $\Gamma$ at 
$R_g=6$\,AU, but our conclusion given in Table~\ref{tbl:n(e)} that $n(e)\lt 4\times 10^4\,{\rm 
cm}^{-3}$ relied on the abundance of carbon (mostly ionized) being in accord with an 
expectation based on solar abundance ratios relative to the neutral forms of oxygen and 
nitrogen.  To make the calculation of $n(e)$ consistent with our findings for the ionization 
equilibrium outcomes for Na and Zn listed in Table~\ref{tbl:n(e)}, and also put us closer to the 
middle of the realm of $n(e)$ that we found from the metastable level populations, we would 
need to reduce our assumed abundance of C by at least a factor of 40.

\subsection{Expulsion by Radiation Pressure?}\label{sec:Expulsion_radiation_pressure}

We now explore one possibility for depleting C in the gas.  A common theme in the study of 
circumstellar atomic gases is the consideration that the star’s outward radiation pressure can 
weaken or overcome the gravitation attraction for atoms (and some dust grains) in orbit (Beust 
et al. 1989 ; Lagrange et al. 1998 ; Olofsson et al. 2001 ; Liseau 2003 ; Brandeker et al. 2004 ; 
Fernández et al. 2006 ; Xie et al. 2013).  The parameter $\beta$ describes the ratio of these two 
forces, $F_{\rm rad}/F_{\rm grav}$, which is independent of the distance from the star since 
they both scale in proportion to $R_g^{-2}$.  For any transition (which we identify with a 
subscript $i$) with an $f$-value equal to $f_i$, the exposure of an atom to a local flux vector 
$(R_*/R_g)^2F_{\lambda,i}\,({\rm erg~cm}^{-2}{\rm s}^{-1}{\rm \AA}^{-1})$ will absorb photons 
at a rate\footnote{Unlike our previous formulations for the effects of starlight, we do not 
consider a dilution factor $W$ for the radiation at some distance from the star because the 
projections of the momentum transfer events along the radial direction are the relevant 
quantities in the equation, rather than the total photon interaction rates of the atoms.}
\begin{equation}\label{eqn: abs_rate}
\gamma_i={\pi e^2\over 9m_ec}g_if_iN_{\nu,i}={\pi e^2\lambda_i^3\over 
9hm_ec^3}\,\left({R_*\over R_g}\right)^2\,g_if_i10^8F_{\lambda,i}~,
\end{equation}
where $e$ and $m_e$ are the charge and mass of the electron, and $N_{\nu,i}$ is the flux of 
photons per unit frequency $\nu$ for the transition being considered.  The factor $g_i/9$ 
represents the probability that the atom is in the appropriate fine-structure level for the 
transition at the wavelength $\lambda_i$.  Each absorption imparts a radial momentum 
impulse $h/\lambda_i$.  Hence, we find that if the absorption lines for all transitions have low 
central optical depths (i.e., there is no self-shielding of the radiation from the star), we obtain 
for neutral carbon
\begin{equation}\label{eqn: beta}
\beta({\rm C}^0)={h R_g^2 \over G M_*m_{\rm C}}\,\sum_i(\gamma_i/\lambda_i) =10^8{\pi 
e^2R_*^2 \over 9m_ec^3GM_*m_{\rm C}}\, \sum_ig_if_i\lambda_i^2F_{\lambda_i}~,
\end{equation}
where $R_*=6.76R_\odot$ is the effective stellar radius (see 
Section~\ref{sec:radiation_dilution}), $M_*=3.3M_\odot$ is the mass of the star (Jamialahmadi 
et al. 2015), and $m_{\rm C}$ is the mass of the carbon atom (12 amu).  We calculate that for 
51~Oph $\beta({\rm C}^0)=666$ when we evaluate the sum in Eq.~\ref{eqn: beta} over all 
\ion{C}{1} transitions listed in the compilation of Morton (2003).  This value is 20,000 times the 
value 0.033 computed by Fernández et al. (2006) for $\beta$~Pic, largely due to the fact that 
UV flux from $\beta$~Pic with $T_{\rm eff}=8000$\,K is about 4 orders of magnitude weaker 
than that of 51~Oph at the wavelengths of the strongest \ion{C}{1} transitions.  A similar 
calculation by Lagrange et al. (1998) for $\beta$~Pic gave $\beta({\rm C}^0)$ that ranged 
between 0.011 and 0.095, depending on the amount of self-shielding in the lines.

We now argue that even though $\beta({\rm C}^0)$ is quite large (considerably in excess of 0.5,  
which is the threshold for becoming unbound), the atoms are unlikely to migrate outward.  At 
6\,AU from the star, each atom experiences an average rate of absorption $\sum_i 
\gamma_i=1.63\,{\rm s}^{-1}$, while concurrently it is exposed to an ionization rate 
$\Gamma=3.4\times 10^{-3}{\rm s}^{-1}$.  After a carbon ion is neutralized, we expect that 
over the survival time of its neutral form, $1/\Gamma$, the radiation pressure accelerates it in 
a radial direction and increases its kinetic energy by $m_{\rm C}\Delta v^2/2$, where
\begin{equation}\label{eqn: Delta v}
\Delta v={F_{\rm rad}\over m_{\rm C}\Gamma}=\beta({\rm C}^0){GM_*\over R_g^2\Gamma}~.
\end{equation}
(Recall from our discussion in Section~\ref{sec:free_electrons} that the ionization of ${\rm 
C}^0$ by charge exchange with protons is negligible compared to $\Gamma$, but $\Gamma$ 
could be enhanced by secondary ionizations arising from energetic electrons produced by the 
ionization of H.)  We note that because $\Gamma$ is proportional to $R_g^{-2}$ (for $R_g\gg 
R_*$), the value of $\Delta v$ does not change with $R_g$.  The kinetic energy of the Keplerian 
motion
\begin{equation}\label{eqn: E_K}
E_{\rm K}={m_{\rm C}v_{\rm K}^2\over 2}={m_{\rm C}GM_*\over 2R_g}
\end{equation}
receives a relative boost
\begin{equation}\label{eqn:Eboost}
{v_{\rm K}^2+\Delta v^2\over v_{\rm K}^2}=1+{\beta({\rm C}^0)^2GM_*\over 
R_g^3\Gamma^2}~.
\end{equation}
Numerically, we find that Eq.~\ref{eqn:Eboost} yields a result for the fractional increase in 
$E_{\rm K}$ equal to $1+2.3\times 10^{-5}$ for $R_g=6$\,AU.  After receiving this small boost 
in kinetic energy, the neutral atom becomes ionized and remains in that state for a time 
interval that averages $[\alpha n(e)]^{-1}=1.1\times 10^7{\rm s}$ for $n(e)=10^5{\rm cm}^{-3}$ 
(or one-tenth that time for $n(e)= 10^6{\rm cm}^{-3}$).  The value that we assign for $\alpha$ 
at $T=8000$\,K still applies, because the suprathermal $\Delta v=0.1\kms$ arising from 
Eq.~\ref{eqn: Delta v} is small compared to the thermal velocities.  During that time interval, 
the strong coupling of the carbon ions to other ions via Coulomb collisions (and possibly an 
embedded magnetic field) will force them to surrender their enhanced velocity $\Delta v$ and 
merge with the flow.  Details on the physics of this ionic braking process are given by Beust et 
al. (1989), Fernández et al. (2006) and Xie et al. (2013).

We now conclude that even though $\beta({\rm C}^0)$ is large, there presently is little chance 
that the carbon atoms can migrate outward.  Instead, they must blend in with the circulating 
river of ions in Keplerian orbits.  At some much earlier time when the protostar was emitting 
strong radiation but not at energies capable of ionizing the medium, the carbon may have had a 
chance to escape.  However, it seems puzzling that carbon in an atomic form is currently not 
being replenished by dissociating CO molecules at the boundary of the thin molecular disk.  
Also, the present-day deficiency of carbon may present a challenge to the notion that the gas is 
composed of byproducts from cometary destruction that we discussed at the end of 
Section~\ref{sec:element_abundances}.  We note, however, that Roberge et al. (2002) found a 
lack of C compared to expectations from N and Fe for transient gas components infalling 
toward 51~Oph.

It is likely that the elements O and N can maintain a neutral state much longer than the dwell 
time for neutral C.  However, we expect that their $\beta$ values are considerably lower than 
that of C$^0$ because they have fewer transitions in the UV and these transitions are strongly 
saturated.  (In computations of $\beta$, it is better to use the actual line equivalent widths that 
were or would have been observed instead of those that apply to unsaturated lines, as was 
done for Eqs.~\ref{eqn: abs_rate} and \ref{eqn: beta}).

\section{Discussion and Summary}\label{sec:discussion}

The spectrum of 51~Oph has a rich assortment of absorption features arising from atomic 
constituents in a region within about 10\,AU of the star.  We have made use of 304 different 
transitions from 16 different elements for which we could measure the column densities for a 
total of 98 different atomic levels, most of which are highly excited and can come only from the 
environment near the star and not the foreground ISM.  For 9 additional species, we searched 
for 14 strong transitions but could only determine upper (or lower) limits for the column 
densities.  After determining column densities from the spectral features, we were able to 
derive a number of conclusions on the nature of the gas in the vicinity of the star.  These 
conclusions are summarized very briefly in Table~\ref{tbl:conclusions}.

\begin{deluxetable}{
C	
c	
l	
}[b]
\tablewidth{0pt}
\tablecolumns{3}
\tablecaption{Summary of Conclusions on the Gas in the Sightline\label{tbl:conclusions}}
\tablehead{
\colhead{Property} & \colhead{Value} & \colhead{Method}
}
\startdata
n(e)&$10^5-3\times 10^6\,{\rm cm}^{-3}$&\ion{N}{1}, \ion{Fe}{2}, and \ion{Ni}{2} metastable 
populations (Sections~\ref{sec:NFeNi_exc} and \ref{sec:outcomes})\\
&&\ion{Na}{1}, \ion{Zn}{1} ionization equilibria (Section~\ref{sec:free_electrons})\\[8pt]
n({\rm H})&$\geq 3.5\times 10^5{\rm cm}^{-3}$&\ion{Cl}{1} and \ion{Cl}{2}  charge exchange 
(Section~\ref{sec:neutral_cl})\\[8pt]
N({\rm H~I})&$1.0\times 10^{21}\,{\rm cm}^{-2}$&$N$(O~I) and $N$(N~I) 
(Section~\ref{sec:element_abundances}) combined with\\
&&charge exchange and protosolar abundances (Section~\ref{sec:HI_HII})\\[8pt]
N({\rm H~II})&$\leq 2.8\times 10^{20}\,{\rm cm}^{-2}$&(Section~\ref{sec:HI_HII})\\[8pt]
T&8000\,K&\ion{N}{1}, \ion{Fe}{2}, and \ion{Ni}{2} metastable populations 
(Sections~\ref{sec:NFeNi_exc} and \ref{sec:outcomes})\\[8pt]
{\rm Element}&A pattern consistent with&Contents of 
Section~\ref{sec:element_abundances}\\ 
{\rm Abundances}&a mildly depleted ISM&and the best matches in 
Fig.~\ref{fig:barplot_comparisons}\\
&or Jupiter-class comet dust\\[8pt]
{\rm Representative~distance}&6\,AU&\ion{Fe}{2}, \ion{Ni}{2} metastable populations 
(Sections~\ref{sec:NFeNi_exc} and \ref{sec:outcomes})\\
{\rm from~the~star}~R_g\\
\enddata
\end{deluxetable}
\newpage
\subsection{Disk Orientation}\label{summarysec:disk_orientation}

In Section~\ref{sec:CO} we noted that the contrast between the absence of both CO and H$_2$ 
absorptions in the UV spectrum of 51~Oph and the detection of CO in emission in the far 
infrared indicates that we must be sampling regions away from the central plane of the disk 
where the molecules are shielded and thus protected from dissociating radiation that can 
reduce the molecular fraction to a very low value.  Thus, while the disk may be seem to be 
nearly edge on from our viewpoint, it is not exactly so.  This slight misalignment could explain 
why we can view resonantly scattered starlight in the bottoms of the saturated absorption 
features of \ion{O}{1}.  Our results, when combined with a future, accurate determination of 
the disk inclination,\footnote{Determinations that we could find in the literature: 
$88\arcdeg^{+2}_{-35}$ (Thi et al. 2005), $\gt 83\arcdeg$ (Berthoud et al. 2007), and 80\arcdeg 
(no uncertainty specified) (Thi et al. 2013).} would help to constrain the scale height $h/r$ of 
the molecular material.
\subsection{Ionization State}\label{summarysec:ionization_state}

Our sight line to 51~Oph consists of significant portions of both neutral and ionized hydrogen, 
which we could not observe directly, but instead could be sensed from the abundances of 
neutral and ionized nitrogen.  Further support for the existence of both kinds of gas arises from 
the metastable level populations of \ion{N}{1} that indicate large electron densities, along with 
the surprisingly strong presence of neutral chlorine that must arise from charge exchange 
reactions with neutral hydrogen.

\subsection{Excitation Mechanisms}\label{summarysec:excitation}

Using atomic data that we could find in the literature for \ion{N}{1}, \ion{Fe}{2}, and \ion{Ni}{2}, 
we can interpret our measurements of their excited metastable levels in a manner that offered 
important clues on the condition and location of the gas.  The populations of these levels are 
maintained by a combination of optical pumping by starlight and collisions with electrons.  The 
findings for the strengths of these two effects are somewhat degenerate with each other, but 
our MCMC analysis allowed us to focus on a restricted range for their possible combinations.  
For the conditions that we found, collisions with neutral hydrogen atoms should have a minimal 
effect on such excitations.

We acknowledge that we have oversimplified our descriptions of the metastable level 
excitations in terms of a gas at a particular distance from the star which has a uniform electron 
density and temperature.  Nevertheless, future interpretations that invoke interpretations of 
more extended gas structures having variable conditions should still be able to make use of our 
simplified conclusions on what we can regard as density-weighted values for these parameters 
as a starting point.  Such modeling may also include a picture where the \ion{N}{1} is located in 
a different region than that which highlights the presence of \ion{Fe}{2} and \ion{Ni}{2}.  Some 
support for a difference in the distributions of ions and neutrals arises from the fact that the 
ions show both a principal, narrow-velocity component (Component~1 described in 
Section~\ref{sec:velocity_components}) and a broader, weaker, slightly displaced one 
(Component~2), while the neutrals show only Component~1 with no evidence for 
Component~2, except perhaps for a one-percent contribution that may be evident in a very 
strong line of \ion{O}{1}. 

\subsection{Electron Density}\label{summarysec:electron_dens}

An important finding of our investigation on the metastable level populations is that the  
electron densities fall somewhere in the range $10^4\lt n(e) \lt 10^7\,{\rm cm}^{-3}$.    The 
neutral to ionized fractions of Na and Zn indicate a somewhat narrower range $10^5\lt n(e) \lt 
3\times 10^6\,{\rm cm}^{-3}$.  Let us now consider that the gas is located at a distance of 
6\,AU from the star, as indicated by the most likely excitations of \ion{Fe}{2} and \ion{Ni}{2}.  
Could an electron density of order $n(e)\approx 10^5\,{\rm cm}^{-3}$ be maintained simply by 
the amount of radiation from the star at wavelengths below the Lyman limit?  We are unsure of 
the geometrical properties of the gas that we are sampling.  Nevertheless, it is instructive to 
imagine a Strömgren sphere that has an inner truncation radius $r_1$ and an outer,
ionization-limited radius $r_2$ and set $r_2=6$\,AU.  In this instance, we find that with a model 
for the ionizing radiation emitted by the star (Fossati et al. 2018), the thickness $r_2-r_1$ of a 
shell containing the fully ionized gas would only be 0.13\,AU if no dust grains were present to 
absorb some of the radiation.  This calculation points to the implausibility that photons from 
the star’s photosphere can maintain the ionization and indicates that some other means of 
ionizing the gas must be operating.  Moreover, the coexistence of neutral nitrogen and 
electrons indicated by the \ion{N}{1} metastable populations disfavors the conventional 
Strömgren sphere that consists of a fully ionized gas bounded by completely neutral gas.

Generally speaking, early A-type stars are not strong emitters of X-ray radiation compared to 
stars of earlier or later spectral types.  Nevertheless,  weak X-ray fluxes have been detected 
from such stars (Schröder \& Schmitt 2007).   A more promising source of EUV and soft X-ray 
radiation is the emission by shocked gas in an accretion column near the star.  The existence of 
such an accretion column is supported by very broad H$\alpha$ emission (Manoj et al. 2006 ; 
Mendigutía et al. 2011 ; Arun et al. 2019).  Ionization of the circumstellar gas by very energetic 
photons is more conducive to the prospect of having both neutral and ionized hydrogen 
coexisting with each other in large regions, rather than being separated by a sharp boundary at 
the point where the opacity to softer ionizing radiation suddenly materializes.

Is there any direct observational evidence for X-ray emission arising from 51~Oph?  The {\it 
Rosat\/} point-source catalog indicates the presence of emission from a source that was 
measured to be 172\arcsec\ from 51~Oph, but the position mismatch of this magnitude 
indicates that it is unlikely (but not impossible) that this source is coincident with the position of 
the star [see, e.g., Figure~1 of Berghöfer et al. (1996)].  A more definitive answer to the 
question of observable X-ray emission may arise from an all-sky survey, eRASS, that will be 
undertaken by the eROSITA instrument on the Russian-German {\it Spektr-RG\/} space 
observatory launched in 2019.

We are unable to test for the presence of X-ray or EUV radiation by looking for evidence of ions 
that can be created by photons with energies significantly greater than 13.6\,eV because they 
will be neutralized by charge exchange with the neutral hydrogen that we know to be present.

\subsection{Gas Temperature}\label{summarysec:gas_temp}

Our interpretation of the populations of the \ion{N}{1} metastable levels yields an outcome for 
the temperature $T=8000$\,K, and this value seems to hold over very broad ranges of possible 
electron densities and distances from the star.  As the level of ionization of the gas is large, we 
can expect that the high temperature is created by a significant heat input from the energetic 
electrons produced by the ionization of hydrogen.  It is interesting to note that this 
temperature coincides with the base of the ``Lyman-$\alpha$ Wall,” which is a location where 
the cooling rate from neutral hydrogen excitation rises steeply with temperature above the 
various other forms of atomic cooling; see, e.g. Figure~2 of Dalgarno \& McCray (1972).  
(However, we point out that the other atomic cooling rates are much lower than indicated in 
that figure, because the gas is substantially denser than the critical densities of the collisionally 
excited levels that are most responsible for cooling at slightly lower temperatures.)

\subsection{Atomic Abundances}\label{summarysec:atomic_abund}

In the circumstellar material, the excitations of different atoms to excited levels is so strong 
that our determinations of the abundances of various elements is heavily dependent on our 
measurements of the column densities of both ground-state and upper level populations, 
together with our estimates for the amounts of atoms in levels that we could not observe.  For 
any particular element, the ratio of amounts of observed to total levels is a driver for the 
reliability of our abundance measurement, and it varied from 5\% to 100\% across different 
elements. The outcomes of our column density measurements range from $\log N = 12.14$ for 
Ti to 17.63 for O.  If we exclude carbon, the pattern of abundances across different elements 
seems to be most consistent with the gas composition of the interstellar medium with a 
moderate level of depletion.  However, we also find reasonably good matches to the 
abundance patterns of solid material associated with Jupiter-family comets.  In the 51~Oph 
system, solid cometary material with a similar composition could have been converted to 
gaseous form from collisions or during close approaches toward the star.  Such an outlook is 
consistent with the observations of transient absorption lines associated with falling 
evaporating bodies (FEBs) detected for 51~Oph and other stars with circumstellar disks
(Vidal-Madjar et al. 1994, 1998 ; Beust et al. 1998, 2001 ; 
Karmann et al. 2001, 2003 ; Thébault \& Beust 2001 ; Roberge et al. 2002 ;
Beust \& Valiron 2007 ; Welsh \& Montgomery 2013 ; Kiefer et al. 2014, 2019 ; Eiroa et al. 2016 ; 
Vidal-Madjar et al. 2017 ; Zieba et al. 2019).  

Debris disks around A-type stars typically span distances from tens to hundreds of AU from the 
central star (Hughes et al. 2018, Fig. 3).  One can imagine that the production of gas arising 
from the collisions of solid bodies strongly favors zones close to the star where the most 
crowding occurs.  This may explain why our characteristic distance $R_g=6$\,AU from the star is 
smaller than the general radial extent of the debris disk (which, as far as we know, has not yet 
been well measured for 51~Oph).

One prominent deviation in the abundances is that of carbon.  From the lack of \ion{C}{1} and 
comparing its ionization equilibrium with those of other species, we concluded that C is 
relatively deficient (by more than one order of magnitude) compared to other volatile elements 
such as O and N.  Even though neutral carbon atoms are subject to a repulsive radiation force 
that is considerably stronger than the gravitational attraction, these atoms become ionized 
before they can acquire a significant outward velocity, and then their migration is halted by 
Coulomb collisions that cause them to blend in and become bound to the sea of circulating 
ions.

\subsection{The Role of UV Spectroscopy}\label{summarysec:UV_spectroscopy}

This paper has highlighted how UV absorption-line spectroscopy can reveal many important 
conclusions on the physical nature and composition of gases in the circumstellar environments 
in a late stage of development. The observations are most easily interpreted when one chooses 
a star that has a radial velocity that differs from that of any foreground material in the ISM.  
Features from metastable levels provide powerful insights on the condition and location of the 
gas, and they frequently appear in the spectra of hot stars with nearly edge-on disk systems.  
The observations of atomic element abundances are complementary to those that investigate 
infrared and millimeter emission, which reveal the dust and molecular constituents.
 
\acknowledgments
This research was supported by an archival research grant nr. HST-AR-15029.001-A provided by 
NASA through the Space Telescope Science Institute (STScI), which is operated by the 
Associations of Universities for Research in Astronomy, Incorporated, under NASA contract 
NAS5-26555.  All of the ultraviolet spectroscopic data analyzed in this paper were obtained 
from the {\it Mikulski Archive for Space Telescopes\/} (MAST) maintained by the STScI.  Specific 
observations used in this paper can be accessed via the following collections of GHRS, STIS and 
FUSE data on the MAST doi: \dataset[10.17909/t9-b6j3-w085]{\doi{10.17909/t9-b6j3-w085}}
Our determination of the distance to 51~Oph came from data from the European Space Agency 
(ESA) mission {\it Gaia} (\url{https://www.cosmos.esa.int/gaia}), processed by the {\it Gaia}
Data Processing and Analysis Consortium (DPAC,
\url{https://www.cosmos.esa.int/web/gaia/dpac/consortium}). Funding for the DPAC
has been provided by national institutions, in particular the institutions
participating in the {\it Gaia} Multilateral Agreement.  The UVES observations were collected 
from European Southern Observatory (ESO) Science Archive Facility from which we downloaded 
observations taken under the ESO observing program 079.C-0789(A).  We obtained information 
on atomic energy levels from the NIST catalog VI/74 available on the Strasbourg VizieR website 
\url{ http://cdsarc.u-strasbg.fr/viz-bin/Cat?VI/74}. We thank B.~T.~Draine for his review and 
comments on a late draft of this paper.

\facilities{HST (STIS, GHRS), FUSE, HUT, VLT:Kueyen (UVES)}
\software{Owens, developed by Martin Lemoine (Institut d’Astrophysique de Paris) and the 
French {\it FUSE\/} Science Team}
\newpage
\appendix
\section{Atomic Absorption Lines}\label{sec:atomic_lines}
Table~\ref{tbl:atomic_lines} lists the atomic lines that we measured to determine column 
densities, along with the $f$-values that we adopted and their references.  All wavelengths are 
vacuum values, regardless of whether they are in the UV or visible.  The logarithmic relative 
uncertainties in the $f$-values listed in Column (5) of the table are from the percentage 
classifications given in the NIST compilation (see note $a$ in the table), except for the elements 
Cr and Ni where uncertainties were specified in the original references.  Strongly saturated lines 
are not included in the table, except for (1) one case where a line from \ion{N}{1} in the ground 
state could be used in checking for the average attenuation of the flux for optical pumping, (2) 
\ion{O}{1} lines featured in the exposition in Section~\ref{sec:allowed_OI}, and (3) absorption 
from a metastable \ion{O}{1} level that was used for determining a lower limit for the column 
density in Section~\ref{sec:HvsE}.
\startlongtable
\begin{deluxetable}{
l	
l	
r	
c	
c	
c	
}
\tablewidth{0pt}
\tablecaption{Properties of Atomic Lines\label{tbl:atomic_lines}}
\tablehead{
\colhead{Wavelength} & \colhead{Species} & \colhead{Excitation} & \colhead{$f$-value} & 
\colhead{$f$-value Relative} & \colhead{$f$-value}\\
\colhead{(\AA)} & \colhead{} & \colhead{(cm$^{-1}$)} & \colhead{} &
\colhead{Uncertainty (dex)} & \colhead{Source}\tablenotemark{a}\\
\colhead{(1)} &
\colhead{(2)} &
\colhead{(3)} &
\colhead{(4)} &
\colhead{(5)} &
\colhead{(6)}\\
}
\startdata

1560.3092 &C~I  &    0&1.28E$-$02&0.01&(1)\\
1656.9283 &C~I  &    0&1.40E$-$01&0.01&(1)\\
1328.8333 &C~I  &    0&6.31E$-$02&0.04&(1)\\
1329.571   &C~I &   43&5.69E$-$02&\nodata&(1)\\
1329.593   &C~I &   43&1.89E$-$02&\nodata&(1)\\
2325.4029&C~II  &    0&4.78E$-$08&0.003&(1)\\
1134.9803\tablenotemark{b}&N~I  &     0&4.15E$-$02&0.04&(1)\\
1159.8168&N~I   &    0&9.95E$-$06&0.01&(1)\\
1492.6250 &N~I  &19224&6.93E$-$02&0.03&(2)\\
1492.8200 &N~I  &19233&1.09E$-$02&0.03&(2)\\
1494.6751 &N~I  &19233&5.80E$-$02&0.03&(2)\\
1310.9431 &N~I  &28838&3.12E$-$02&0.04&(2)\\
1318.9980 &N~I  &28838&1.20E$-$02&0.03&(2)\\
1319.6695 &N~I  &28838&8.90E$-$03&0.03&(2)\\
1327.9172 &N~I  &28838&2.50E$-$03&0.04&(2)\\
1411.9310 &N~I  &28838&2.67E$-$02&0.01&(2)\\
1742.7192 &N~I  &28838&1.93E$-$02&0.04&(2)\\
1745.2485 &N~I  &28838&3.82E$-$02&0.04&(2)\\
1310.5403 &N~I  &28839&2.97E$-$02&0.04&(2)\\
1310.9498 &N~I  &28839&4.51E$-$03&0.04&(2)\\
1319.0050 &N~I  &28839&2.97E$-$03&0.03&(2)\\
1319.6762 &N~I  &28839&1.50E$-$02&0.03&(2)\\
1326.5709 &N~I  &28839&1.77E$-$03&0.04&(2)\\
1411.9387 &N~I  &28839&3.03E$-$03&0.01&(2)\\
1411.9483 &N~I  &28839&2.39E$-$02&0.01& (2)\\
1742.7309 &N~I  &28839&4.80E$-$02&0.04&(2)\\
1745.2603 &N~I  &28839&9.16E$-$03&0.04&(2)\\
1083.9937\tablenotemark{b} &N~II &0&1.11E$-$01&0.01&(1)\\
1084.5659\tablenotemark{b} &N~II &49&2.72E$-$02&0.01&(1)\\
1084.5841\tablenotemark{b} &N~II &49&8.30E$-$02&0.01&(1)\\
1085.5328\tablenotemark{b} &N~II &131&1.06E$-$03&0.01&(1)\\
1085.5511\tablenotemark{b} &N~II &131&1.61E$-$02&0.01&(1)\\
1085.7096\tablenotemark{b} &N~II &131&9.21E$-$02&0.01&(1)\\
1302.1685\tablenotemark{c} &O~I  &    0&4.80E$-$02&0.01&(1)\\
1355.5977 &O~I  &    0&1.16E$-$06&0.04&(1)\\
1304.8576\tablenotemark{c} &O~I  &  158&4.78E$-$02&0.01&(1)\\
1358.5123 &O~I  &  158&6.27E$-$07&0.04&(1)\\
1306.0286\tablenotemark{c} &O~I  &  227&4.78E$-$02&0.01& (1)\\
1152.1512 &O~I & 15868 &1.08E$-$01&0.04&(3)\\ 
5897.5581 &Na~I &    0&3.201E$-$01&$<$0.01&(1)\\
5891.5833 &Na~I &    0&6.408E$-$01&$<$0.01&(1)\\
2026.4768 &Mg~I&    0&1.13E$-$01 &0.01&(1)\\
1845.5205 &Si~I&      0&2.70E$-$01 &0.07&(1)\\
1304.3702 &Si~II&    0&9.17E$-$02&0.08&(1)\\
1808.0126 &Si~II&    0&2.08E$-$03&0.12&(1)\\
1309.2758 &Si~II&  287&9.13E$-$02&0.08&(1)\\
1816.9285 &Si~II&  287&1.66E$-$03&0.12&(1)\\
1817.4512 &Si~II&  287&1.29E$-$04&0.12&(1)\\
1381.4760 &P~I  &    0&3.16E$-$01&\nodata&(1)\\
1152.8180 &P~II &    0&2.45E$-$01&\nodata&(1)\\
1155.0137 &P~II & 165&6.10E$-$02&\nodata&(1)\\
1153.9951 &P~II & 469&1.86E$-$01&\nodata&(1)\\
1425.0300 &S~I &      0&1.25E$-$01&0.03&(1)\\
1335.7258\tablenotemark{d} &Cl~I &    0&3.13E$-$02&0.12&(1)\\
1347.2396 &Cl~I &    0&1.53E$-$01&0.08&(1)\\
1351.6561 &Cl~I &  882&1.21E$-$01&0.08&(1)\\
1363.4475 &Cl~I &  882&5.50E$-$02&0.08&(1)\\
1071.0358&Cl~II&      0&1.50E$-$02&0.02&(1)\\
4227.918&Ca~I&         0&1.77E+00&0.01&(1)\\
3934.7750&Ca~II&     0&1.267E$-$01&$\lt 0.01$&(1)\\
3969.5901&Ca~II&     0&3.116E$-$01&$\lt 0.01$&(1)\\
3384.7304&Ti~II&       0&3.58E$-$01&0.03&(1)\\
2056.2568 &Cr~II&    0&1.03E$-$01&0.04&(1)\\
2062.2359 &Cr~II&    0&7.59E$-$02&0.04&(1)\\
2066.1638 &Cr~II&    0&5.12E$-$02&0.04&(1)\\
2744.4530 &Cr~II&11961&1.730E$-$01&0.04&(4)\\
2669.5010 &Cr~II&12032&7.744E$-$02&0.04&(4)\\
2679.5840 &Cr~II&12032&1.294E$-$01&0.03&(4)\\
2742.8430 &Cr~II&12032&5.104E$-$02&0.04&(4)\\
2749.7965 &Cr~II&12032&1.318E$-$01&0.04&(4)\\
2666.8120 &Cr~II&12147&1.306E$-$01&0.11&(4)\\
2672.5980 &Cr~II&12147&7.778E$-$02&0.03&(4)\\
2751.5400 &Cr~II&12147&1.084E$-$01&0.04&(4)\\
2758.5350 &Cr~II&12147&7.583E$-$02&0.04&(4)\\
2664.2140 &Cr~II&12303&7.013E$-$02&0.04&(4)\\
2677.9560\tablenotemark{e} &Cr~II&12303&1.282E$-$01&\nodata&(4)\\
2752.6780 &Cr~II&12303&6.323E$-$02&0.04&(4)\\
2763.4050 &Cr~II&12303&1.364E$-$01&0.04&(4)\\
2844.0850 &Cr~II&12303&2.870E$-$01&0.01&(4)\\
2677.9540\tablenotemark{e} &Cr~II&12496&2.244E$-$01&\nodata&(4)\\
2767.3480 &Cr~II&12496&2.051E$-$01&0.04&(4)\\
2836.4630\tablenotemark{f} &Cr~II&12496&3.614E$-$01&0.01&(4)\\
3422.1906 &Cr~II&19528&9.53E$-$02&0.03&(4)\\
3404.2968 &Cr~II&19631&5.42E$-$02&0.07&(4)\\
3343.5421 &Cr~II&19797&2.575E$-$02&0.02&(4)\\
3359.4649 &Cr~II&19797&3.571E$-$02&0.06&(4)\\
3423.7209 &Cr~II&19797&6.29E$-$02&0.02&(4)\\
3369.0170 &Cr~II&20024&8.549E$-$02&0.02&(4)\\
3409.7432 &Cr~II&20024&4.71E$-$02&0.04&(4)\\
3380.7907 &Cr~II&25046&3.40E$-$02&0.05&(4)\\
2299.6631 &Mn~II&    0&4.81E$-$04&0.12&(1)\\
2305.7141 &Mn~II&    0&1.15E$-$03&0.08&(1)\\
2576.8770 &Mn~II&    0&3.61E$-$01&0.01&(1)\\
2594.4990\tablenotemark{g} &Mn~II&    0&2.80E$-$01&0.01&(1)\\
2606.4620 &Mn~II&    0&1.98E$-$01&0.01&(1)\\
3442.9710 &Mn~II&14325&5.000E$-$02&0.03&(5)\\
3461.3050 &Mn~II&14593&3.340E$-$02&0.03&(5)\\
3489.6730 &Mn~II&14901&3.85E$-$02&0.03&(5)\\
2610.9810 &Mn~II&27547&3.500E$-$01&0.12&(6)\\
3720.9928    &Fe~I&0&4.11E$-$02&0.01&(1)\\
1611.2004 &Fe~II&    0&1.38E$-$03&0.12&(1)\\
2249.8768 &Fe~II&    0&1.82E$-$03&0.04&(1)\\
2260.7805 &Fe~II&    0&2.44E$-$03&0.04&(1)\\
1618.4680 &Fe~II&  384&2.14E$-$02&0.01&(1)\\
2146.7218\tablenotemark{h} &Fe~II&  384&2.35E$-$04&\nodata&(1)\\
2252.2537 &Fe~II&  384&5.58E$-$04&$>$0.3&(1)\\
2253.8254 &Fe~II&  384&3.23E$-$03&0.08&(1)\\
2269.5248\tablenotemark{h} &Fe~II&  384&3.06E$-$04&0.08&(1)\\
2280.6202 &Fe~II&  384&4.37E$-$03&0.03&(1)\\
2333.5156 &Fe~II&  384&7.78E$-$02&0.08&(1)\\
2365.5518 &Fe~II&  384&4.95E$-$02&0.03&(1)\\
2383.7884 &Fe~II&  384&5.57E$-$03&0.08&(1)\\
2389.3582 &Fe~II&  384&8.25E$-$02&0.03&(1)\\
2396.3559\tablenotemark{i} &Fe~II&  384&2.88E$-$01&0.03&(1)\\
2599.1465 &Fe~II&  384&1.08E$-$01&0.03&(1)\\
2612.6542 &Fe~II&  384&1.26E$-$01&0.03&(1)\\
2626.4511 &Fe~II&  384&4.41E$-$02&0.03&(1)\\
1625.9123 &Fe~II&  667&6.08E$-$03&0.08&(1)\\
1629.1596\tablenotemark{j} &Fe~II&  667&3.67E$-$02&0.01&(1)\\
2251.6338 &Fe~II&  667&2.20E$-$03&0.08&(1)\\
2268.2878 &Fe~II&  667&3.62E$-$03&0.08&(1)\\
2328.1212 &Fe~II&  667&3.45E$-$02&0.03&(1)\\
2349.0223\tablenotemark{k} &Fe~II&  667&8.98E$-$02&0.08&(1)\\
2381.4887 &Fe~II&  667&3.38E$-$02&0.04&(1)\\
2396.1497\tablenotemark{i} &Fe~II&  667&1.53E$-$02&0.03&(1)\\
2607.8664 &Fe~II&  667&1.18E$-$01&0.04&(1)\\
2618.3991 &Fe~II&  667&5.05E$-$02&0.04&(1)\\
2632.1081 &Fe~II&  667&8.60E$-$02&0.03&(1)\\
2250.8727 &Fe~II&  862&1.35E$-$03&0.08&(1)\\
2255.1048 &Fe~II&  862&2.11E$-$04&$>$0.3&(1)\\
2261.5588 &Fe~II&  862&2.25E$-$03&0.08&(1)\\
2338.7233 &Fe~II&  862&8.97E$-$02&0.03&(1)\\
2359.8266 &Fe~II&  862&6.79E$-$02&0.08&(1)\\
2614.6044 &Fe~II&  862&1.08E$-$01&0.03&(1)\\
2621.1905 &Fe~II&  862&3.93E$-$03&0.04&(1)\\
2631.8312 &Fe~II&  862&1.31E$-$01&0.03&(1)\\
2256.6869 &Fe~II&  977&1.17E$-$03&\nodata &(1)\\
2345.0011 &Fe~II&  977&1.53E$-$01&0.03&(1)\\
2622.4518 &Fe~II&  977&5.60E$-$02&0.03&(1)\\
2629.0777 &Fe~II&  977&1.73E$-$01&0.04&(1)\\
2332.0233 &Fe~II& 1872&2.07E$-$02&0.03&(7)\\
2348.8346\tablenotemark{k} &Fe~II& 1872&4.30E$-$02&0.03&(7)\\
2360.7210 &Fe~II& 1872&3.00E$-$02&0.04&(7,8)\\
2361.0159 &Fe~II& 2430&3.90E$-$02&0.03&(8)\\
2362.7434 &Fe~II& 2430&1.18E$-$02&0.08&(7,8)\\
2392.2069 &Fe~II& 2430&4.05E$-$03&0.04&(7,8)\\
2355.6103 &Fe~II& 2837&1.48E$-$02&0.04&(7)\\
2367.3166 &Fe~II& 2837&8.50E$-$03&0.04&(7)\\
2369.3195 &Fe~II& 2837&3.40E$-$02&0.03&(7)\\
2383.9718 &Fe~II& 2837&3.06E$-$02&0.03&(8)\\
2385.7331 &Fe~II& 2837&4.09E$-$03&0.08&(7,8)\\
2371.2226 &Fe~II& 3117&1.46E$-$02&0.04&(7,8)\\
2375.9187 &Fe~II& 3117&4.20E$-$02&0.04&(7)\\
2385.1148 &Fe~II& 3117&2.75E$-$02&0.08&(7)\\
1562.2689 &Fe~II& 7955&2.85E$-$03&0.11&(9)\\
1635.4003 &Fe~II& 7955&7.22E$-$02&0.2\phn &(10)\\
2715.2171 &Fe~II& 7955&4.70E$-$02&0.03&(10)\\
2740.3577 &Fe~II& 7955&2.49E$-$01&0.03&(7,8)\\
2756.5507 &Fe~II& 7955&3.06E$-$01&0.04&(7,8)\\
1641.7630 &Fe~II& 8391&4.85E$-$02&0.2\phn &(11)\\
2592.3176 &Fe~II& 8391&5.80E$-$02&0.08&(8)\\
2725.6909 &Fe~II& 8391&1.07E$-$02&0.04&(7)\\
2728.3465 &Fe~II& 8391&6.98E$-$02&0.03&(7,8)\\
2747.7940 &Fe~II& 8391&1.91E$-$01&0.03&(8)\\
2750.1341\tablenotemark{l} &Fe~II& 8391&3.27E$-$01&0.03&(7)\\
2583.3560 &Fe~II& 8680&8.80E$-$02&0.03&(8)\\
2611.8523 &Fe~II& 8680&1.12E$-$02&0.08&(8)\\
2731.5424 &Fe~II& 8680&3.13E$-$02&0.04&(7)\\
2737.7758 &Fe~II& 8680&6.90E$-$02&0.04&(7)\\
2747.2957 &Fe~II& 8680&3.48E$-$01&0.03&(7)\\
2749.9936\tablenotemark{l} &Fe~II& 8680&1.37E$-$01&0.03&(7,8)\\
2769.7520 &Fe~II& 8680&8.20E$-$03&0.04&(10)\\
2578.6936 &Fe~II& 8846&1.20E$-$01&0.2\phn &(10)\\
2594.5034\tablenotemark{g} &Fe~II& 8846&3.30E$-$02&0.04&(8)\\
2744.0081 &Fe~II& 8846&4.50E$-$01&0.03&(7)\\
2750.2987\tablenotemark{l} &Fe~II& 8846&1.32E$-$01&0.04&(7)\\
2762.6287 &Fe~II& 8846&3.20E$-$02&0.04&(7)\\
2019.4288 &Fe~II&15844&1.9E$-$02&0.2\phn &(11)\\
2041.3455 &Fe~II&15844&2.9E$-$02&0.2\phn &(11)\\
2064.3371 &Fe~II&15844&1.0E$-$02&0.2\phn &(11)\\
2162.7011 &Fe~II&15844&1.8E$-$02&0.2\phn &(10)\\
1876.8391 &Fe~II&16369&2.4E$-$02&0.2\phn &(11)\\
1878.3873 &Fe~II&16369&1.8E$-$03&0.3\phn &(11)\\
2033.0634 &Fe~II&16369&2.1E$-$02&0.2\phn &(11)\\
2051.6908 &Fe~II&16369&2.3E$-$02&0.2\phn &(11)\\
2176.1348 &Fe~II&16369&1.5E$-$02&0.2\phn &(11)\\
2088.2055 &Fe~II&18360&4.5E$-$02&0.2\phn &(11)\\
1877.4657 &Fe~II&20340&4.3E$-$02&0.08&(11)\\
2001.0254 &Fe~II&20340&4.8E$-$02&0.08&(11)\\
2221.0720\tablenotemark{m} &Fe~II&20340&3.10E$-$02&0.2\phn &(10)\\
2383.6237 &Fe~II&20340&1.61E$-$02&0.2\phn &(10)\\
2037.0908 &Fe~II&20516&2.6E$-$02&0.2\phn &(11)\\
1880.9722 &Fe~II&20805&2.9E$-$02&0.2\phn &(11)\\
1888.7342 &Fe~II&20805&7.6E$-$02&0.08&(11)\\
2011.3468 &Fe~II&20805&4.0E$-$02&0.2\phn &(11)\\
2234.6115 &Fe~II&20805&3.0E$-$02&0.2\phn &(11)\\
2214.3450 &Fe~II&21251&2.39E$-$02&0.2\phn &(10)\\
2346.0553 &Fe~II&21251&5.48E$-$02&0.11&(9)\\
2224.1775 &Fe~II&21711&2.2E$-$02&0.2\phn &(10)\\
2362.4468 &Fe~II&21711&2.01E$-$02&0.08&(10)\\
2619.8555 &Fe~II&22637&2.78E$-$02&0.2\phn&(10)\\
2632.3929 &Fe~II&22637&8.20E$-$02&0.08&(10)\\
2621.4769 &Fe~II&22810&3.40E$-$02&0.2\phn&(10)\\
2624.5078 &Fe~II&22939&1.99E$-$02&0.2\phn &(9)\\
2630.3724 &Fe~II&22939&8.57E$-$02&0.08&(10)\\
2627.2838 &Fe~II&23031&5.4E$-$02&0.2\phn &(10)\\
2630.8554 &Fe~II&23031&8.0E$-$02&0.2\phn & (10)\\
1785.2720 &Fe~II&23317&7.40E$-$01&0.2\phn&(9)\\ 
1786.7520 &Fe~II&23317&5.60E$-$01&0.2\phn&(9)\\ 
1788.0039 &Fe~II&23317&2.10E$-$01&0.2\phn&(11, 12)\\ 
1788.0780 &Fe~II&23317&1.50E$-$01&0.2\phn&(9)\\ 
2712.6451 &Fe~II&25428&5.6E$-$02&0.08&(10)\\
2728.1907 &Fe~II&25428&2.9E$-$02&0.08&(10)\\
2770.1736\tablenotemark{n} &Fe~II&25428&2.78E$-$02&0.08&(9)\\
2709.8587 &Fe~II&25787&6.4E$-$02&0.2\phn &(10)\\
2832.3938 &Fe~II&25787&1.38E$-$01&0.08&(10)\\
2836.5452\tablenotemark{f} &Fe~II&25787&9.3E$-$02&0.08&(10)\\
2713.1943 &Fe~II&25805&1.72E$-$02&0.12&(13)\\ 
2768.3197\tablenotemark{o} &Fe~II&26170&2.12E$-$01&0.03&(10)\\
2784.5119 &Fe~II&26170&1.02E$-$01&0.03&(10)\\
2754.1013 &Fe~II&26352&2.58E$-$01&0.12&(13)\\
2780.1194 &Fe~II&26352&9.3E$-$02&0.08&(10)\\
2638.4295 &Fe~II&26932&1.4E$-$01&0.2\phn &(10)\\
2775.5049 &Fe~II&26932&6.3E$-$02&0.2\phn &(10)\\
2841.4847 &Fe~II&26932&1.85E$-$01&0.08&(10)\\
2364.5330\tablenotemark{o} &Fe~II&27314&2.5E$-$02&0.2\phn &(10)\\
2665.4560 &Fe~II&27314&2.54E$-$01&0.03&(10)\\
2704.7908 &Fe~II&27314&1.51E$-$01&0.03&(10)\\
2667.4286\tablenotemark{o} &Fe~II&27620&2.65E$-$01&0.03&(10)\\
2717.0220 &Fe~II&27620&1.27E$-$01&0.08&(10)\\
2693.4008 &Fe~II&30388&1.83E$-$01&0.03&(10)\\
2685.5506 &Fe~II&30764&2.12E$-$01&0.03&(10)\\
2593.5595 &Fe~II&32875&3.16E$-$01&0.03&(10)\\
2626.2722 &Fe~II&32909&3.08E$-$01&0.03&(10)\\
2577.6353 &Fe~II&33466&1.58E$-$01&0.03&(10)\\
2588.7233 &Fe~II&33501&2.13E$-$01&0.03&(10)\\
2698.2611 &Fe~II&36126&9.0E$-$02&0.08&(10)\\
2607.2945 &Fe~II&36252&2.35E$-$01&0.03&(10)\\
2770.1502\tablenotemark{n} &Fe~II&36252&1.50E$-$02&0.2\phn &(10)\\
2286.8640 &Co~II& 3350&3.06E$-$01&0.03&(15)\\
2308.5700 &Co~II& 4028&2.54E$-$01&0.04&(15)\\
2354.1423 &Co~II& 4560&1.58E$-$01&0.03&(15)\\
1317.217  &Ni~II&     0&5.71E$-$02&0.05&(16)\\
1370.1323 &Ni~II&    0&5.88E$-$02&0.05&(16)\\
1741.5500 &Ni~II&    0&4.27E$-$02&0.04&(17)\\
1751.9100 &Ni~II&    0&2.77E$-$02&0.04&(17)\\
1381.2860\tablenotemark{p}  &Ni~II& 1506&1.15E$-$01&$>$0.3&(18)\\
1748.2820\tablenotemark{p}  &Ni~II& 1506&1.42E$-$01&$>$0.3&(18)\\
1754.8100 &Ni~II& 1506&1.58E$-$02&0.05&(19)\\
1788.4858 &Ni~II& 1506&2.50E$-$02&0.05&(19)\\
2131.9350 &Ni~II& 8393&4.27E$-$03&0.05&(19)\\
2166.2300 &Ni~II& 8393&1.66E$-$01&0.03&(19)\\
2217.1680 &Ni~II& 8393&3.16E$-$01&0.03&(19)\\
2223.6420 &Ni~II& 8393&7.59E$-$02&0.03&(19)\\
2316.7480 &Ni~II& 8393&1.86E$-$01&0.03&(19)\\
2169.7720 &Ni~II& 9330&1.06E$-$01&0.03&(19)\\
2175.3500 &Ni~II& 9330&1.37E$-$01&0.03&(19)\\
2211.0680 &Ni~II& 9330&4.14E$-$02&0.03&(19)\\
2225.5560 &Ni~II& 9330&1.22E$-$01&0.03&(19)\\
2270.9140 &Ni~II& 9330&1.54E$-$01&0.03&(19)\\
2303.7010 &Ni~II& 9330&1.65E$-$01&0.03&(19)\\
2159.4150 &Ni~II&10115&1.70E$-$02&0.03&(19)\\
2175.8240 &Ni~II&10115&1.26E$-$01&0.03&(19)\\
2207.4000 &Ni~II&10115&1.49E$-$01&0.03&(19)\\
2227.0200 &Ni~II&10115&1.00E$-$01&0.03&(19)\\
2265.1610 &Ni~II&10115&1.55E$-$01&0.03&(19)\\
2297.8490 &Ni~II&10115&1.42E$-$01&0.03&(19)\\
2346.1630 &Ni~II&10115&1.70E$-$02&0.03&(19)\\
2185.2860 &Ni~II&10663&2.03E$-$01&0.03&(19)\\
2202.0920 &Ni~II&10663&1.51E$-$01&0.03&(19)\\
2254.5460 &Ni~II&10663&2.23E$-$01&0.03&(19)\\
2298.1970 &Ni~II&10663&1.17E$-$01&0.03&(19)\\
2279.4740 &Ni~II&13550&1.94E$-$01&0.04&(19)\\
2297.2580 &Ni~II&13550&1.73E$-$01&0.03&(19)\\
2335.3020 &Ni~II&13550&6.56E$-$02&0.03&(19)\\
2395.2520 &Ni~II&13550&1.85E$-$01&0.03&(19)\\
2287.7860 &Ni~II&14995&1.46E$-$01&$>$0.3&(18)\\
2298.9750 &Ni~II&14995&2.20E$-$01&0.04&(19)\\
2376.1450 &Ni~II&14995&7.28E$-$02&0.03&(19)\\
2221.0900\tablenotemark{m} &Ni~II&23108&2.27E$-$01&$>$0.3&(18)\\
2300.3610 &Ni~II&23108&1.19E$-$01&$>$0.3&(18)\\
2300.8040 &Ni~II&23108&2.43E$-$01&$>$0.3&(18)\\
2277.9840 &Ni~II&23796&3.50E$-$01&$>$0.3&(18)\\
2175.7810 &Ni~II&24788&1.50E$-$02&$>$0.3&(18)\\
2181.1550 &Ni~II&24788&2.73E$-$01&$>$0.3&(18)\\
2186.1870 &Ni~II&25036&3.03E$-$01&$>$0.3&(18)\\
2288.3500 &Ni~II&25036&1.87E$-$01&$>$0.3&(18)\\
2211.7830 &Ni~II&29070&1.26E$-$01&$>$0.3&(18)\\
2277.1410 &Ni~II&29070&1.13E$-$01&$>$0.3&(18)\\
2309.2290 &Ni~II&29070&1.27E$-$01&$>$0.3&(18)\\
2341.9260 &Ni~II&29070&3.98E$-$01&$>$0.3&(18)\\
2256.8380 &Ni~II&29593&2.13E$-$01&$>$0.3&(18)\\
2304.5620 &Ni~II&29593&2.26E$-$01&$>$0.3&(18)\\
2337.4340\tablenotemark{q} &Ni~II&29593&3.06E$-$01&$>$0.3&(18)\\
2313.6290 &Ni~II&32499&3.48E$-$01&$>$0.3&(18)\\
2304.4610 &Ni~II&32523&2.46E$-$01&$>$0.3&(18)\\
2345.9890 &Ni~II&32523&3.98E$-$01&$>$0.3&(18)\\
1358.7730 &Cu~II&   0&2.63E$-$01&0.08&(1)\\
1367.9509 &Cu~II&   0&1.79E$-$01&0.08&(1)\\
1472.3951 &Cu~II&   0&2.17E$-$02&$>$0.3&(1)\\
2136.6545 &Cu~II&21928&3.89E$-$01&0.01&(20)\\
2192.9531 &Cu~II&22847&3.02E$-$01&0.04&(20)\\
2139.2477 &Zn~I &    0&1.47E00&0.01&(1)\\
2026.1370 &Zn~II&    0&6.30E$-$01&\nodata&(21)\\ 
2062.6604 &Zn~II&    0&3.09E$-$01&\nodata&(21)\\ 

\enddata
{\tabletypesize{\scriptsize}
\tablerefs{
\tablenotetext{a}{Except for reference (1), most of the transition $f$-values were
obtained from either the National Institute of Standards and Technology (NIST)
website, \url{https://physics.nist.gov/PhysRefData/ASD/lines_form.html} or
The Atomic Line List v 2.05b21 \url{http://www.pa.uky.edu/~peter/newpage/}. Original
determinations are as follows: 
(1) Values and references listed in Morton (2003),
(2) Tachiev \& Froese Fischer (2002),
(3) Butler \& Zeippen (1991)
(4) Nilsson et al. (2006),
(5) Den Hartog et al. (2011),
(6) Kurucz (1990) with additional data downloaded
from \url{http://kurucz.harvard.edu/linelists.html} on December 11, 2012. 
(7) Bergeson et al. (1996),
(8) Schnabel et al. (2004),
(9) Tayal \& Zatsarinny (2018),
(10) Fuhr \& Wiese (2006),
(11) Raassen \& Uylings (1998),
(12) Johansson et al. (1995) 
(13) Sikström et al. (1999),
(14) Mullman et al. (1998),
(15) Salih et al. (1985),
(16) Jenkins \& Tripp (2006) 
(17) Fedchak et al. (2000), 
(18) Kurucz (2012),
(19) Fedchak \& Lawler (1999),
(20) van Hoof (2017, 2018): no original reference specified.
(21) Kisielius et al. (2015)}}
\tablenotetext{b}{These strong transitions for nitrogen are very strongly saturated and not 
properly resolved in the FUSE spectrum.  The equivalent width of the \ion{N}{1} feature was 
used only to verify the derivation of the saturation factors (s.f., see Eq.~\protect\ref{eqn: s.f.}) 
for optical pumping, and the very strong absorptions by \ion{N}{2} indicated the importance of 
ionization processes beyond that provided by photons with energies below the ionization 
potential of hydrogen.}
\tablenotetext{c}{Very strongly saturated lines shown in Figure~\protect\ref{fig:OI_wideplot}.  
The very weak intersystem line at 1355.598\,\AA\ was used to derive $N$(\ion{O}{1}).}
\tablenotetext{d}{This line is not useful in the present study, since it occurs near the bottom of 
a very strong absorption by the $\lambda\lambda$1335.649 1335.708 lines of excited 
\ion{C}{2}.  However, in future investigations of other stars, this \ion{Cl}{1} line may not suffer 
from this interference.}
\tablenotetext{e}{The \ion{Cr}{2} lines at 2677.956 and 2677.954\,\AA\ are severely blended.  
However, the simultaneous fitting of the various lines from different levels performed by the 
Owens analysis allows for this superposition.}
\tablenotetext{f}{The \ion{Cr}{2} line at 2836.463\,\AA\ and the \ion{Fe}{2} line at 
2836.5452\,\AA\ are blended, but our analysis fitted these features simultaneously.}
\tablenotetext{g}{The \ion{Mn}{2} line at 2594.4990\,\AA\ and \ion{Fe}{2} line at 
2594.5034\,\AA\  are blended, but our analysis fitted these features simultaneously.  These fits 
are substantiated by 3 other lines that are free from interference for both \ion{Mn}{2} and 
\ion{Fe}{2} in the same energy levels.} 
\tablenotetext{h}{This line is too weak to see in our spectrum.}
\tablenotetext{i}{This line was not considered in our analysis because it was too close to the 
edge of our spectrum.}
\tablenotetext{j}{This line coincided with a detector flaw, so it was not considered in our 
analysis.}
\tablenotetext{k}{The \ion{Fe}{2} lines at 2348.835 and 2349.022\,\AA\ are blended, but our 
analysis fitted these features simultaneously.}
\tablenotetext{l}{The \ion{Fe}{2} lines at 2749.994, 2750.134, and 2750.299\,\AA\ overlap each 
other, but our analysis fitted these features simultaneously.}
\tablenotetext{m}{The \ion{Fe}{2} line at 2221.072\,\AA\ and the \ion{Ni}{2} line at 
2221.090\,\AA\ are blended, but our analysis fitted these features simultaneously.  Other lines 
of \ion{Fe}{2} in this level substantiate the fit.}
\tablenotetext{n}{The \ion{Fe}{2} lines at 2770.1736 and 2770.1502\,\AA\ are blended, but our 
analysis fitted these features simultaneously.  Added confidence in these fits comes from other 
lines of \ion{Fe}{2} from the same levels.}
\tablenotetext{o}{The lines of \ion{Fe}{2} at 2364.533, 2667.429, and 2768.320\,\AA\ are 
blended with lines of \ion{Fe}{2} originating from much higher levels, which are too weak to 
matter.}
\tablenotetext{p}{These two lines for \ion{Ni}{2} were not used because two other lines with 
more accurate $f$-values were available for this level.}
\tablenotetext{q}{An absorption line is present at this location, but it indicates an unreasonably 
high column density for this level.  There must be some other unidentified transition that 
coincides with this one.}
}

\end{deluxetable}

\end{document}